\documentclass[a4paper,11pt]{article}
\pdfoutput=1 

\usepackage{jheppub} 

\usepackage[T1]{fontenc} 
\usepackage{amsfonts}
\usepackage{braket}
\usepackage{bbold}
\usepackage{amsmath}
\usepackage{amssymb}
\usepackage{tcolorbox}
\usepackage{mathtools}
\usepackage{caption}
\usepackage{subcaption}
\newsavebox{\imagebox}
\usepackage{ulem}

\def\myfloor#1{\left\lfloor {#1} \right \rfloor}

\newcommand{\textchg}{}
\newcommand{\textdel}[1]{}

\makeatletter
\newcommand{\subalign}[1]{%
  \vcenter{%
    \Let@ \restore@math@cr \default@tag
    \baselineskip\fontdimen10 \scriptfont\tw@
    \advance\baselineskip\fontdimen12 \scriptfont\tw@
    \lineskip\thr@@\fontdimen8 \scriptfont\thr@@
    \lineskiplimit\lineskip
    \ialign{\hfil$\m@th\scriptstyle##$&$\m@th\scriptstyle{}##$\hfil\crcr
      #1\crcr
    }%
  }%
}
\makeatother

\title{\boldmath Quantum Information in Holographic Duality}
\author{Bowen Chen,}
\emailAdd{chenbw95@gmail.com}
\author{Bart{\l}omiej Czech, and}
\emailAdd{bartlomiej.czech@gmail.com}
\author{Zi-zhi Wang}
\emailAdd{wang-zz17@mails.tsinghua.edu.cn}
\affiliation{Institute for Advanced Study, Tsinghua University, Beijing 100084, China}

\abstract{We give a pedagogical review of how concepts from quantum information theory build up the gravitational side of the AdS/CFT correspondence. The review is self-contained in that it only presupposes knowledge of quantum mechanics and general relativity; other tools---including holographic duality itself---are introduced in the text. We have aimed to give researchers interested in entering this field a working knowledge sufficient for initiating original projects.

The review begins with the laws of black hole thermodynamics, which form the basis of this subject, then introduces the Ryu-Takayanagi proposal, the JLMS relation, and subregion duality. We discuss tensor networks as a visualization tool and analyze various network architectures in detail. Next, several modern concepts and techniques are discussed: R{\'e}nyi entropies and the replica trick, differential entropy and kinematic space, modular Berry phases, modular minimal entropy, entanglement wedge cross-sections, bit threads, and others. We discuss the extent to which bulk geometries are fixed by boundary entanglement entropies, and analyze the relations such as the monogamy of mutual information, which boundary entanglement entropies must obey if a state has a semiclassical bulk dual. We close with a discussion of black holes, including holographic complexity, firewalls and the black hole information paradox, islands, and replica wormholes.
}

\begin{document} 
\maketitle
\flushbottom
\normalem

\section{Preliminaries}
\label{sec:intro}

The modern age has shown a remarkable momentum for deconstruction. We deconstruct social mores and works of literature, take a {\it longue dur{\`e}e} view of history and ask {\it why} questions about biological organisms. But nowhere is the trend more evident than in physics, where breaking successive levels of putative indivisibility of matter (from the atom to nucleons to quarks) has been complemented with an understanding of forces and heat as manifestations of particle exchanges and collisions. It is perhaps unsurprising, therefore, that the very stage on which physics plays out---space and time---has also become subject to deconstruction.

Yet the ingredients from which space and time seem to emerge certainly merit surprise. They are concepts in quantum information theory, most notably quantum entanglement and entropy. The information-theoretic toolkit initially made its way to gravitation because of a remarkable parallel between the laws of thermodynamics and the behavior of black holes \cite{Bardeen:1973gs}, which identified the entropy of a black hole with its horizon area \cite{Bekenstein:1973ur, Bekenstein:1974ax, Hawking:1974rv, Hawking:1974sw}. This is a powerful hint about the fundamental nature of gravity because thermodynamics is a forerunner of microscopic-level statistical mechanics and because black holes are, loosely speaking, bound states of gravity, which is the dynamics of space and time. Half a century after the formulation of black hole thermodynamics, we are still deciphering the hint.

Remarkable advances have been achieved, primarily in the last fifteen years. A key catalyst of progress was the formulation of the AdS/CFT correspondence \cite{Maldacena:1997re, witten98, Aharony:1999ti}---a {\it holographic} duality between a theory of gravity and a non-gravitational theory in one less macroscopic dimension---which in principle translates any gravitational question to the language of a well-defined and well-understood quantum field theory. On a technical level, holographic duality revealed a concrete set of microscopic degrees of freedom---those of the dual conformal field theory---whose statistical mechanics could in a thermodynamic limit produce gravity. Conceptually, it delivered several key insights:
\begin{itemize}
\item Identifying areas with entropies, as in black hole thermodynamics, is natural in holographic duality: ordinary, field-theoretic entanglement entropy in the CFT manifests itself as area on the AdS side of the duality \cite{Ryu:2006bv, Ryu:2006ef, Hubeny:2007xt}; see Section~\ref{sec:rt}.
\item The identification is not endemic to black holes, but applies in empty space and in horizonless geometries sourced by generic matter configurations.
\item Therefore, an AdS geometry is a map of entanglement of the CFT quantum state \cite{marksessay, bianchimyers, erepr}. This view is sometimes dubbed {\it geometrization of entanglement}; see also Section~\ref{sec:tns} for a parallel with tensor networks pioneered in condensed matter physics.
\end{itemize}
These insights and their consequences---the subject of the present review---have now become an integral part of the AdS/CFT correspondence. A central question for the future is the extent to which they carry over to gravitational systems with non-AdS boundary conditions.

We assume familiarity with quantum mechanics and general relativity. Readers versed in the AdS/CFT correspondence will have an easier time reading the text, though we have made an effort to make it accessible to those unfamiliar with holography. The review necessarily skims the surface of what has now become an enormous subject. We hope to pique the reader's interest enough so she will delve further. Other reviews with closely related scope include \cite{Iqbal:2016qyz, Rangamani:2016dms, mattsreview, Harlow:2018fse}. Throughout the text we set the speed of light $c \equiv 1$.

\subsection{Black hole thermodynamics}
\label{sec:bhthermo}

Classically, black holes eat up everything that comes in contact with them and grow forever. This seems to suggest that the mass of a black hole should be a non-decreasing function of time, but that assertion is incorrect. A series of works in the early 1970s \cite{Penrose:1969pc, Christodoulou:1970wf, Penrose:1971uk, Christodoulou:1972kt} revealed that in the presence of other charges one can in fact extract energy from a black hole using only classical operations. It is instructive to inspect the simplest example, which occurs for a rotating (Kerr) black hole \cite{Penrose:1971uk}. The following review is adapted from Section~1.5.3 of \cite{Jacobson:1996utr}.

The black hole has (at least) two Killing (symmetry-generating) vector fields, which at infinity generate time translation and rotation about the black hole's axis of symmetry. We will call them $\psi_t$ and $\psi_\theta$, though we should remember that in any reasonable coordinate system they will not equal $\partial_t$ or $\partial_\theta$ except at infinity. In the interior of spacetime, the presence of the black hole affects their behavior drastically, such that $\psi_t$ is not even timelike inside the event horizon. Most pertinent for energy extraction, the rotation of the black hole renders $\psi_t$ spacelike even outside the event horizon, in a locus called {\it ergoregion}.

The fact that the black hole is rotating means that some linear combination $\chi = \psi_t - \Omega \psi_\theta$ is future-pointing and light-like at the horizon; it generates translations along null rays that never leave the horizon. (When the black hole's angular momentum is in the direction of increasing $\theta$ at infinity, the coefficient of $\psi_\theta$ is negative so $\Omega > 0$.) We will drop a particle with momentum 
\begin{equation}
p_{\rm tot} = p_1 + p_2,
\end{equation}
which carries energy $p_{\rm tot} \cdot \psi_t = E_0$ and angular momentum $p_{\rm tot} \cdot \psi_\theta = \Delta J$. By an ingenuous mechanism, we will arrange for our particle to split into two just outside the black hole. One decay product with momentum $p_1$ will fall into the black hole and change its energy by $\Delta M = p_1 \cdot \psi_t$ and angular momentum by $\Delta J = p_1 \cdot \psi_\theta$ while the other one will carry out to infinity energy $p_2 \cdot \psi_t = E_0 - \Delta M$ and zero angular momentum ($p_2 \cdot \psi_\theta = 0$). The outside world extracts energy $-\Delta M$ from the black hole, so we want $\Delta M$ to be negative.

Now $p_1 \cdot \chi = \Delta M - \Omega \Delta J \geq 0$ or else particle 1 will be tachyonic. This imposes a limit on how negative $\Delta M$ can become, relative to the angular momentum dropped in the black hole:
\begin{equation}
- \!\Delta M \leq - \Omega \Delta J
\end{equation}
The most efficient process, in terms of energy extraction per deposited angular momentum, is when $\Delta M = \Omega \Delta J$. In this case, particle 1 is lightlike and travels along an orbit of $\chi$---that is, remains forever on the event horizon. Note that energy extraction relies on decreasing the angular momentum of the black hole ($\Delta J \leq 0$) and can only continue until the black hole ceases to rotate: $\Omega = 0$ implies $-\Delta M \leq 0$.

A similar analysis for charged and rotating (Kerr-Newman) black holes reveals the following condition for optimal energy extraction per charge and angular momentum deposited:
\begin{equation}
\Delta M = \Omega \Delta J + \Phi \Delta Q
\label{reversible}
\end{equation}
Here $\Phi$ is the potential difference between the black hole horizon and infinity and $\Delta Q$ is the deposited charge. It turns out that this is also the condition for the horizon area of the black hole not to change under an infinitesimal variation of its charges. In the Kerr case, this is easy to see from the first order (shear-free) Raychaudhuri's equation
\begin{equation}
\frac{d}{d\lambda} \rho = \frac{1}{2} \rho^2 + 8 \pi G T_{ab} k^a k^b
\label{raychaudhuri}
\end{equation}
for the expansion of an infinitesimal cross-sectional area $\rho = d \log (\delta A) / d\lambda$ of a lightfront generated by $k^a$. Applying (\ref{raychaudhuri}) to the event horizon, we see that $\rho=0$ (the horizon area remains constant) when the stress-energy tensor is that of a massless particle traveling along the horizon ($T_{ab}k^a k^b \propto k_a k_b k^a k^b = 0$), i.e. when $\Delta M = \Omega \Delta J$. 

Assuming the null energy condition ($T_{ab} k^a k^b \geq 0$ for all null vectors $k^a$) and no naked singularities, equation~(\ref{raychaudhuri}) also tells us that all other classical processes increase the area of the black hole. Any growth in horizon area therefore marks a (classical) irreversibility. Conversely, the only classically reversible processes obey (\ref{reversible}) and keep the horizon area constant. This is a striking analogy with the second law of thermodynamics, wherein the horizon area plays the role of entropy:
\begin{equation}
dS \geq 0 \quad \leftrightarrow \quad dA \geq 0 
\qquad\qquad \textrm{(2nd law)}
\label{2ndlaw}
\end{equation}
We could augment the reversibility condition~(\ref{reversible}) to a full-fledged first law of thermodynamics by including an entropy term, represented by a change in horizon area:
\begin{equation}
\Delta E = T \Delta S + \sum \textrm{(force)} \textrm{(displacement)}
\quad \leftrightarrow \quad 
\Delta M = \kappa \Delta A + \Omega \Delta J + \Phi \Delta Q
\qquad \textrm{(1st law)}
\label{1stlaw}
\end{equation}
This identification will make sense if we can interpret:
\begin{equation}
T\Delta S \quad \leftrightarrow \quad \kappa \Delta A
\equiv \left( \frac{\partial A}{\partial M} \right)^{-1} \Delta A
\label{hawkingt}
\end{equation}
With appropriate units for $\Delta A$, $\kappa$ becomes {\it surface gravity}---the acceleration (relative to the horizon-generating Killing time) of a particle that hovers just outside the horizon. The same $\kappa$ also turns out to be the temperature at which black holes radiate, as famously discovered by Hawking \cite{Hawking:1974rv, Hawking:1974sw}. This discovery strongly suggests that the relation between the kinematics of black holes and the laws of thermodynamics \cite{Bardeen:1973gs} is more than an accidental analogy.

Hawking's calculation of black hole radiation is important for this review, so we briefly sketch it below.

\subsection{Entangled states and Unruh and Hawking temperatures}
\label{sec:unruhhawking}

A sufficiently small neighborhood of a black hole horizon looks flat. We will first identify a notion of temperature in flat space and only then relate the calculation to black holes. Since we are using the more exotic black holes to motivate an information-theoretic inquiry into all gravitating systems, of which flat space seems the most mundane, it is ironic that understanding (\ref{hawkingt}) requires a U-turn back to flat space. Indeed, the requisite phenomenon in flat space---the Unruh effect \cite{Davies:1974th, Unruh:1976db}---was discovered later than Hawking radiation \cite{Hawking:1974rv, Hawking:1974sw} even though the latter is a special case of the former. This strengthens the significance of black hole thermodynamics as a hint: black holes merely command our attention to universal aspects of spacetime, which we are prone to overlook in more familiar settings.

\paragraph{Vacuum as an entangled state}
\begin{figure}
     \centering
     \savebox{\imagebox}{\includegraphics[width=.3\textwidth]{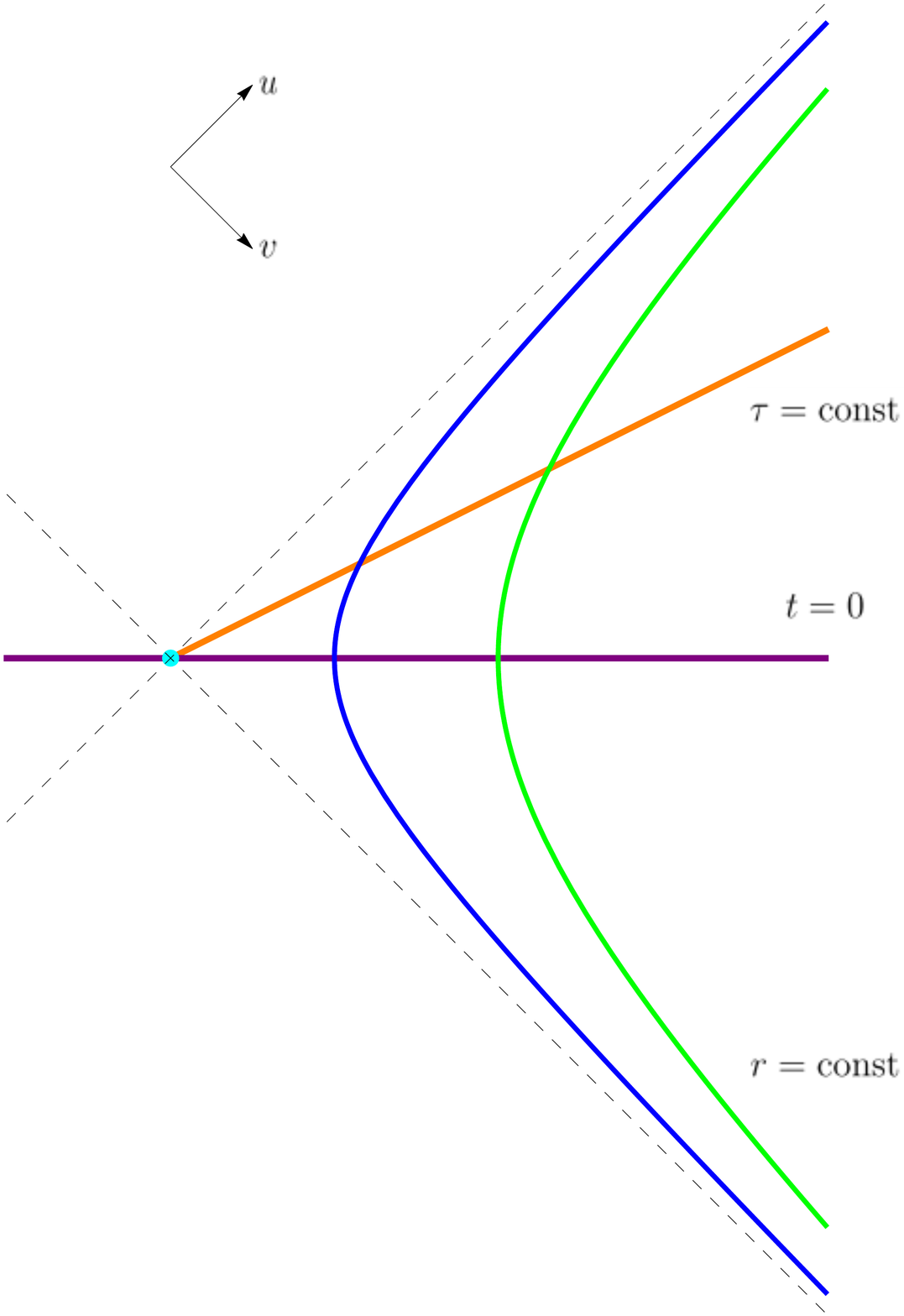}}
     \begin{subfigure}[t]{0.5\textwidth}
         \centering
         \usebox{\imagebox}
         \label{flatrindler2D}
     \end{subfigure}
\hfill
     \begin{subfigure}[t]{0.45\textwidth}
         \centering
         \raisebox{\dimexpr.5\ht\imagebox-0.50\height}{
         \includegraphics[width=0.7\textwidth]{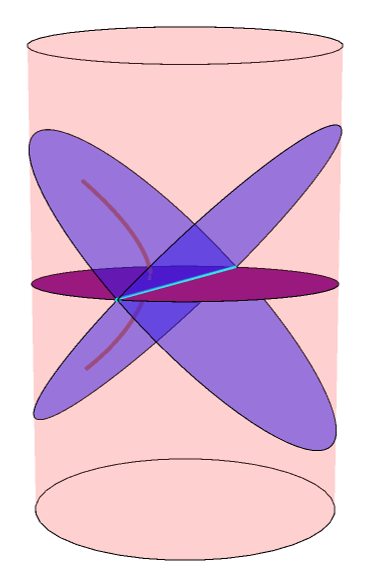}}
         \label{flatrindler3D}
     \end{subfigure}
        \caption{Rindler wedges in (left) 1+1-dimensional flat space and (right) 2+1-dimensional AdS; compare with Figure~\ref{fig:cylinder}. The colored hyperbolae are trajectories of accelerating observers. The Rindler horizons---the origin in 1+1 dimensions and a line in AdS$_3$---are marked in cyan.}
        \label{flatrindler}
\end{figure}
Consider two-dimensional flat space in light-like coordinates $ds^2 = -dt^2 + dx^2 \equiv du dv$. The coordinates divide spacetime into four quadrants, distinguished by the signs of $u$ and $v$; see the left panel of Figure~\ref{flatrindler}. The top (bottom) regions are the future (past) wedges; they will play the roles of the black (white) hole interiors when we return to black hole settings. More relevant for us are the left and right regions, which are called Rindler wedges; they will eventually become the two exterior regions of a maximally extended black hole spacetime. 

To stay in a Rindler wedge, you have to accelerate---same as for black holes, if you don't wish to fall in. Uniformly accelerated trajectories in flat space take the form:
\begin{equation}
x = a^{-1} \cosh {a\tau} + x_0 \qquad {\rm and} \qquad t = a^{-1} \sinh {a\tau} + t_0
\label{accelerated}
\end{equation}
With $(t_0, x_0)$ at the origin, the trajectories become
\begin{equation}
u = (a\sqrt{2})^{-1} e^{a\tau} \qquad {\rm and} \qquad v = (a\sqrt{2})^{-1} e^{-a\tau}\,,
\end{equation}
and attain a nice set of properties. They always stay inside the right Rindler wedge, trajectories with different $a$ do not intersect, and their union covers the whole wedge. This means that we can use 
\begin{equation}
r = a^{-1} \label{flatrindlerr}
\end{equation}
and $\tau$ as coordinates, which cover only one Rindler wedge: 
\begin{equation}
ds^2 = -r^2 d\tau^2 + dr^2
\label{flatrindlerm}
\end{equation}
(We display equal $\tau$ and equal $r$ slices in Figure~\ref{flatrindler}.) 
It also means that for observers moving along constant $r$ slices---uniformly accelerated observers---the origin appears as a horizon; they can neither see nor impact what happens in the left Rindler wedge. Unruh explained \cite{Unruh:1976db} that such an observer, if moving in the global vacuum $|0\rangle$, would detect a temperature proportional to her acceleration:
\begin{equation}
T = \hbar a /2 \pi \approx 3.98 \cdot 10^{-20} K \cdot ({a}/{g})
\label{unruheff}
\end{equation} 
The first step in a rough argument for (\ref{unruheff}) starts by recognizing $|0\rangle$ as an entangled state of $\mathcal{H}_L$ and $\mathcal{H}_R$---the Hilbert spaces of the left and right Rindler wedges.

\begin{figure}
     \centering
     \savebox{\imagebox}{\includegraphics[width=.45\textwidth]{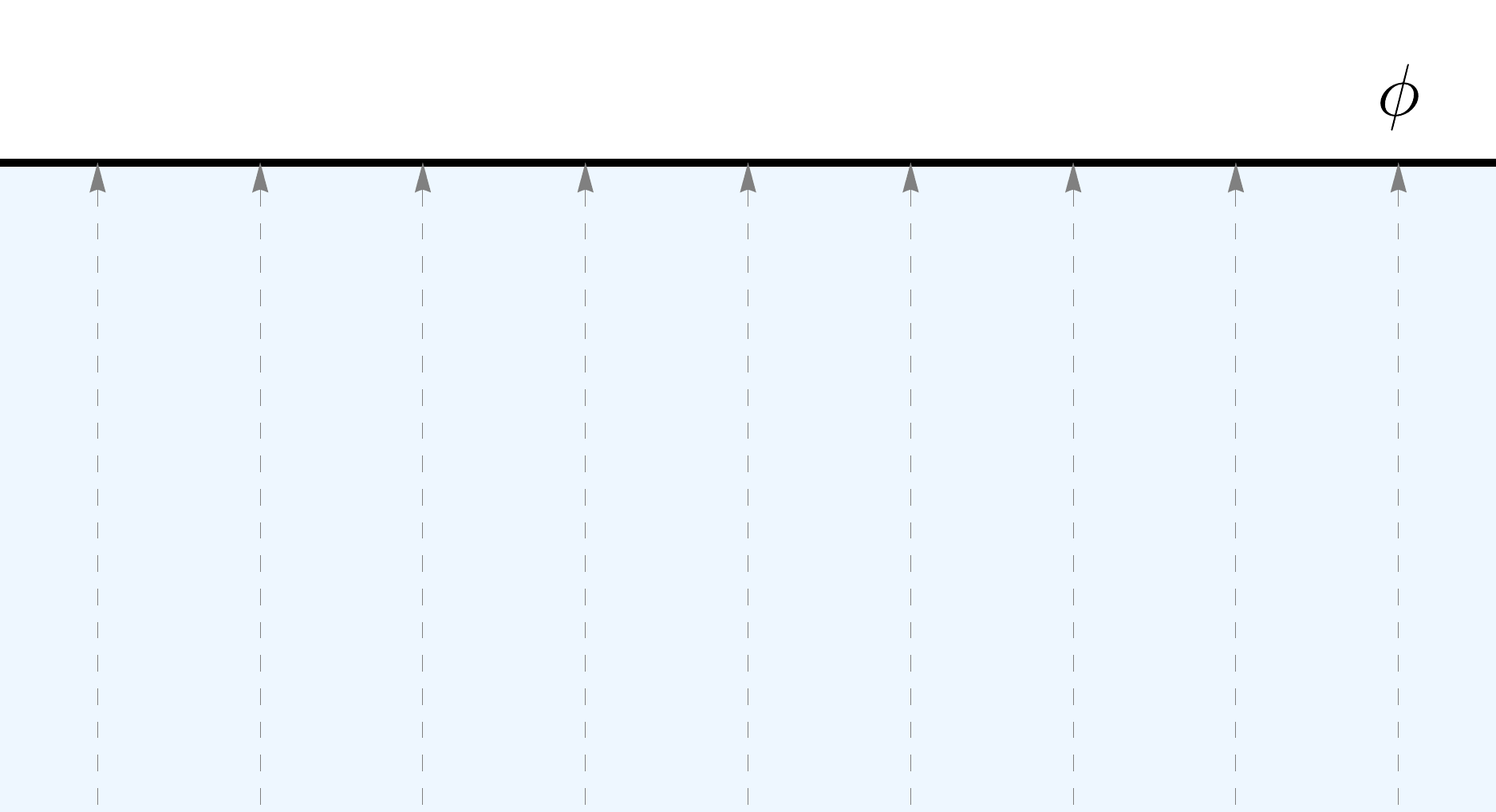}}
     \begin{subfigure}[t]{0.45\textwidth}
         \centering
         \raisebox{\dimexpr.5\ht\imagebox-0.5\height}{
         \includegraphics[width=\textwidth]{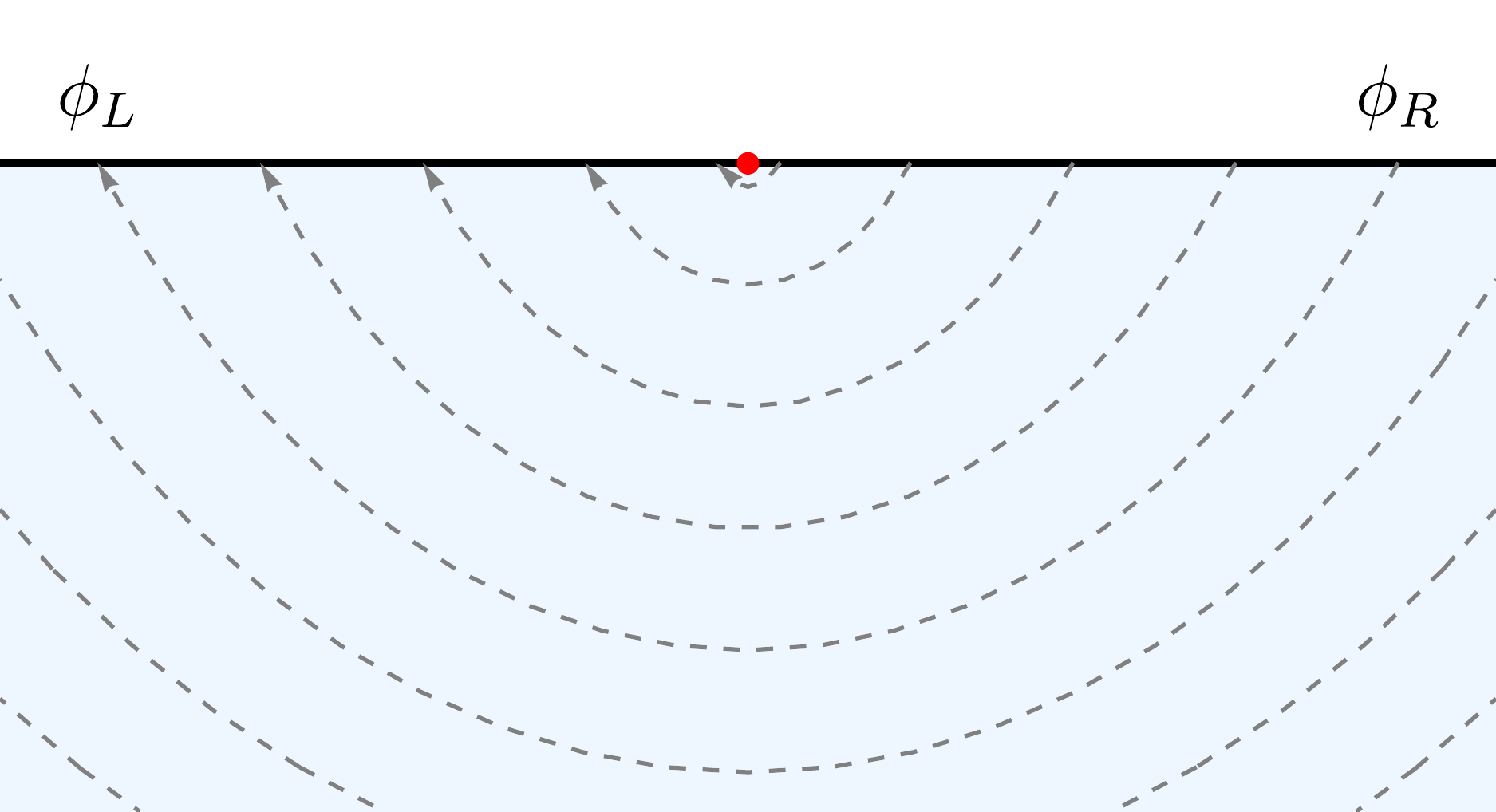}}
         \caption{}
         \label{pathintslicing1}
     \end{subfigure}
     \hfill
     \begin{subfigure}[t]{0.5\textwidth}
         \centering
         \usebox{\imagebox}
         \caption{}
         \label{pathintslicing2}
     \end{subfigure}
        \caption{Two ways of slicing the same Euclidean path integral over the lower half-plane in equation~(\ref{rindlerdecomp}).}
        \label{pathintslicing}
\end{figure}

Recall that the vacuum is prepared by a Euclidean path integral over the lower half-plane. From equation~(\ref{accelerated}), we see that the analytic continuation $\tau \to i\tau$ makes $t$ imaginary and takes us exactly to the Euclidean plane. But the generator of accelerated trajectories $\partial_\tau$ now becomes a generator of rotations about the origin, as is easily seen from the sines and cosines that replace hyperbolic functions in (\ref{accelerated}). The analytic continuation turns boosts into rotations. Suppose we wanted to compute some matrix element of the Euclidean 180$^\circ$ rotation. In path integral language, we would write
\begin{equation}
\langle \phi_L | \exp{(-\pi H_\tau)} | \phi_R \rangle = 
\int_{\phi(\tau=0) = \phi_R}^{\phi(\tau=-\pi) = \phi_L} \mathcal{D}\phi\, e^{-S_E} \propto 
\int_\textrm{lower half-plane}^{\phi(t=0) = (\phi_L, \phi_R)} \mathcal{D}\phi\, e^{-S_E} =
\langle \phi_L \phi_R | 0 \rangle
\label{rindlerdecomp}
\end{equation}
The last equality recognizes the path integral as one that prepares the global vacuum; see Figure~\ref{pathintslicing}. 

Equation~(\ref{rindlerdecomp}) rewrites the global vacuum $|0\rangle$ as an {\it entangled state} between the Hilbert spaces of the left and right Rindler wedges:
\begin{equation}
|0\rangle = \sum_{\tilde\phi_L, \tilde\phi_R} \langle \tilde\phi_L \tilde\phi_R | 0 \rangle \, 
|\tilde\phi_L \rangle \otimes | \tilde\phi_R \rangle\,,
\label{vacuument}
\end{equation}
where $|\tilde{\phi}_{L,R}\rangle$ is some basis for $\mathcal{H}_{L,R}$. Of course, such a rewriting presumes that the global Hilbert space can be factorized as $\mathcal{H} = \mathcal{H}_L \otimes \mathcal{H}_R$, an assumption that is often untrue \cite{chr, djordje, donnellywall1, donnellywall2, ronaksandip} and always a source of subtleties \cite{yuji, marksrednicki, jenniferdjordje}. We will cavalierly ignore this issue except for a brief comment in Section~\ref{sec:jlms}.

\paragraph{Quantum entanglement}
A lot has been written about the meaning of quantum entanglement, which we are encountering here for the first time. It effects `spooky action at a distance' \cite{spooky}, enables quantum computing \cite{jozsalinden, guifreqc}, and reveals that scientists `know less than nothing' \cite{lessthan0}. It is also the central concept of the present review. For a minimal discussion of entanglement, consider how an observer with access only to Hilbert space $\mathcal{H}_R$ probes a general state
\begin{equation}
|W\rangle = \sum_{ij} w_{ij} |\tilde{i}\rangle_L \otimes |{j}\rangle_R
\qquad \qquad 
{\rm with} \quad |\tilde{i}\rangle \in \mathcal{H}_L 
\quad {\rm and} \quad |{j}\rangle \in \mathcal{H}_R
\label{entangledstate}
\end{equation}
Access only to $\mathcal{H}_R$ means that all operators wielded by the observer are of the form $\mathbb{1}_L \otimes \mathcal{O}_R$, so the observer's experimental outcomes can be rewritten as:
\begin{align}
\!\!\langle W | \mathbb{1}_L \otimes \mathcal{O}_R | W \rangle = 
{\rm Tr}\, \mathbb{1}_L \otimes \mathcal{O}_R | W \rangle \langle W | =
{\rm Tr}_R\, {\rm Tr}_L \mathbb{1}_L \otimes \mathcal{O}_R | W \rangle \langle W | & =
{\rm Tr}_R\, \mathcal{O}_R {\rm Tr}_L | W \rangle \langle W | \nonumber \\
& \equiv {\rm Tr}_R\, \mathcal{O}_R\, \rho_R
\end{align}
Because of correlations with the unobserved $\mathcal{H}_L$, the experimenter effectively interacts with a mixed state
\begin{equation}
\rho_R = {\rm Tr}_L | W \rangle \langle W | = 
\sum_{jk} |j \rangle_R \left( \sum_i w_{ij} w_{ik}^* \right)
{}_R\langle k |,
\label{traceout}
\end{equation}
which is called the reduced density matrix of $R$. Note that if we treat the amplitudes in expansion~(\ref{entangledstate}) as a matrix $(W)_{ij} = w_{ij}$ then $\rho_R = W W^\dagger$. We might also form the reduced density matrix $\rho_L = {\rm Tr}_R | W \rangle \langle W |$ and find $\rho_L= W^\dagger W$. 

For future convenience, we mention here two important concepts in studying quantum entanglement. The first one is {\it purification}. While in equations~(\ref{entangledstate}-\ref{traceout}) we went from $|W\rangle$ to $\rho_R$, it is often useful to go in reverse and find a $|W\rangle \in \mathcal{H}_L \otimes \mathcal{H}_R$ such that ${\rm Tr}_L | W \rangle \langle W |$ equals a given $\rho_R$; this $|W\rangle$ is a purification. Note that finding a purification amounts to solving $\rho_R = W W^\dagger$, but if $W$ solves $\rho_R = W W^\dagger$ then so does $W V$ with $V$ a unitary matrix. We see that purifications are not unique. When the spectrum of $\rho_R$ has no degeneracies, its purifications are parameterized by unitary transformations acting on $\mathcal{H}_L$.

The second useful concept is a particularly important way to quantify the magnitude of bipartite quantum entanglement. It is the von Neumann entropy of $\rho_R$:
\begin{equation}
S(R) = - {\rm Tr}_R\, \rho_R \log \rho_R
\label{vonneumann}
\end{equation}
This quantity will play a fundamental role in our discussion of AdS/CFT.

\paragraph{Unruh effect} We are now ready to return to equation~(\ref{unruheff}). The accelerating observers moving along trajectories~(\ref{accelerated}) have access only to the Hilbert space of the right Rindler wedge $\mathcal{H}_R$. The global state in which they perform their experiments is of the form~(\ref{entangledstate}). Therefore, they effectively interact with a reduced density matrix, which in terms of the matrix of amplitudes $W$ takes the form $WW^\dagger$. Looking back to equation~(\ref{rindlerdecomp}), we recognize that $W \propto \exp{(-\pi H_\tau})$ and the mixed state seen by the accelerated observers is:
\begin{equation}
\rho_R = \frac{1}{\mathcal{Z}} \exp{(-2 \pi H_\tau}) 
\label{thermal}
\end{equation}
This is a thermal state with {\it temperature} $1/(2\pi)$. We call the coefficient of proportionality $\mathcal{Z}$ because the trace of the thermal state is the partition function; including it in $\rho_R$ ensures that $\langle 0 | \mathbb{1} | 0 \rangle = {\rm Tr}\, \rho_R = 1$. 

Accelerating observers feel a temperature because they move along trajectories generated by $H_\tau$, with respect to which the ground state is thermal. Although the `temperature' in (\ref{thermal}) is dimensionless, any one observer feels a physical (dimensionful) temperature. Its magnitude is set by the proper time along the trajectory, which is related to $\tau$ as:
\begin{equation}
d(\textrm{proper time}) = r d\tau = a^{-1} d\tau 
\end{equation}
This explains equation~(\ref{unruheff}). The factor $\hbar$ simply converts an inverse length into energy; unlike $c$, we keep it explicit because the limit $\hbar \to 0$ is illuminating. Steam engineers might want to include a factor of $k_B$.

\paragraph{Hawking effect}
We now return to black holes. A sufficiently small neighborhood of the event horizon is flat. For example, substituting 
\begin{equation}
r = \sqrt{8GM (R - 2GM)} \qquad {\rm and} \qquad \tau = t / 4GM
\end{equation}
in the flat space Schwarzschild metric 
\begin{equation}
ds^2 = - \left(1 - \frac{2GM}{R}\right) dt^2 + \left(1 - \frac{2GM}{R}\right)^{-1} dR^2 + R^2 d\Omega^2
\label{schwarzschild}
\end{equation} 
gives
\begin{equation}
ds^2 = - r^2 d\tau^2 + dr^2 + 
\left( 2GM + \frac{r^2}{8GM}\right)^2 d\Omega^2\,,
\label{schwarzrindler}
\end{equation}
which is (\ref{flatrindlerm}) with extra dimensions. The only source of non-flatness is the $r$-dependence of $g_{\Omega\Omega}$. If the state of quantum fields near the horizon is the vacuum, it must be entangled just like (\ref{vacuument}). Observers hovering at sufficiently small $R=R_0$ (sufficiently so $g_{\Omega\Omega}\approx const.$ in equation~\ref{schwarzrindler}) must therefore feel a temperature just like accelerating observers in Rindler space do. This temperature diverges when $R_0 \to 2GM$ as:
\begin{equation}
T(R_0) = \frac{\hbar a(R_0)}{2\pi} = \frac{\hbar}{2\pi r(R_0)} = \frac{\hbar}{8\pi GM \sqrt{R_0/2GM - 1}}
\label{localdivhawking}
\end{equation}
Away from the horizon, the temperature dependence can be read off by redshifting (\ref{localdivhawking}) using
\begin{equation}
T(R) = T(R_0)\, \frac{\sqrt{1-2GM/R_0}}{\sqrt{1-2GM/R}}\,,
\label{redshift}
\end{equation}
which gives:
\begin{equation}
T(R) 
= \frac{\hbar}{8\pi GM} \frac{1}{\sqrt{1-2GM/R}} 
\end{equation}
Applying~(\ref{redshift}) implements the phrase `relative to the horizon-generating Killing time' in our description of $\kappa$ below equation~(\ref{hawkingt}). This means that, in retrospect, we could have read off $\kappa$ (the Hawking temperature) automatically after identifying the relation between the near-horizon Rindler time and time at infinity:
\begin{equation}
\tau = T / 4GM \qquad \Rightarrow \qquad 
T_H = T(\infty) = \frac{\hbar}{8\pi G M} = \frac{\hbar}{2\pi} \cdot \frac{1}{4GM} = \kappa
\label{tempshortcut}
\end{equation}
We remind the reader that the specific formula $\hbar/(8\pi G M)$ applies to Schwarzschild black holes in flat space. Black holes with other charges and in other asymptotic boundary conditions (such as AdS) will have different expressions for their Hawking temperatures.

Note that in Rindler space a redshifting of the Unruh temperature out to infinity dilutes it to zero---a foregone conclusion because flat space has no scale that could set a non-vanishing result. In the black hole case, the mass (and other charges) of the hole set the scale for a non-zero temperature at infinity. On a conceptual level, the Unruh (Hawking) temperature is ushered in by the acceleration of the observer---that is, by her efforts to stay away from the future wedge (the black hole interior) and never look inside the complementary Rindler wedge (the other asymptotic region). One distinguishing feature of a black hole is that you must accelerate even when you are infinitely far away. 

Now that the Hawking temperature has fixed the overall magnitude of $\kappa$, we have natural units for relating horizon area to entropy:
\begin{equation}
S_{BH} = \frac{A}{4G\hbar} \qquad {\rm and} \qquad T = \kappa \,\propto\, \hbar
\label{bhentropy}
\end{equation}
The subscript BH stands for Bekenstein and Hawking \cite{Bekenstein:1973ur, Bekenstein:1974ax, Hawking:1974rv, Hawking:1974sw}, or is it Black Hole? Note the way $\hbar$ enters these identifications. When writing $\kappa \Delta A$ in (\ref{1stlaw}), we cannot turn off quantum mechanics unless we are willing to lose contact with the Hawking-Unruh effect. The fact that canceling powers of $\hbar$ enter the two factors of $\kappa \Delta A$ is deeply significant: it is the source and solution of the black hole information paradox \cite{juaneternalAdS, vijaygrf}, it underlies applications of the AdS/CFT correspondence to condensed matter physics \cite{subirsean}, and is essential for the matter of the present review. 

\subsection{Holographic duality}
\label{sec:adscft}

Identification~(\ref{bhentropy}) has a delightful corollary: it is impossible to fit more entropy than $A / 4G\hbar$ in a region bounded by area $A$. To understand this {\it holographic principle}, try to imagine a region $R$ which violates it. The most promising scenario is when $R$ is a ball so we get the most volume to contain the large entropy. What mass can $R$ contain? It cannot be larger than the mass of the black hole with area $A$, or else the mass would have collapsed into a black hole before it were ever squeezed into $R$. But the possibility that the mass would be smaller than that of the black hole is not much better. Adding the missing mass to effect black hole collapse would decrease the entropy! Thus, violating the {\it holographic principle} requires imagining that a flux of positive mass brings with it negative entropy!

The idea that the maximal entropy content of a region is bounded by its surrounding area and not volume is a radical departure from our usual experience with thermodynamics. It is an essentially gravitational effect: non-gravitating systems do not suffer black hole collapse and can accumulate arbitrarily high entropy densities. As general as it is, the argument is non-constructive in that it gives no statistical mechanics-type mechanism to prevent a faster-than-area growth of entropy. One route toward such a mechanism is to envision that any gravitating system has another, more fundamental description in terms of degrees of freedom living on its surrounding area, that is in one less dimension \cite{tHooft:1993dmi,Susskind:1994vu}. This {\it holographic dual} may be a non-local and superbly exotic theory, but if its internal entropy density is bounded by volume, we would have an area-wise bound from the higher-dimensional point of view. 

Unfortunately, we do not know a general way to construct holographic duals of arbitrary gravitational systems, and we cannot be sure that the speculation about holographically dual theories is robustly true. Fortunately, we have one class of examples where the speculation turns out to be spectacularly correct! It is the family of conformal field theories dual to Einstein gravity (with controllable corrections) with anti-de Sitter boundary conditions, known as the AdS/CFT correspondence. An especially rewarding feature of this class of examples is that the holographic duals are not-at-all exotic, local quantum field theories.

\paragraph{AdS/CFT---original argument}
Though we shall not need details of Maldacena's argument for AdS/CFT \cite{Maldacena:1997re, witten98, Aharony:1999ti}, we briefly sketch its salient features to avoid the appearance of magic. Maldacena considered $N \gg 1$ coincident D3-branes of type IIB string theory, which is a 10-dimensional theory. D3-branes are 3+1-dimensional tensionful membranes and, by Einstein's equations, their tension deforms the surrounding spacetime. In the present case, the D3-branes source the so-called extremal black brane geometry: 
\begin{equation}
ds^2 
= \left(1 + \frac{N \tilde{l}_P^{4}}{r^{4}} \right)^{-1/2}\!\! (-dx_0^2 + dx_1^2 + dx_2^2 + dx_3^2)
\, + \left(1 + \frac{N \tilde{l}_P^{4}}{r^{4}} \right)^{1/2} (dr^2 + r^2 d\Omega_{5}^2)\,, \,\,\,
\label{pbrane}
\end{equation}
where $\tilde{l}_P$ is the Planck length with some extra numerical constants absorbed. The first summand in~(\ref{pbrane}) runs parallel to the D-branes while the other part is transverse to them. We now take the limit of small $r$ (which sheds the `$1+$'s) and reset the metric in terms of $z = \sqrt{N} \tilde{l}_P^2 / r$:
\begin{equation}
ds^2 
= \sqrt{N}\tilde{l}_P^2 \left( 
\frac{-dx_0^2 + dx_1^2 + dx_2^2 + dx_3^2 + dz^2}{z^2} + d\Omega_5^2
\right)
\label{nearhorizon}
\end{equation}
This geometry is a product of five-dimensional anti-de Sitter space AdS$_5$---a homogeneous space with constant negative curvature---and a five-dimensional sphere $S^5$, both with curvature radii:
\begin{equation}
L_{AdS} = N^{1/4} \tilde{l}_P
\label{adsradius}
\end{equation}
It turns out that in the limit of small $r$ the D-branes decouple from the rest of the world. But the internal dynamics of D-branes is known in string theory to be that of a (super)conformal gauge field theory with $N$ colors. This yields the conclusion that {\bf super-Yang-Mills theory on $\mathbb{R}^{3,1}$ is equivalent to string theory on spacetime (\ref{nearhorizon}).} The low energy limit of the latter is ordinary Einstein gravity.

Similar derivations, which start from different-dimensional membranes in various combinations \cite{Aharony:2008ug, deBoer:1998kjm, Seiberg:1999xz},
sometimes on non-trivial 10-dimensional background geometries \cite{Douglas:1997de, Kachru:1998ys, Witten:1998xy, Klebanov:1998hh}, lead to similar equivalences. In each case, a string theory in anti-de Sitter space (AdS) whose low energy limit is Einstein gravity turns out to be equivalent to the internal dynamics of a D-brane configuration, which is a conformal field theory (CFT). We now list the features of the AdS/CFT correspondence, which are important for the present review:

\paragraph{AdS/CFT---important features:}

\begin{itemize}
\item AdS is one higher-dimensional than the CFT. Note how the transversal $r = L_{AdS}^2/z$ direction combined with the brane directions $x_{0,1,2,3}$ to form the AdS$_5$ metric in (\ref{nearhorizon}). This is generic.
\item Brane constructions of AdS/CFT produce AdS$_{d+1}$ times a compact $(9-d)$-dimensional space (such as the $S^5$ in \ref{nearhorizon}), which is left over from the 10-dimensional parent. This space {\it geometrizes} internal symmetries of the dual conformal field theory. We will ignore this aspect of AdS/CFT in this review, except for mentioning that an information-theoretic study of it was initiated in \cite{Mollabashi:2014qfa, karchinternal}. (See also \cite{rtcs1, rtcs2} for setups when a CFT is coupled to other fields with extra symmetries, or \cite{wcftrt, wcftrt2} where conformal symmetry is partly broken or modified.) 
\item There is a large parameter $N$, which on the gravity side of the duality sets a hierarchy between the macroscopic length scale (the curvature radius of anti-de Sitter space) and a fundamental scale such as the Planck length; see (\ref{adsradius}). The field theory must similarly have some large parameter $N$; in the case above it is the number of colors in a Yang-Mills theory but in other setups it may play a different role. Most of the review will concern leading (and sometimes next-to-leading) order terms in $1/N$ expansions. 
\item Symmetries: AdS$_{d+1}$ is a completely homogeneous space, that is any point can be mapped to any other by an isometry. On distance scales smaller than $L_{AdS}$, this homogeneous space looks approximately like $R^{d,1}$ so the isometries in question must comprise $d+1$ translations, $d(d-1)/2$ rotations, and $d$ boosts. We may think of time-like translation as a decompactified rotation and the $d$ space-like translations as boosts, in which case we have $d(d-1)/2 + 1$ rotations and $2d$ boosts. These symmetries make up $SO(d,2)$, which is the isometry group of AdS$_{d+1}$. 

This is also the (global) conformal group in $(d-1)+1$ dimensions. Thus, isometries of AdS$_{d+1}$ are conformal symmetries from the field theory point of view. 
\item The extra dimension of AdS$_{d+1}$ corresponds to scale in the field theory \cite{lennyedward}. When we compare the isometry group of AdS$_{d+1}$ to the conformal group on $\mathbb{R}^{d-1,1}$, we discover that \textchg{the symmetry which in the bulk acts like a translation in the $z$-direction} is a boundary dilatation. (The \textchg{bulk} $z$-$x_{1,2,3}$ rotations and the $z$-$x_0$ boost map to special conformal transformations.) Therefore, a change in $z$ is a change in scale from the field theory point of view. Conformal symmetry is a symmetry between scales, and on the AdS side it manifests itself as a symmetry between different $z=const.$ slices (radial slices) of the geometry.
\item Any $z = const.$ slice of the geometry is, roughly, a snapshot of the CFT at a certain scale. Smaller values of $z$ correspond to smaller length scales (higher energy scales) in the CFT, but on the AdS side the metric goes as $z^{-2}$, so small $z$ means large spatial sizes on the gravity side. This is sometimes called the UV/IR correspondence \cite{joeamanda}.
\item Imposing a radial cutoff $z \geq z_0$ therefore corresponds to truncating the ultraviolet physics from the CFT. To recover the actual CFT (from which no scales have been truncated), we must send the cutoff $z_0 \to 0$, which means going to unboundedly large length scales on the gravity side. For this reason, some people say that the CFT `lives on the asymptotic boundary,' which is $z \to 0$. For a non-compact spacetime such as AdS, the asymptotic boundary is the geometry obtained after stripping off a divergent factor from the successively larger transversal slices of the geometry. In the case of metric~(\ref{nearhorizon}), this is:
\begin{equation}
ds_\partial^2 = \lim_{z_0 \to 0} \frac{z_0^2}{L_{\rm AdS}^2}\, ds^2 \big|_{z=z_0} 
= -dx_0^2 + dx_1^2 + dx_2^2 + dx_3^2
\end{equation}
This is the $\mathbb{R}^{3,1}$ on which the holographic field theory lives.
\item When a conformal field theory is in a thermal state, the temperature sets a thermal scale. In this situation, some object must break the $z$-translational symmetry of AdS. This object is a black hole horizon. Thermal states in the CFT describe black holes with AdS boundary conditions. This gives a natural holographic interpretation of the Hawking temperature. 
\end{itemize}

\begin{figure}[tbp]
\centering 
\includegraphics[width=.5\textwidth]{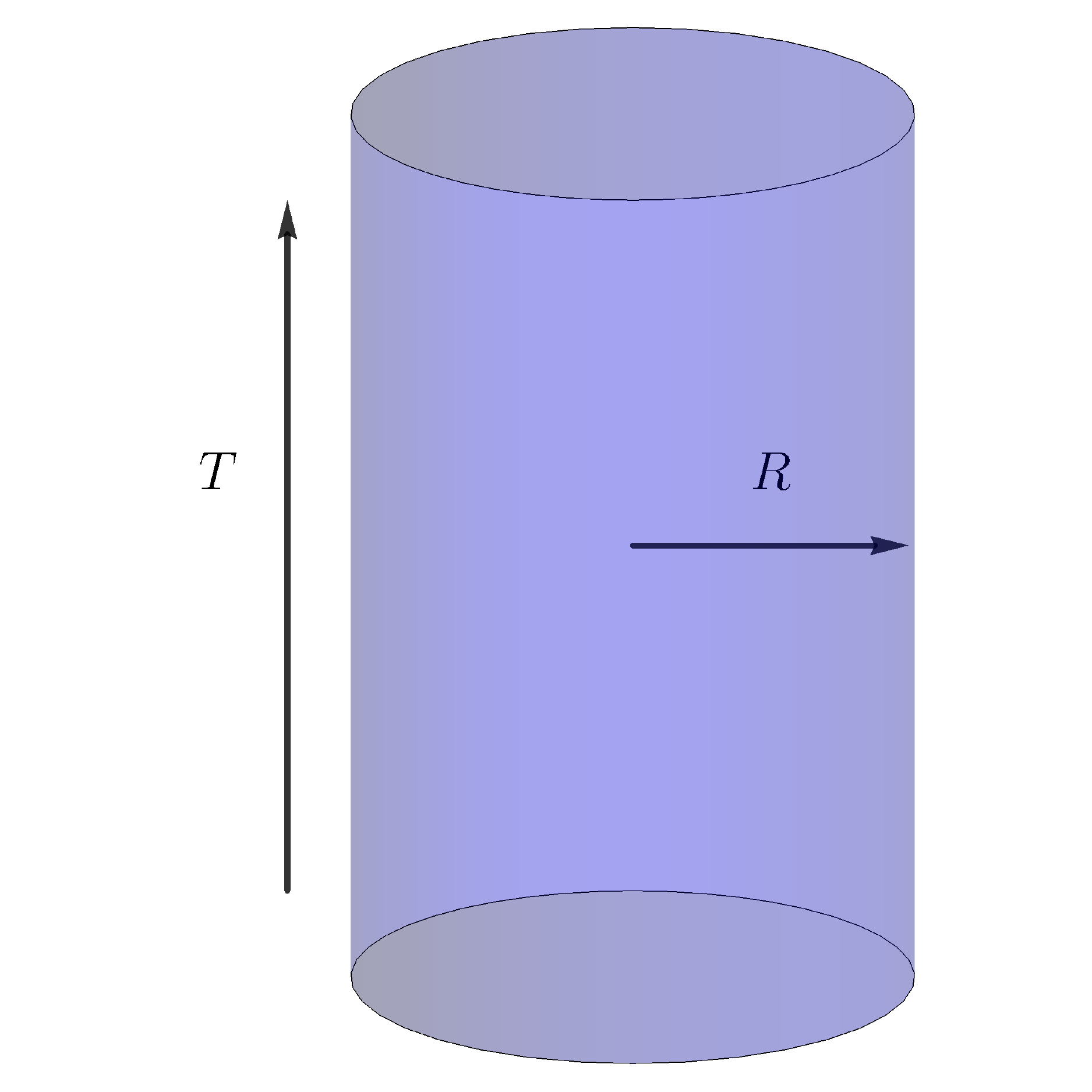}
\hfill
\caption{Global AdS space is presented in equation~(\ref{globaladshyperb}) as a solid cylinder. The dual CFT lives on its asymptotic boundary, which is the hollow cylinder.}
\label{fig:cylinder}
\end{figure}

One might question whether AdS/CFT, as presented here, actually realizes the holographic principle. We do have a lower-dimensional description of a gravitational system, but does it surround the system in question? It turns out that we can compactify the conformal boundary of AdS space so it becomes $S^{d-1} \times \mathbb{R}_{\rm time}$, in which case the holographically dual field theory truly surrounds the bulk spacetime. The coordinates of the resulting {\it global} AdS space, of which the first piece in (\ref{nearhorizon}) covers only a part, are:
\begin{align}
ds^2 & = 
- \left( \frac{{R^2}+{L_{\rm AdS}^2}}{L_{\rm AdS}^2} \right) dT^2 
+ \left( \frac{L_{\rm AdS}^2}{R^2+{L_{\rm AdS}^2}}\right) dR^2  + R^2 d\Omega_{d-1}^2 
\nonumber \\
& = L_{\rm AdS}^2 (- \cosh^2 \rho\, dt^2 + d\rho^2 + \sinh^2 \rho\, d\Omega_{d-1}^2 )
\label{globaladshyperb}
\end{align}
with $\sinh \rho = R / L_{\rm AdS}$ and $t = T / L_{AdS}$. The asymptotic boundary is
\begin{equation}
ds_\partial^2 = \lim_{\rho_0 \to \infty} 4e^{-2\rho_0}\, ds^2\big|_{\rho = \rho_0} 
= L_{\rm AdS}^2 (-dt^2 + d\Omega_{d-1}^2)
\label{boundaryglobal}
\end{equation}
The asymptotic $S^{d-1} \times \mathbb{R}_{\rm time}$ is often drawn as a hollow cylinder, which is filled inside by global AdS$_{d+1}$; see Figure~\ref{fig:cylinder}.

The final point is addressed to a reader who may be skeptical of string theory as a description of Nature. Because we do not live in anti-de Sitter space, AdS/CFT should be understood as a toy model whose main virtue is that it is explicit and calculable. The dependability of the toy model relies on the mathematical manipulations and physics intuitions employed in its derivation, \textchg{which certainly includes the rigorous technical backing provided by string theory.} Yet it does not hinge on whether \textchg{string theory} actually describes Nature, \textchg{partly because skeptics are entitled to treat all of string theory as but a toy model of gravity.} Because the main distinguishing feature of \textchg{AdS/CFT}---its holographic character---was anticipated from other points of view (black hole thermodynamics and many others, e.g.~\cite{Witten:1988hf,tHooft:1993dmi,Susskind:1994vu,Brown:1986nw}\footnote{
%
%
There is another type of bulk-boundary duality, which relates 2+1-dimensional Chern-Simons theory to rational two-dimensional conformal field theories \cite{frs}, and which was first studied in \cite{Witten:1988hf}. In a certain precise sense \cite{gukovetal}, this class of bulk-boundary dualities also subsumes the AdS$_3$/CFT$_2$ correspondence, though not higher-dimensional cases of gauge-gravity duality.}),
%
%
we can safely assume that the \textchg{holographic} model is not a cooky artefact of string theory. In particular, Reference~\cite{Heemskerk:2009pn} established that correlation functions of {\it any} large $N$ conformal field theory that satisfies one additional technical assumption can be understood as resulting from propagation of local quantum fields on an AdS spacetime, which is an alternative, bottom-up proof of the AdS/CFT correspondence. \textchg{Conversely, any gravitational theory with a negative cosmological constant and AdS boundary conditions defines a CFT because it defines CFT correlation functions according to the so-called extrapolate holographic dictionary \cite{lennyedward, banksetal, extrapolate}.} In summary, we believe that lessons about gravity and information theory harbored by AdS/CFT are equally reliable independently of the correctness of string theory as a fundamental description of Nature.

\section{Entanglement in AdS/CFT}

\subsection{AdS black holes, CFT thermal and thermofield double states}
\label{sec:adsbhs}

In field theory, studying time-dependent processes in a thermal state in Hilbert space $\mathcal{H}$ is often done by considering a special pure state in a duplicated Hilbert space $\mathcal{H} \otimes \bar{\mathcal{H}}$:
\begin{equation}
|{\rm TFD}\rangle = \frac{1}{\sqrt{\mathcal{Z}(\beta)}} \sum_i e^{-\beta E_i/2} |i\rangle \otimes |\bar{i}\rangle ,
\label{tfdstate}
\end{equation}
where $|i\rangle$ are Hamiltonian eigenstates. This {\it thermofield double state} is an entangled state of our Hilbert space and a fictitious partner space $\bar{\mathcal{H}}$. By equation~(\ref{traceout}), probing $|{\rm TFD}\rangle$ only by acting on $\mathcal{H}$ produces the same results as does the reduced density matrix
\begin{equation}
\rho_\mathcal{H} = {\rm Tr}_{\bar{\mathcal{H}}}\, |{\rm TFD}\rangle \langle {\rm TFD}| = \frac{1}{\mathcal{Z}} e^{-\beta H} = \rho(\beta)\,,
\label{thermalstate}
\end{equation}
which is the thermal state. In effect, we may view the thermal state as a reduced density matrix of a fictitious pure state in $\mathcal{H} \otimes \bar{\mathcal{H}}$. We have encountered such states below equation~(\ref{traceout}) and called them purifications. The thermofield double state is a canonical purification where $W = \sqrt{\rho}$, so in the notation (\ref{entangledstate}) we can write $|{\rm TFD}\rangle = |\rho^{1/2}\rangle$. For the present purposes, the crucial fact is that the thermal entropy of $\rho(\beta)$ is an entanglement entropy~(\ref{vonneumann}) in $|{\rm TFD}\rangle$.

In the AdS/CFT correspondence a CFT thermal state describes an AdS black hole. When studying a black hole spacetime such as Schwarzschild, it is often convenient and always instructive to zoom in on a neighborhood of the horizon. We did so in equation~(\ref{schwarzrindler}) and discovered that the black hole exterior locally looks like one Rindler wedge cut out of flat space---a reasoning that revealed Hawking's temperature, as well as the existence of a second asymptotic region. In studying stationary black holes, it is a standard exercise to find so-called Kruskal coordinates, which always reveal a second exterior. This second exterior plays the role of the auxiliary $\bar{\mathcal{H}}$ in the construction of the thermofield double state.

\paragraph{Simplest case: BTZ black hole}
The non-rotating BTZ black hole \cite{btzref}, which is an AdS$_3$ analogue of the Schwarzschild solution, is described by the metric:
\begin{align}
ds^2 & = 
- \left( \frac{R^2-R_0^2}{L_{\rm AdS}^2} \right) dT^2 
+ \left( \frac{L_{\rm AdS}^2}{R^2-R_0^2}\right) dR^2  + R^2 d\theta^2 
\label{btzmetric}
\\
& = -(R_0 / L_{\rm AdS})^2 \sinh^2 \rho\, dT^2 + L_{\rm AdS}^2 d\rho^2 + R_0^2 \cosh^2\rho\, d\theta^2
\qquad {\rm with}~\theta \equiv \theta + 2\pi
\label{btzhyperb}
\end{align}
where $R_0$ is the horizon radius and in the last line we set $R = R_0 \cosh \rho$. Using 
\begin{equation}
u = \tanh(\rho/2)\, e^{T R_0/L_{\rm AdS}^2} \qquad {\rm and} \qquad v = \tanh(\rho/2)\, e^{-T R_0 / L_{\rm AdS}^2}\,,
\label{btzlight}
\end{equation}
the metric becomes:
\begin{equation}
ds^2 = 4 L_{\rm AdS}^2 \frac{du dv}{(1 - uv)^2} + R_0^2 \left(\frac{1+uv}{1-uv}\right)^2 d\theta^2
\label{btzkruskal}
\end{equation}
As advertised, this looks like flat space near the horizon, which now falls at $uv = 0$. By the logic above equation~(\ref{tempshortcut}), coordinate change~(\ref{btzlight}) neatly reveals the temperature of the black hole to be $R_0 / 2\pi L_{\rm AdS}^2$.

More pertinently, all values of $u$ and $v$, positive and negative, are on an equal footing in metric~(\ref{btzkruskal}). The exterior region of the black hole in metric~(\ref{btzhyperb}) is covered by $u, v \geq 0$ subject to $0 \leq uv \leq 1$, with $uv=0$ marking the horizon and $uv = 1$ the asymptotic boundary. But $u, v \leq 0$, subject to the same $0 \leq uv \leq 1$ restriction, is also present in metric~(\ref{btzkruskal}), and it is an identical copy of the black hole exterior. Moreover, that second exterior region has its own asymptotic boundary, which now falls at $uv = 1$ with $u,v \leq 0$. In AdS/CFT, an asymptotic boundary is where the dual CFT lives. Evidently, the dual description of the maximally extended BTZ metric~(\ref{btzkruskal}) is a state not of a single CFT, but of a product of two CFTs, which live at $uv = 1$ with $u, v \gtrless 0$.

Reference~\cite{juaneternalAdS} pointed out that the CFT procedure of going from $\rho(\beta) = e^{-\beta H}/\mathcal{Z}$ to the pure state $|{\rm TFD}\rangle$ is in perfect analogy to going from metric~(\ref{btzmetric}) to its maximally extended version~(\ref{btzkruskal}). In both cases, we discover or invent a second Hilbert space---$\bar{\mathcal{H}}$ and $\bar{\mathcal{H}}_{\rm CFT}$ of the second asymptotic region. On the CFT side, the procedure equates the thermal entropy of $\rho(\beta) = e^{-\beta H}/\mathcal{Z}$ with the entanglement entropy of $|{\rm TFD}\rangle$. Translating this to the gravity side tells us that the horizon entropy of the BTZ black hole represents the entanglement between its two exteriors. 

Observe how this argument unifies smoothness of the horizon, the black hole entropy formula, and quantum entanglement. If the geometry does not have a second asymptotic region, it cannot be smooth. (Smoothness is the same as flatness at sufficiently small scales.) In that case, the entropy of the black hole cannot be entanglement entropy because $\bar{\mathcal{H}}_{\rm CFT}$ never materializes. But then Hawking's argument for black hole temperature also falls apart, and it is less clear whether the identification $S = A/4G\hbar$ is still justified. It appears that quantum entanglement is an essential prerequisite for a quantitative relation between horizon areas and entropies. 

\paragraph{Intermediate case: AdS-Rindler space, a.k.a. hyperbolic black hole} The above reasoning seems restricted to the context of black holes, but it is in fact robust enough to include empty space. This is because a Rindler wedge in AdS is a very special black hole.  

In \cite{hyperbbh}, Emparan wrote down a family of asymptotically AdS$_{d+1}$ black hole solutions whose horizons are not spheres $S^{d-1}$ but hyperbolic spaces $H^{d-1}$. In the $d=2$ case, the `hyperbolic space' $H^1 \sim \mathbb{R}$ and the metric becomes:
\begin{equation}
ds^2 = -(R_0 / L_{\rm AdS})^2 \sinh^2 \rho\, dT^2 + L_{\rm AdS}^2 d\rho^2 + R_0^2 \cosh^2\rho\, d\theta^2
\quad {\rm with}~\theta \in (-\infty, \infty)\,\,\,\,
\label{hyperbh}
\end{equation}
This is identical to the BTZ geometry (\ref{btzhyperb}) except the horizon is decompactified. That difference has no effect on equation~(\ref{btzlight}), so the temperature is still $R_0 / 2\pi L_{\rm AdS}^2$.

In a seemingly unrelated exercise, we may want to adapt the construction of Rindler space (\ref{flatrindlerm}) from flat space to anti-de Sitter space. We would again consider a family of accelerated observers, whose accelerations and initial conditions are chosen so that they have the same `blind spot': an identical region of AdS space, which they neither influence nor observe. These trajectories would again be non-intersecting, so we could use proper time $\tau$ along the trajectories, their accelerations $a$, and $d-1$ transversal directions, to describe a {\it Rindler wedge} of AdS$_{d+1}$. For $d=2$, the resulting metric becomes:
\begin{equation}
ds^2 = L_{\rm AdS}^2 
\big( - \sinh^2 \rho\, d\tau^2 + d\rho^2 + \cosh^2\rho\, d\theta^2\big)
\label{adsrindler}
\end{equation}
Just as in equation~(\ref{flatrindlerr}) in flat space, we conveniently reparameterized the acceleration:
\begin{equation}
\rho \equiv 
\frac{1}{2} \log \frac{aL_{\rm AdS}+1}{aL_{\rm AdS}-1}
\label{adsrindlerr}
\end{equation}
This parameterization only covers $a > L_{\rm AdS}^{-1}$ because accelerated trajectories with $a < L_{\rm AdS}^{-1}$ are oscillating. Rather than escaping to infinity, observers accelerating with $a < L_{\rm AdS}^{-1}$ fall back to the interior of AdS and inadvertently see all of it, so they cannot be part of the Rindler construction. Note how the flat space limit $L_{\rm AdS} \to \infty$ removes this restriction on $a$ and recovers equation~(\ref{flatrindlerr}). 

Now compare the hyperbolic black hole~(\ref{hyperbh}) to the AdS-Rindler metric~(\ref{adsrindler}). When $R_0 = L_{\rm AdS}$, these metrics are identical! Evidently, a Rindler wedge of AdS space is a hyperbolic black hole with temperature $(2\pi L_{\rm AdS})^{-1}$! Although equations~(\ref{hyperbh}) and (\ref{adsrindler}) are for $d=2$, the conclusion is the same in all dimensions. 

In what sense is AdS-Rindler space a black hole? Solutions~(\ref{hyperbh}) behave as ordinary black holes: they cloak singularities, which are timelike or spacelike depending on whether $R_0 \gtrless L_{\rm AdS}$. But the cross-over value $R_0 = L_{\rm AdS}$ is so special that the singularity disappears altogether and the black hole metric becomes Rindler. AdS-Rindler space is a black hole because it belongs to a parametric family of black hole solutions. This is enough to ensure that black hole thermodynamics with all its consequences applies equally well to AdS-Rindler wedges.

We may now repeat the BTZ reasoning from the previous page, this time applied to Rindler wedges. Since metrics~(\ref{btzhyperb}) and (\ref{adsrindler}) differ only by an irrelevant identification in $\theta$, literally nothing changes. A single `black hole' = Rindler geometry is holographically dual to the thermal state $\rho\,(\beta = 2\pi L_{\rm AdS})$ of a CFT, which lives on the asymptotic boundary of~(\ref{adsrindler}), that is on 1+1-dimensional flat space $ds^2 = L_{\rm AdS}^2 (-d\tau^2 + d\theta^2)$. We duplicate the Hilbert space of this CFT and find a thermofield double purification of $\rho(\beta)$. In the bulk, this means going from metric~(\ref{adsrindler}) to its Kruskal coordinates, which is equation~(\ref{btzkruskal}) with $R_0 = L_{\rm AdS}$ and $\theta \in (-\infty, \infty)$. But now the `black hole' in question is Rindler space and the thermofield double state describes empty AdS$_3$! (A coordinate change takes the $R_0 = L_{\rm AdS}$ metric~(\ref{btzkruskal}) to (\ref{globaladshyperb}), which the reader is welcome to verify.) In other words, the thermofield double is the global vacuum! The second exterior of the `black hole' is the complementary Rindler wedge, and the partner Hilbert space $\bar{\mathcal{H}}$ lives on its asymptotic boundary, which is $uv = 1$ with $u,v < 0$ in (\ref{btzkruskal}). Our discussion showcases $d=2$ to exploit the connection with BTZ, but the conclusions are unchanged in higher dimensions \cite{chm}. 

Most importantly, we again recognize that the horizon entropy of a black hole quantifies the entanglement between its two asymptotic boundaries. But in the present case, the horizon in question is a Rindler horizon in empty space! Therefore, the relation between entanglement and area is not limited to prototypical black holes such as Schwarzschild or BTZ. Even in ordinary, singularity-free geometries, {\bf if a surface is an appropriate type of acceleration horizon}, its area in units of $4G\hbar$ is entanglement entropy for the Hilbert spaces describing its two sides. 

Section~\ref{sec:rt} quantifies this boldfaced proviso; this is the famous Ryu-Takayanagi (RT) proposal. The arguments above are not enough to establish the validity of the RT proposal, whose scope extends far beyond stationary two-sided black holes. We will sketch a proof of the Ryu-Takayanagi proposal in Section~\ref{sec:renyi}.

\subsection{The Ryu-Takayanagi proposal}
\label{sec:rt}

\begin{figure}
     \centering
     \begin{subfigure}[t]{0.3\textwidth}
         \centering
         \includegraphics[width=.98\textwidth]{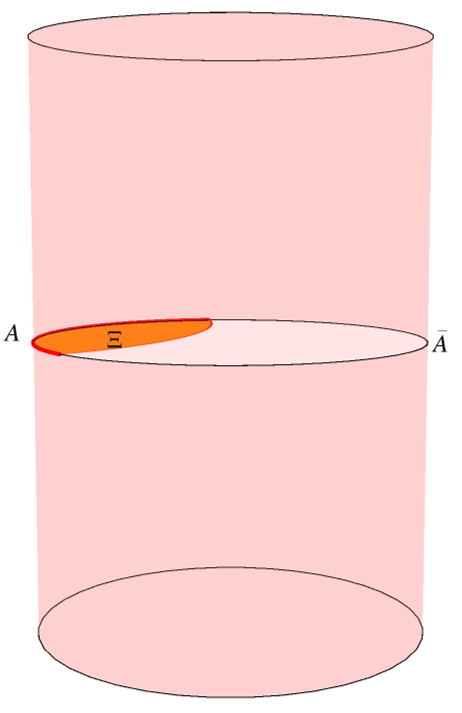}
     \end{subfigure}
     \hspace{2cm}
     \begin{subfigure}[t]{0.45\textwidth}
        \centering
        \vspace*{-6cm}
         \includegraphics[width=\textwidth]{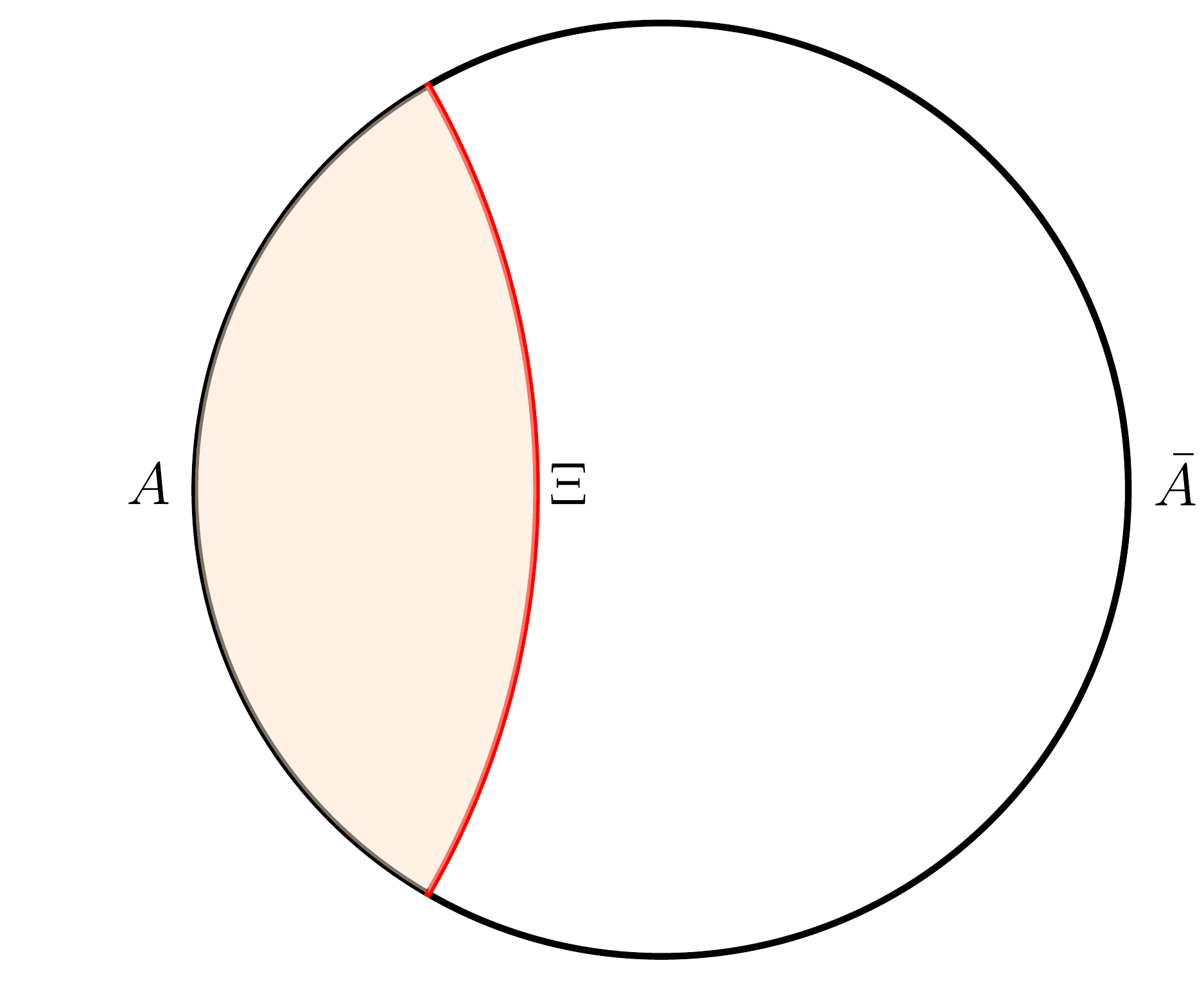}
     \end{subfigure}
        \caption{An equal time slice of an asymptotically AdS geometry, assumed static and horizonless. A Cauchy slice of the boundary CFT---the circle---is divided into regions $A$ and $\bar{A}$. Proposal~(\ref{rtsimplest}) considers surfaces $\Xi$ that asymptote to the border of $A$ and $\bar{A}$, and sets $S(A) = S(\bar{A})$ to the minimal area achieved by such a $\Xi$.}
        \label{RTslice}
\end{figure}

Consider a setup from Figure~\ref{RTslice}. Choose a Cauchy slice of the asymptotic boundary; this is where quantum states $|\phi\rangle_{\rm global}$ of the holographically dual CFT live. Now divide this slice into two regions: $\mathcal{H}_{\rm CFT} = \mathcal{H}_A \otimes \mathcal{H}_{\bar{A}}$. (Once again, we ignore obstructions to factorization.) The global state is an entangled state between $\mathcal{H}_A$ and $\mathcal{H}_{\bar{A}}$ as in (\ref{entangledstate}). We may form its reduced density matrices $\rho_A$ and $\rho_{\bar{A}}$ as in (\ref{traceout}), and compute the entanglement entropy $S(A)$ as in equation~(\ref{vonneumann}). So long as we start from a pure state $|\phi\rangle_{\rm global} \in \mathcal{H}_{\rm CFT}$, the reasoning below equation~(\ref{traceout}) guarantees that $S(A) = S(\bar{A})$. 

For simplicity, first assume that the state $|\phi\rangle_{\rm global}$ is time-reversal symmetric. We may always choose coordinates for the dual bulk geometry, which respect this time-reversal symmetry; this choice selects a privileged Cauchy slice for the gravitational bulk. Working on this slice, find the surface of minimal area, which extends to the boundary and asymptotes to the division between $A$ and $\bar{A}$. Such a minimal surface is codimension-1 on the equal time slice, so codimension-2 in the full spacetime. Reference~\cite{Ryu:2006bv} posited that the area of the minimal surface (also called the Ryu-Takayanagi or RT surface) in units of $4G\hbar$ equals $S(A)$:

\begin{tcolorbox}
{\bf Ryu-Takayanagi proposal for pure, time reversal-symmetric states:}
\begin{equation}
S(A) = \min_{\partial \Xi \to \partial A} \frac{{\rm Area}(\Xi)}{4 G \hbar}
\label{rtsimplest}
\end{equation}
Here we minimize over surfaces $\Xi$, which live on the time reversal-symmetric slice of the bulk geometry and which asymptote to the border of region $A \subset {\rm CFT}$ on the asymptotic boundary.
\end{tcolorbox}

Our AdS-Rindler analysis is a special case of this proposal. The factorization $\mathcal{H}_{\rm CFT} = \mathcal{H}_A \otimes \mathcal{H}_{\bar{A}}$ refers to the CFT Hilbert spaces, which are dual to the individual Rindler wedges. The minimal surface invoked by Ryu and Takayanagi is the AdS-Rindler horizon. Indeed, the locus $u=v=0$ in metric~(\ref{btzkruskal}) is a spacelike geodesic, i.e. a minimal surface on a spatial slice of the bulk geometry; see the cyan line in the right panel of Figure~\ref{flatrindler}. A minimal surface on a static bulk slice is `an appropriate type of an acceleration horizon.' 

As stated, the above discussion excludes the BTZ geometry (\ref{btzhyperb}) on two counts. First, it assumes that the global geometry is holographically dual to a pure state while the one-sided BTZ spacetime is dual to the thermal state, which is mixed. We could try to get around this by going to the thermofield double state~(\ref{tfdstate}) and its dual geometry~(\ref{btzkruskal}), but a well-defined proposal should not depend on a choice of purification, which is not unique. Second, the entanglement responsible for the entropy of the BTZ black hole was between two completely disjoint CFTs, each of which lives on $S^1 \times \mathbb{R}_{\rm time}$. Therefore, the regions $A$ and $\bar{A}$ in question have no border on the asymptotic boundary: $\partial A = \emptyset$. Of course, the surface we want in the present case is the BTZ horizon, but equation~(\ref{rtsimplest}) seems to allow $\Xi = \emptyset$ and $S(A) = 0$. 

The full statement of the Ryu-Takayanagi proposal fixes these problems:
\begin{tcolorbox}
{\bf Ryu-Takayanagi proposal for time reversal-symmetric states:}
\begin{equation}
S(A) = \min_{\Xi \sim A} \frac{{\rm Area}(\Xi)}{4 G \hbar}
\label{rtfull}
\end{equation}
Here we minimize over surfaces $\Xi$ on the time reversal-symmetric slice of the bulk geometry, which are {\it homologous} to $A \subset {\rm CFT}$ on the asymptotic boundary. `Homologous,' denoted $\sim$, means that $\Xi$ and $A \subset {\rm CFT}$ fully enclose some region of the bulk slice, i.e. $\Xi \cup A = \partial \tilde{\Sigma}$ for some $\tilde{\Sigma} \subset \textrm{bulk slice}$. 
\end{tcolorbox}

Observe how this version of the proposal applies to the BTZ spacetime. The homology condition ensures that the minimal surface separates the boundary regions, which excludes $\Xi = \emptyset$. But it does so without mentioning $\partial A$, so if the CFT region at hand has no border on the asymptotic boundary, the minimal surface needs not go near the boundary. This occurs when we have multiple asymptotic regions entangled together as in the two-sided geometry~(\ref{btzkruskal}). From a single-sided point of view, the Ryu-Takayanagi surface acts like a Rindler horizon for observers who accelerate so as not to see the purifying regions, like in our BTZ and AdS-Rindler discussions. This fact motivates the first point in a list of...

\paragraph{Direct consequences of the Ryu-Takayanagi proposal:}
\begin{itemize}
\item When a CFT state $\rho$ is mixed, its dual geometry has a horizon. In the Ryu-Takayanagi proposal~(\ref{rtfull}), this horizon becomes the minimal surface $\Xi$ and its area is $-(4G\hbar) {\rm Tr}\, \rho \log \rho$.
\item Unless region $A$ covers an entire CFT, the entanglement entropy of $A$ is divergent. From the bulk point of view, this is because the minimal surface must extend to $\partial A$ at spatial infinity, so the divergence is proportional to ${\rm vol}(\partial A)$. It is well known that entanglement entropies in finite energy states in field theory have precisely this type of divergence \cite{area1, areasrednicki}. This is known as the \emph{area law}, where the word `area' connotes codimension-2 in field theory and so refers to $\partial A$. This divergence tells us that neighboring field theory regions are entangled at arbitrarily small scales near their common border, which accords with interpreting radial AdS slices as a hierarchy of CFT scales. We typically present such entanglement entropies subject to a UV regulator in the CFT, which is implemented on the AdS side as a large scale cutoff. 
\item When disconnected CFTs are in an entangled state, they produce finite entanglement entropies. In the bulk, the corresponding minimal surfaces wrap horizons of black holes, i.e. localize at AdS radial scales dual to thermal CFT scales. This is how entanglement entropies encompass thermal entropies as a special case. One example is our BTZ discussion.
\item Quantum entanglement is the glue that holds space together \cite{marksessay}. To understand this, consider a one-parameter family of time reversal-symmetric, holographic states $|\psi \rangle \in \mathcal{H}_1 \otimes \mathcal{H}_2$ whose quantum entanglement approaches zero. By the Ryu-Takayanagi proposal, the area of the minimal surface that separates the asymptotic boundaries where $\mathcal{H}_{1,2}$ live also approaches zero, i.e. the bulk geometry pinches off. For explicit calculations that showcase this qualitative expectation, see \cite{rindlerqg}.
\item Under suitable circumstances, the Ryu-Takayanagi formula (and generalizations) imply Einstein's equations (and generalizations) in the bulk. This includes Einstein's equations \cite{eeq1}
and equations of motion of higher curvature theories \cite{eeq2} linearized about pure AdS and about backreacted spacetimes with matter fields turned on \cite{eeq3, eeq4}. Deriving Einstein's equations from the behavior of entanglement entropy and/or areas of horizons has a long history \cite{eeqted, entropicgr, maxentropyballs, Czech:2017ryf}.
\item Entanglement entropies in large $N$ theories undergo first order phase transitions. The generic  reason for this is that the homology condition can give rise to distinct topological classes of surfaces $\Xi$, which exchange dominance. Phase transitions in holographic entanglement entropy were studied in depth in \cite{Headrick:2010zt} and recently in \cite{xihuajia}. Their significance runs deep. For example, they underlie the novel semiclassical calculations that reconcile black hole evaporation with unitarity, which are reviewed in Section~\ref{sec:replicas}. The intricate pattern of such phase transitions is also believed to encode the answer to the question: which theories and states have holographic duals? (see Section~\ref{sec:hec}.) We illustrate the phase transitions with two examples below.
\end{itemize}
All statements above must be interpreted at leading order in a $1/N$ expansion; see discussion below equation~(\ref{adsradius}). We discuss subleading corrections in Section~\ref{sec:genentropy}.

\begin{figure}
     \centering
     \savebox{\imagebox}{\includegraphics[width=.45\textwidth]{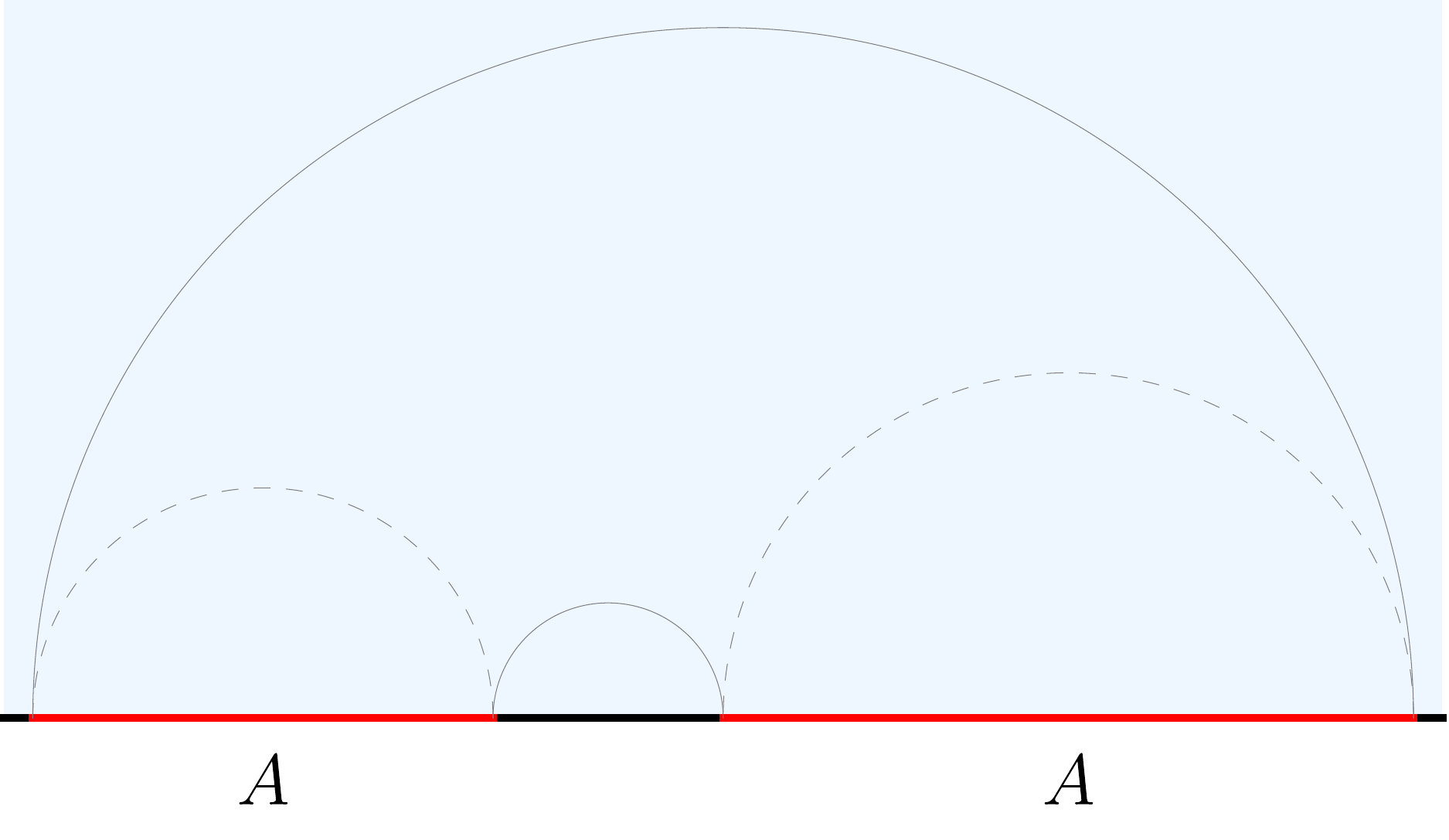}}
     \begin{subfigure}[t]{0.45\textwidth}
         \centering
         \raisebox{\dimexpr.5\ht\imagebox-0.45\height}{
         \includegraphics[width=0.7\textwidth]{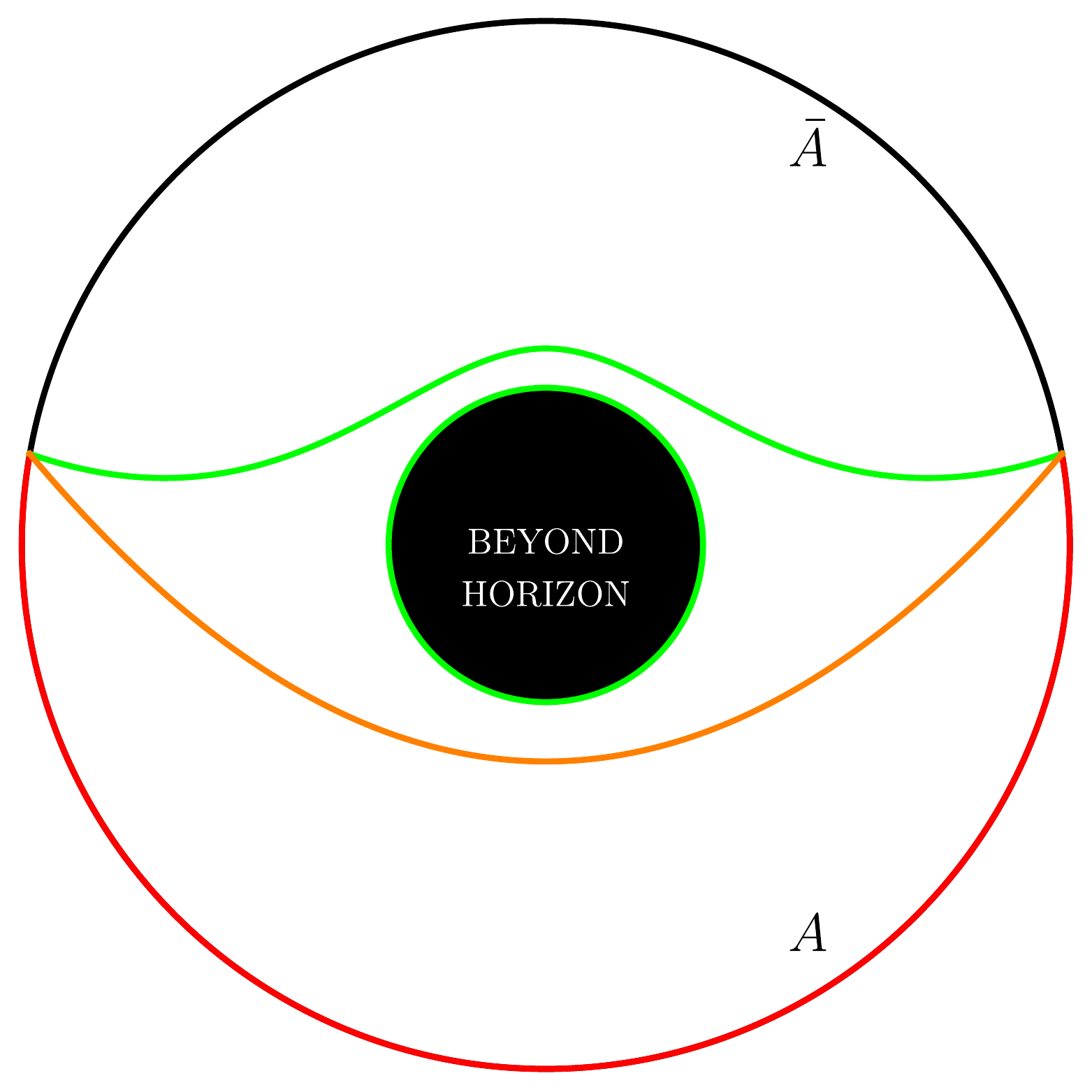}}
         \label{bhtransition}
     \end{subfigure}
     \hfill
     \begin{subfigure}[t]{0.5\textwidth}
         \centering
         \usebox{\imagebox}
         \label{2inttransition}
     \end{subfigure}
        \caption{Two examples of phase transitions in holographic entanglement entropy. Left: the entanglement entropy of a single interval $A$ in a CFT$_2$ thermal state can be realized holographically either as an individual geodesic homologous to $A$ (orange), or as the union of a geodesic homologous to $\bar{A}$ together with the black hole horizon (green). Right: the holographic entanglement entropy of two intervals in the CFT$_2$ vacuum can be in the connected (continuous lines) or disconnected (dashed lines) phase.}
        \label{phase}
\end{figure}

\paragraph{Phase transitions in holographic entanglement entropies}
Consider a single interval $(-\alpha, \alpha)$ in a holographic two-dimensional CFT in the thermal state. The relevant bulk geometry is a static slice of the BTZ black hole~(\ref{btzhyperb})
\begin{equation}
ds^2 = d\rho^2 + \tilde{R}_0^{\,2} \cosh^2\rho\, d\theta^2
\qquad {\rm with}~\theta \equiv \theta + 2\pi,
\label{staticbtzmetric}
\end{equation}
where we rescaled the metric by an overall constant and set $\tilde{R}_0 = R_0 / L_{\rm AdS}$. There are two locally minimal surfaces, which are homologous to $(-\alpha, \alpha)$ on the asymptotic boundary; see Figure~\ref{phase}. Option~1 is a single geodesic, which is individually homologous to the interval $(-\alpha, \alpha)$. Option~2 is the union of the geodesic homologous to the complement $(\alpha, 2\pi-\alpha)$ together with the BTZ horizon $\rho = 0$. The Ryu-Takayanagi proposal picks the minimum of their two lengths. The length of the geodesic in Option~1
\begin{equation}
\cosh\rho = 
\frac{\cosh (\tilde{R}_0 \alpha)}{\sqrt{\sinh^2 (\tilde{R}_0 \alpha)-\sinh^2 (\tilde{R}_0\theta)}}
\label{btzgeodesic}
\end{equation}
regulated with a bulk cutoff $\rho < \log{\mu^{-1}}$ is $2 \log \big(2\tilde{R}_0 \sinh(\tilde{R}_0 \alpha)/\mu\big)$, where we assume $\mu \ll \tilde{R}_0$ as befits a cutoff. Therefore, the lengths of the two options are
\begin{equation}
\label{abcde}
2 \log \frac{2 \tilde{R}_0 \sinh(\tilde{R}_0 \alpha)}{\mu} 
\qquad {\rm versus} \qquad
2 \log \frac{2 \tilde{R}_0 \sinh(\tilde{R}_0 (\pi-\alpha))}{\mu} + 2\pi \tilde{R}_0
\end{equation}
and the phase transition occurs when these two equal. When $R_0 \sim L_{\rm AdS}$, the critical interval leaves out an $\mathcal{O}(1)$ angle from the full circle, e.g. $\alpha_{\rm critical}(\tilde{R}_0=2) \approx 0.94\pi$. Note that the cutoff drops out from calculating the critical interval size because both options must have the same CFT ultraviolet divergences.

For a second example, consider $A = (x_L^1, x_R^1) \cup (x_L^2, x_R^2)$ in a holographic two-dimensional CFT on the plane. We take the CFT to be in the ground state, so the bulk geometry is pure AdS$_3$ in coordinates that asymptote to a plane. These are called Poincar{\'e} coordinates and we have seen them in equation~(\ref{nearhorizon}). The static slice on which to look for $\Xi$ is simply:
\begin{equation}
ds^2 = \frac{dx^2 + dz^2}{z^2}
\label{poincareh2}
\end{equation}
Again, there are two topologically distinct minimal surfaces $\Xi$: a pair of geodesics homologous to $(x_L^1, x_R^1)$ and $(x_L^2, x_R^2)$ versus a pair of geodesics homologous to $(x_R^1, x_L^2)$ and $(x_R^2, x_L^1)$. A geodesic with endpoints $x_L$ and $x_R$ is given by
\begin{equation}
(x-x_L) (x-x_R) + z^2 = 0,
\label{poincaregeod}
\end{equation}
which conveniently looks like a semi-circle on a Euclidean $x$-$z$ plane. Its length, regulated with $z > \mu$, is $2 \log (x_R - x_L) / \mu$ for $\mu \ll 1$. Therefore the lengths of the two options are:
\begin{equation}
2 \log \frac{(x_R^1-x_L^1)(x_R^2-x_L^2)}{\mu^2}
\qquad {\rm versus} \qquad
2 \log \frac{(x_L^2-x_R^1)(x_R^2-x_L^1)}{\mu^2}
\label{2intoptions}
\end{equation}
Note that their difference is again cutoff-independent and turns out to be a function of the conformal cross-ratio, which is conformally invariant. We could have guessed so because conformal transformations are AdS isometries and a bulk isometry cannot change the phase of holographic entanglement entropy. The phase transition occurs when ${(x_R^1-x_L^1)(x_R^2-x_L^2)} = {(x_L^2-x_R^1)(x_R^2-x_L^1)}$.

\paragraph{The covariant Hubeny-Rangamani-Takayanagi (HRT) proposal} When the quantum state is not time-reversal symmetric, we need a more involved prescription for holographic entanglement entropy. First formulated in \cite{Hubeny:2007xt}, it was later recast as the following maximin presciption \cite{Wall:2012uf}:

\begin{tcolorbox}
{\bf Hubeny-Ryu-Takayanagi proposal (maximin version):}
\begin{equation}
S(A) = \max_{\Sigma\, | \, A \subset \partial \Sigma} 
\left(\min_{\Xi \subset \Sigma\,{\&}\,\Xi \sim A} \frac{{\rm Area}(\Xi)}{4 G \hbar}\right)
\label{hrt}
\end{equation}
We first minimize the area over surfaces $\Xi$ homologous to $A$, which live on a fixed bulk spatial slice $\Sigma$. The bulk slices $\Sigma$ must asymptote to the boundary region $A \subset {\rm CFT}$ on the asymptotic boundary and the homology condition is imposed on $\Sigma$, that is $\Xi \cup A = \partial \tilde{\Sigma}$ with $\tilde{\Sigma} \subset \Sigma$. The holographic entanglement entropy is the maximal such area, with the maximum taken over all bulk spatial slices $\Sigma$ that asymptote to $A$.
\end{tcolorbox}

\begin{figure}[tbp]
\centering 
\includegraphics[width=.5\textwidth]{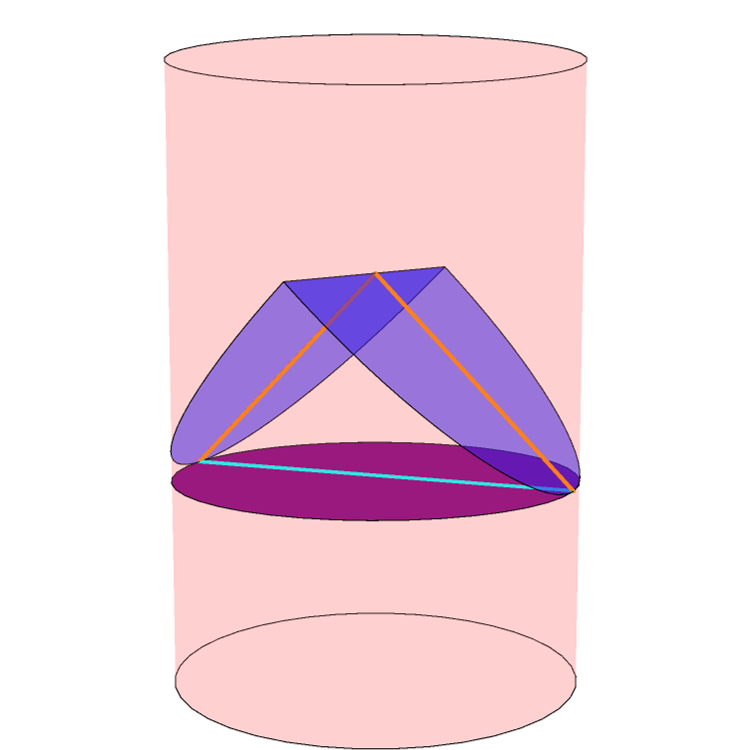}
\hfill
\caption{\label{maximin} This illustrates the maximization step in the maximin construction~(\ref{hrt}). The cyan and orange surfaces are minimal with respect to their underlying Cauchy surfaces. Compared to all other minimal surfaces, the cyan surface achieves maximal area.}
\end{figure}

The minimization in~(\ref{hrt}) is the same as in (\ref{rtfull}), but the subsequent maximization merits comment. Apply the HRT prescription to the semi-circle $(-\pi/2, \pi/2)$ of a holographic CFT on $S^1 \times \mathbb{R}_{\rm time}$ in its ground state; see Figure~\ref{maximin}. The bulk geometry is pure AdS$_3$ with cylindrical boundary: 
\begin{equation}
ds^2 = L_{\rm AdS}^2 (- \cosh^2 \rho\, dt^2 + d\rho^2 + \sinh^2 \rho\, d\theta^2 )
\label{globalads3}
\end{equation}
The RT proposal looks for minimal surfaces (curves) on the $t={\rm const.}$ slice and settles on the spacelike geodesic that is its diameter. But in the covariant prescription, we consider all allowed slices $\Sigma$, including ones that contain null rays shot radially from the semi-circle endpoints toward the center of AdS$_3$. Their length vanishes because they are null rays, but the maximization discards this option and again selects the RT geodesic. We recognize that the maximization is designed to discard lightlike shortcuts in the surface $\Xi$, which is now not a minimal but an {\it extremal} surface.

\subsection{Structure and symmetry of spacetime---an entanglement view}

The discussions in Sections~\ref{sec:adsbhs} and \ref{sec:rt} may seem disconnected. Section~\ref{sec:adsbhs} recognizes that bifurcation horizons of black holes manifest quantum entanglement. The reasoning heavily exploits the Rindler-like character of the near-horizon geometry, with the four wedges---left and right Rindler wedges and the past and future wedge---meeting on its surface. Section~\ref{sec:rt} emphasizes that quantum entanglement manifests itself geometrically on surfaces, which are extremal.

In fact, the discussions are not disjoint. A bifurcation horizon of a stationary black hole is an extremal surface, which is how Section~\ref{sec:rt} subsumes Section~\ref{sec:adsbhs}. On the other hand, an extremal surface $\Xi$ is codimension-2, i.e. everywhere orthogonal to 1+1-dimensional planes, which can be divided into four wedges. The extremization in (\ref{hrt}) selects $\Xi$ that is minimal with respect to spatial deformations and maximal with respect to timelike deformations in the orthogonal 1+1-space. This feature must somehow endow Ryu-Takayanagi surfaces with the same thermodynamic properties, which are enjoyed by black hole bifurcation horizons.

The subject of {\it subregion duality} formalizes this intuition. It does so by reimagining two important aspects of stationary black holes, which were conspicuously absent in Section~\ref{sec:rt}. One is the division of spacetime into black (white) hole interiors and two exterior regions. The other one is the Killing symmetry that generates time-translations at infinity, from which the Hawking temperature is derived (viz. equation~\ref{tempshortcut}). Both these concepts carry over to discussions of general Ryu-Takayanagi surfaces, albeit in insightfully modified forms. 

We begin with defining an RT-surface analogue of a black hole exterior, which is called an entanglement wedge.

\subsubsection{Subregion duality and entanglement wedges}
\label{sec:ew}

We first have to invent a question, which a black hole exterior-like region will answer. 

Our analysis of black holes in Section~\ref{sec:adsbhs} started from the statement that the thermal state~(\ref{thermalstate}) of a single CFT is dual to a one-sided black hole such as (\ref{btzhyperb}). We then said that the thermofield double state~(\ref{tfdstate})---an entangled state in a product of two CFTs---is dual to the maximally extended black hole~(\ref{btzkruskal}), which has two exteriors and two asymptotic boundaries. A key fact here is that tracing out the second Hilbert space from $\mathcal{H} \otimes \bar{\mathcal{H}}$ wipes out from $|{\rm TFD}\rangle$ the information about the second exterior whose asymptotic boundary supports $\bar{\mathcal{H}}$. Note that if we had traced out $\bar{\mathcal{H}}$ from some other purification $|W\rangle$ of $\rho(\beta)$ instead of $|{\rm TFD}\rangle$, we would have---by definition---obtained back the same $\rho(\beta)$, which describes the same single black hole exterior. We see that the reduced density matrix $\rho_{\mathcal{H}} = {\rm Tr}_{\bar{\mathcal{H}}} |{\rm TFD}\rangle \langle{\rm TFD}| = {\rm Tr}_{\bar{\mathcal{H}}} |W\rangle \langle W|$ retains the information only about the exterior region whose boundary hosts $\mathcal{H}$ and forgets everything about the choice of purification $W$. 

To summarize: Individual exterior regions of black holes remain unchanged when we alter the purification of the thermal density matrix $\rho(\beta)$. We may ask a similar question about a general reduced density matrix \cite{rhodual}: 
\begin{tcolorbox}
{\bf Subregion duality---statement of problem:}
\newline
Consider a state $\rho$ (which may or may not be pure), which is the holographic dual of a semiclassical, asymptotically AdS spacetime. Divide the Hilbert space of the holographic CFT into $\mathcal{H}_A \otimes \mathcal{H}_{\bar{A}}$ and form the reduced density matrix $\rho_A = {\rm Tr}_{\bar{A}}\, \rho$. Which part of the original spacetime is dual to $\rho_A$? 
\end{tcolorbox}
\noindent
Here are a few synonymous questions: What is the largest bulk region, which is independent of the choice of purification of $\rho_A$? What is the largest bulk region, which is unaffected by unitary transformations on $\bar{A}$? (This is equivalent to the former since different purifications are related by unitary transformations on the purifying Hilbert space; see text below equation~\ref{traceout}.) And finally: what is the closest analogue of a single black hole exterior for a general Ryu-Takayanagi surface? 

For completeness, we mention another rephrasing of the problem, which has been an important driver of understanding bulk reconstruction, and which is often the most convenient in explicit calculations. In its low energy limit, a generic AdS/CFT setup involves perturbative fields propagating on a semiclassical, asymptotically AdS spacetime. For the holographic description to make sense, every perturbative bulk field $\phi(x)$ must have an expansion in terms of CFT operators. Assuming such an expansion only involves local operators $\mathcal{O}(X)$, we write:
\begin{equation}
\phi(x) = \int_{{\rm CFT}} dX\, K(x;X)\, \mathcal{O}(X)
\label{hkllreconstr}
\end{equation}
This is often called the HKLL prescription after \cite{hkll1, hkll2, Hamilton:2006fh, hkll3}; see also \cite{banksetal, holoprobes, Bena:1999jv} for earlier versions. The kernel $K(x;X)$, also called a smearing function, is not unique---a fact that reflects deep conceptual properties of the holographic map, some of which we discuss in Section~\ref{sec:error}. The criterion for finding $K(x;X)$ is that $\phi(x)$ must solve bulk equations of motion, which means that it is essentially a Green's function, sensitive to the choice of boundary conditions.

To proceed, we need a definition. We will use it extensively, both in the boundary field theory and in the bulk:
\begin{tcolorbox}
{\bf Domain of dependence---definition:}
\newline
The  domain of dependence $D(A)$ of a region $A$ is that region, which is spacelike separated from all points on a Cauchy slice of the theory, which live outside $A$. In a causal theory, it is the region where the physics can be calculated from initial data inside $A$ by using equations of motion.

{\bf Example:} In a 1+1-dimensional field theory, the domain of dependence of a spacelike interval is a causal diamond. 
\end{tcolorbox}
\noindent

One sufficient condition for subregion duality is the following: if $x$ belongs to the part of spacetime dual to $\rho_A$ then there exists $K(x; X)$ in equation~(\ref{hkllreconstr}) whose support in the CFT is confined to the CFT domain of dependence of $A$. We allow the smearing function to occupy the CFT domain of dependence of $A$, rather than restricting it to $A$ itself, because $D(A)$---unlike $A$ itself---is a covariant concept. 

Many authors have studied formula~(\ref{hkllreconstr})---a smearing of \emph{local} CFT operators---and found explicit kernels $K(x;X)$ for AdS-Rindler~\cite{hkll2} and BTZ \cite{Hamilton:2006fh} spacetimes. These are case-by-case realizations of subregion duality. However, as was emphasized to us by Patrick Hayden, the existence of a formula~(\ref{hkllreconstr}) is sufficient but not necessary for an instance of subregion duality to hold. It is possible that a bulk operator can be rewritten as a combination of boundary operators, and yet every such rewriting necessarily involves non-local boundary operators. There is evidence that this is the generic situation: References~\cite{tomaitor, subregionchannel} have given general, HKLL-like reconstructions that rely on modular flow (see equation~\ref{modularshift}), which in almost all situations is non-local. Yet another conceptually tricky aspect of (\ref{hkllreconstr}) is the extent to which a foreknowledge of the bulk geometry is used in finding $K(x;X)$. If one needs to know the bulk geometry ahead of time, before solving for a Green's function and finding $K(x;X)$, then formula~(\ref{hkllreconstr}) is not sufficient for demonstrating subregion duality. This issue was emphasized e.g. in~\cite{virasoroope, knowsitsplace}. We will not study explicit smearing functions in this review. Instead, in Section~\ref{sec:error}, we sketch a more formal, explicit realization of bulk reconstruction subregion by subregion.

Having stated the problem of subregion duality, we now state its solution. The idea of its proof \cite{ewrproof} is explained at the end of Section~\ref{sec:jlms}.
\begin{tcolorbox}
{\bf Entanglement Wedge of CFT region $A$:}
\newline
Find an extremal surface $\Xi$ in proposal~(\ref{hrt}), then identify the subregion $\tilde{\Sigma}$ of a bulk Cauchy slice, which is bound by $\Xi$ and $A \subset {\rm CFT}$: $\partial \tilde{\Sigma} = \Xi \cup A$. The entanglement wedge of $A$ is the bulk domain of dependence of $\tilde{\Sigma}$:
\begin{equation}
\mathcal{W}(A) = D_{\rm bulk}(\tilde{\Sigma})
\label{ewdef}
\end{equation}
The entanglement wedge $\mathcal{W}(A)$ is dual $\rho_A$. It solves the subregion duality problem.
\end{tcolorbox}

\begin{figure}[tbp]
\centering 
\includegraphics[width=.5\textwidth]{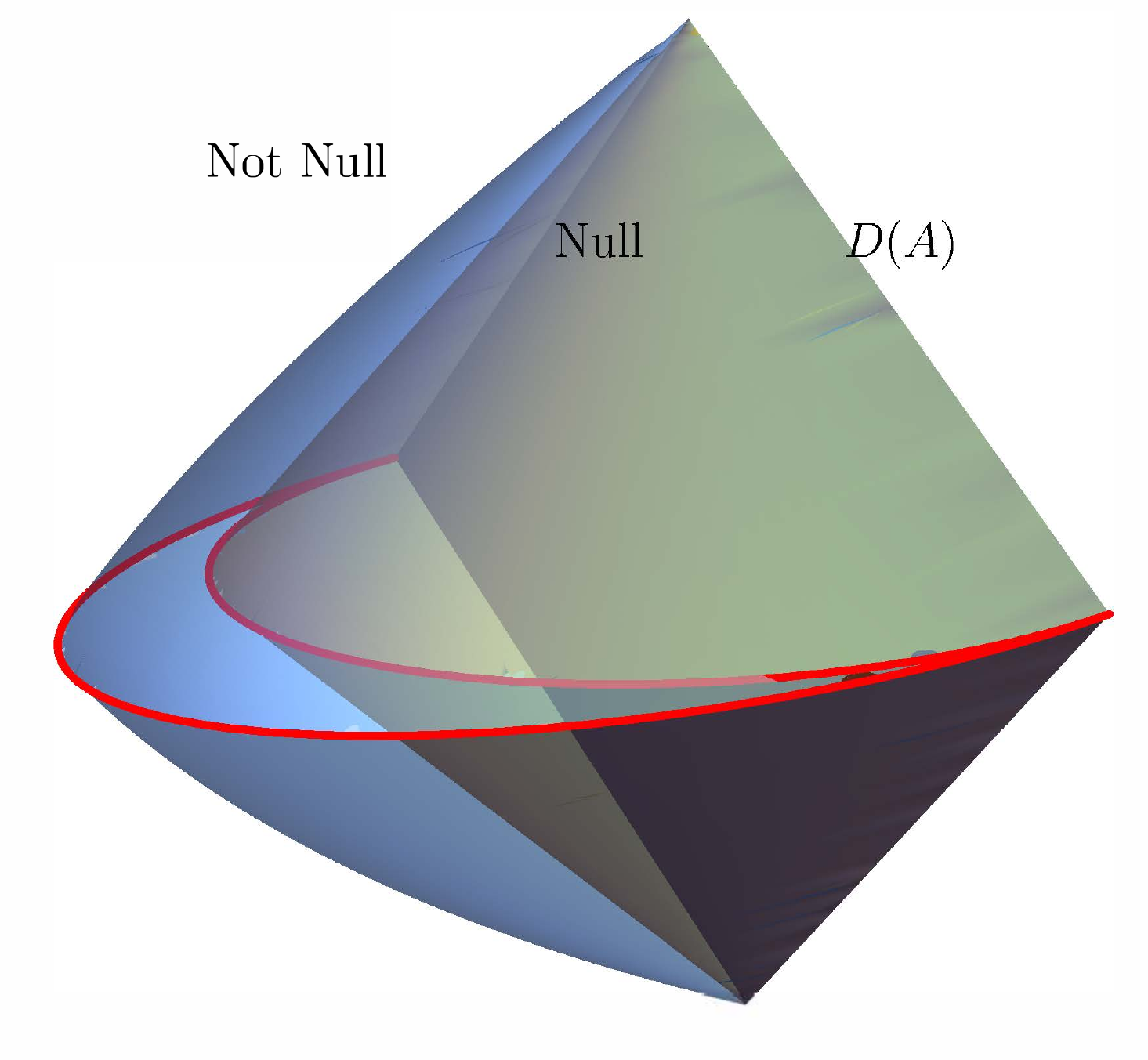}
\hfill
\caption{\label{wedge} Comparison between \textchg{the} entanglement wedge (blue) and causal wedge (green) with respect to the same boundary causal diamond $D_{\rm CFT}(A)$. The boundary of the causal wedge is \textchg{a smooth null surface. The boundary of the entanglement wedge is not generally smooth due to caustics though it is null wherever it is smooth.} \textdel{generically not null due to caustics.}}
\end{figure}

\paragraph{Salient properties of entanglement wedges:}
\begin{itemize}
\item The maximizing Cauchy slice ${\Sigma}$ in (\ref{hrt}) is not unique, and neither is $\tilde{\Sigma}$. But they all identify the same surface $\Xi$ and therefore all choices of $\tilde{\Sigma}$ have the same boundary $\partial \tilde{\Sigma} = \Xi \cup A$. Consequently, they all have the same domain of dependence, so $\mathcal{W}(A)$ is well-defined. 
\item As long as the Quantum Focusing Conjecture (a quantum generalization of the null energy condition) is satisfied in the bulk, nested CFT regions have nested entanglement wedges \cite{Akers:2016ugt}: 
\begin{equation}
D_{\rm CFT}(A) \subset D_{\rm CFT}(B) 
\quad \Rightarrow \quad \mathcal{W}(A) \subset \mathcal{W}(B)
\label{nesting}
\end{equation} 
\item The entanglement wedge of a region asymptotes to its boundary domain of dependence; see Figure~\ref{wedge} for illustration.
\begin{equation}
\mathcal{W}(A) \cap {\rm CFT} = D_{\rm CFT}(A)
\label{boundarylimit}
\end{equation}
If this were violated, a boundary point $X$ that is in $\mathcal{W}(A)$ but outside $D_{\rm CFT}(A)$ would be timelike-separated from at least one point $Y$ on the CFT Cauchy slice $\Sigma \cap {\rm CFT}$, which is outside $A$. (Here $\Sigma$ is the bulk slice stipulated in definition~\ref{hrt}.) By virtue of being on the asymptotic boundary, this $Y$ should be part of $\partial \tilde{\Sigma}$ in equation~(\ref{ewdef}), yet it is neither in $A$ nor in $\Xi$.
\item Entanglement wedges generally reach deeper than bulk causality would na{\"\i}vely suggest. To make this statement rigorous, imagine what it would mean to solve subregion duality using bulk causality alone. Because the region dual to $\rho_A$ must asymptote to the boundary domain of dependence of $A$ as in (\ref{boundarylimit}), the simplest option is the {\it causal wedge} $\mathcal{C}(A)$: the bulk region where the future and the past of $D_{\rm CFT}(A)$ intersect, see Figure~\ref{wedge}. So long as the null energy condition holds, this region is a subset of the entanglement wedge \cite{Wall:2012uf, causalwedges1, causalwedges2}
\begin{equation}
\mathcal{C}(A) \subset \mathcal{W}(A)
\label{notcausalwedge}
\end{equation}
and it is {\it not} the answer to the subregion duality problem.
\item \textchg{In spacetimes obeying the null energy condition,} inclusion~(\ref{notcausalwedge}) is a consequence of the Gao-Wald theorem \cite{gaowald}, which says that if two boundary points $X$ and $Y$ are null-separated on boundary, their separation through the bulk is generically spacelike, null in marginally fine-tuned cases, and never timelike. Therefore, a geodesic segment that goes from the bottom of $D_{\rm CFT}(A)$ to the HRT surface $\Xi$, as well as one that goes from $\Xi$ to the top of $D_{\rm CFT}(\bar{A})$, must be spacelike. This is equivalent to (\ref{notcausalwedge}). 
\item The entanglement wedge of $A$ contains bulk points, which are spacelike-separated from $D_{\rm CFT}(A)$, i.e. not in causal contact with anywhere in the CFT where the physics could be calculated from initial data on $A$. See Figure~\ref{wedge} for illustration. This fact is synonymous with (\ref{notcausalwedge}), but we reiterate this point because it is one of the most surprising and counterintuitive features of subregion duality.
\item The cases discussed in Section~\ref{sec:adsbhs} are marginal in that the Gao-Wald gravitational time delay vanishes and, as a result, the entanglement wedge agrees with the causal wedge. A set of sufficient conditions for this circumstance was characterized from the CFT point of view in \cite{localhmods}. 
\item The boundary of the entanglement wedge has caustics: points where initially parallel lightlike rays meet and where the expansion from equation~(\ref{raychaudhuri}) diverges. This is the only way that a part of $\partial \mathcal{W}(A)$ can be spacelike (as equation~\ref{notcausalwedge} implies) even though $\mathcal{W}(A)$ is the domain of dependence of a region. The pattern was caustics was studied and illustrated in \cite{causalwedges2}.
\end{itemize}

\subsubsection{Modular Hamiltonians}
\label{sec:jlms}
The other aspect of stationary black holes, which we wish to creatively implant onto general entanglement wedges, is their timelike Killing symmetry. In the near-horizon region where the metric is approximately Rindler, this symmetry generates transversal boosts; we have seen examples in equations~(\ref{schwarzrindler}) and (\ref{btzkruskal}). On the asymptotic boundary, it generates global time translations. Recall that the Hawking temperature $\beta^{-1}$ measures the relative factor between the two: how fast proper time of accelerating observers in the near-horizon Rindler region passes relative to global time at infinity; see equations~(\ref{tempshortcut}) and (\ref{btzlight}) for examples. This temperature then enters the holographic description of the black hole as a thermal state
\begin{equation}
\rho(\beta) = e^{-\beta H}\,,
\label{rhothermal}
\end{equation}
where $H$ is the Hamiltonian of the boundary CFT.

A typical entanglement wedge does not have a global Killing symmetry. Therefore, in going from black holes to generic entanglement wedges, we will be forced to discard the global, far-from-horizon considerations: that $H$ generates time translations and sets a temperature readable at infinity. But other aspects of $H$ generalize well. Because every reduced density matrix $\rho_B$ of a CFT region $B$ is a trace class non-negative operator, we can take its logarithm and define the {\it modular Hamiltonian} $H_{\rm mod} \equiv - \log \rho_B$. 
To the extent that the near-horizon (or near-Ryu-Takayanagi surface) geometry is the same for all entanglement wedges, this operator should---like $H$ in geometries dual to (\ref{rhothermal})---generate orthogonal boosts near the RT surface \cite{Faulkner:2013ana, Jafferis:2014lza, refjlms}:
\begin{equation}
H_{\rm mod} \equiv - \log \rho_B =  \hat{A} + H_{\rm boost}^\perp + \ldots
\label{boostgen}
\end{equation}
Here $\hat{A}$ is a possible operator that localizes on the RT surface in the bulk and therefore does not affect the action of $H_{\rm mod}$ as a boost. The ellipses include corrections suppressed by powers of $N$ and (signatures of) deviations from the assumed Rindler-like geometry, which are controlled by local curvature invariants near the Ryu-Takayanagi surface and the dominant eigenvalue of its extrinsic curvature tensor \cite{tomaitor}. Failing to define a global Killing field, further away from the RT surface $H_{\rm mod}$ becomes an intractable, non-local operator, which means that the ellipses in (\ref{boostgen}) dominate. But even in that case $H_{\rm mod}$ always identifies a one-parameter family of operators with identical correlation functions in $\rho_B$: 
\begin{equation}
{\rm Tr}\, e^{-H_{\rm mod}} \mathcal{O} 
= {\rm Tr}\, e^{-H_{\rm mod}} e^{-i s H_{\rm mod}} \mathcal{O} e^{i s H_{\rm mod}}
\equiv {\rm Tr}\, e^{-H_{\rm mod}} \mathcal{O}(s)
\label{modularshift}
\end{equation}
We call the transformation $\mathcal{O} \to \mathcal{O}(s)$ modular time translation, in analogy to similar statements involving time translations in the thermal state where $H_{\rm mod} = \beta H$. 

\paragraph{Consequences and extensions of (\ref{boostgen}):} 
\begin{itemize}
\item The modular Hamiltonian is derived from, and contains all the information about, the quantum entanglement between regions $B$ and $\bar{B}$ in the CFT. Identification~(\ref{boostgen}) is a striking example of geometrization of entanglement.
\item The entanglement entropy (\ref{vonneumann}) is $S = {\rm Tr}\, \rho_B H_{\rm mod} \equiv \langle H_{\rm mod} \rangle_{\rho_B}$.
\item Ordinarily, we expect that the boost $H_{\rm boost}^\perp \subset H_{\rm mod}(B)$ transforms a neighborhood of an RT surface on either side of it: the $B$-facing side, and the $\bar{B}$-facing side. This is in some tension with how we defined $H_{\rm mod}(B)$---as an operator acting on $\mathcal{H}_B$ and not on $\mathcal{H}_{\bar{B}}$. One might try to make sense of (\ref{boostgen}) by restricting $H_{\rm boost}^\perp$ to act only on the $B$-facing side of the RT surface, but such an operator is necessarily singular. This pathology is holographically dual to the fact that in quantum field theory defining the modular Hamiltonian as $H_{\rm mod}(B) = - \log \rho_B$ is not rigorous. As mentioned in Section~\ref{sec:unruhhawking}, the CFT Hilbert space may not factorize as $\mathcal{H}_B \otimes \mathcal{H}_{\bar{B}}$, in which case we cannot even take a partial trace to define $\rho_B$ \cite{chr, djordje, donnellywall1, donnellywall2, ronaksandip}. Even if the Hilbert space does factorize, the modular Hamiltonian depends on a largely arbitrary choice of boundary conditions at the entangling surface \cite{yuji, marksrednicki, jenniferdjordje}. The way around this problem is to consider the \emph{two-sided} or \emph{full} modular Hamiltonian \cite{blancocasini, anec}:
\begin{equation}
H^{\rm total}_{\rm mod}(B) \equiv H_{\rm mod}(B) - H_{\rm mod}(\bar{B})
\label{hmod2sided}
\end{equation} 
While the individual terms on the right hand side may not be well defined, the combination on the left always is \cite{araki76, haagbook}. There is a relative minus sign on the right hand side because a boost that acts `toward the future' in the right Rindler wedge acts `toward the past' in the left Rindler wedge; see the left panel of Figure~\ref{flatrindler}. The same relative minus sign cancels off ultraviolet pathologies in $H_{\rm mod}(B)$ and $H_{\rm mod}(\bar{B})$. We ignore this complication elsewhere in the text. 
\item When we consider how perturbations $\rho_B \to \rho_B + \delta \rho$ affect the entanglement entropy, the term ${\rm Tr}\, \rho_B \delta H_{\rm mod} = - {\rm Tr}\, \delta \rho \subset \delta {\rm Tr}\, \rho_B H_{\rm mod}$ vanishes lest the perturbed state be improperly normalized. Therefore:
\begin{equation}
\delta S = {\rm Tr}\, (\delta \rho) H_{\rm mod} = \langle H_{\rm mod} \rangle_{\rm \delta \rho}
\equiv \Delta \langle H_{\rm mod} \rangle_\rho
\label{ent1stlaw}
\end{equation}
This should be distinguished from $\langle \Delta H_{\rm mod} \rangle_\rho$ from equation~(\ref{relativeentropy}). Equation~(\ref{ent1stlaw}) is often called the \emph{entanglement first law}. 
\item A bulk counterpart of (\ref{ent1stlaw}) is that for infinitesimal changes in the metric, one can compute the change in holographic entanglement entropy by integrating $\delta \sqrt{h}$ (the induced metric) on the old surface $\Xi$. Finding a new extremal surface corresponds to a shift in $H_{\rm mod}$ because the surface $\Xi$ is preserved by $H_{\rm mod}$: $\Xi$ won't shift if $H_{\rm mod}$ doesn't. Shifting the location of an RT surface is necessarily a second order effect.
\item Consider bulk field theory on a semiclassical background geometry dual to $\rho_B$. We may then ask about the bulk modular Hamiltonian, which describes the entanglement of perturbative bulk fields across the RT surface. This $H_{\rm mod}^{\rm bulk}$ acts near the RT surface as the orthogonal boost---a result we obtained in our Rindler analysis leading to~(\ref{thermal}). Therefore, it is at least plausible that equation~(\ref{boostgen}) may be rewritten as:
\begin{equation}
H_{\rm mod}^{\rm CFT} =  \hat{A} + H_{\rm mod}^{\rm bulk} + \mathcal{O}(N^{<0})
\label{jlms}
\end{equation}
The superscript on the left emphasizes that that modular Hamiltonian is defined in the CFT and includes effects to all orders in $N^{-1}$. In contrast, the bulk modular Hamiltonian $H_{\rm mod}^{\rm bulk}$ is $\mathcal{O}(N^0)$, which is how perturbative bulk fields are defined in AdS/CFT. The authors of \cite{refjlms} confirmed equation~(\ref{jlms}) with careful reasoning, a result known by the acronym JLMS.
\item In (\ref{jlms}), the term $\hat{A}$ is leading order in an expansion in $N^{-1}$ since it is the only term which can produce $\langle H_{\rm mod}^{\rm bulk} \rangle = {\rm Area}/4G\hbar$. (Recall that $G \propto l_P^{d-1}$ carries a negative power of $N$ relative to $L_{\rm AdS}$.) We interpret $\hat{A}$ as the area operator in the semiclassically quantized bulk theory; it is diffeomorphism-invariant because the surface $\Xi$ is. In cases where the CFT Hilbert space fails to factorize into $\mathcal{H}_B \otimes \mathcal{H}_{\bar{B}}$, $\hat{A}$ might involve extra ingredients, which reflect the scheme-dependence of computing entanglement entropy in such settings \cite{chr}.
\item A measure of distinguishability of two density matrices $\rho$ and $\sigma$ is the relative entropy:
\begin{equation}
S(\rho|\sigma) = {\rm Tr}\,(\rho \log \rho - \rho \log \sigma) 
\equiv \langle \Delta H_{\rm mod} \rangle_\rho
\label{relativeentropy}
\end{equation}
When $\sigma$ differs from $\rho$ by terms at most of order $\mathcal{O}(N^0)$, this expression is only sensitive to the bulk modular Hamiltonians \cite{refjlms, ewrproof}. 
We have:
\begin{align}
S(\rho^{\rm CFT} | \sigma^{\rm CFT}) 
& = {\rm Tr}\, \rho^{\rm CFT} \big(H_{\rm mod}^{{\rm CFT}, \sigma} - H_{\rm mod}^{{\rm CFT}, \rho}\big)
\label{equalrel} \\
& = {\rm Tr}\, \rho^{\rm bulk} \big(H_{\rm mod}^{{\rm bulk}, \sigma} - H_{\rm mod}^{{\rm bulk}, \rho}\big)
= S(\rho^{\rm bulk} | \sigma^{\rm bulk}) 
\nonumber
\end{align}
where in the last line we refer to the reduced density matrices of the bulk perturbative fields on one side of $\Xi$.
\item Taking $\rho^{\rm CFT}$ and $\sigma^{\rm CFT}$ to be different states on some CFT region $B$, we see that their distinguishability is entirely captured by the distinguishability of perturbative fields in the entanglement wedge of $B$, which is where $\rho^{\rm bulk}$ and $\sigma^{\rm bulk}$
are defined.
\end{itemize}
This last point is so momentous that it deserves its own graybox:

\begin{tcolorbox}
{\bf Proof of subregion duality \cite{ewrproof}:}
\newline
The bulk region $\mathcal{W}(B)$ is dual to the CFT state $\rho_B^{\rm CFT}$ because distinguishing $\rho_B^{\rm CFT}$ from another state $\sigma_B^{\rm CFT}$ can always be done by inspecting $\mathcal{W}(B)$ and using (\ref{equalrel}). 
\end{tcolorbox}
\noindent
But for a complete proof, Reference~\cite{ewrproof} had to tackle an additional subtlety, to which we turn presently.

\subsection{Error correction}
\label{sec:error}
\begin{figure}
     \centering
     \begin{subfigure}[t]{0.4\textwidth}
         \centering
         \includegraphics[width=\textwidth]{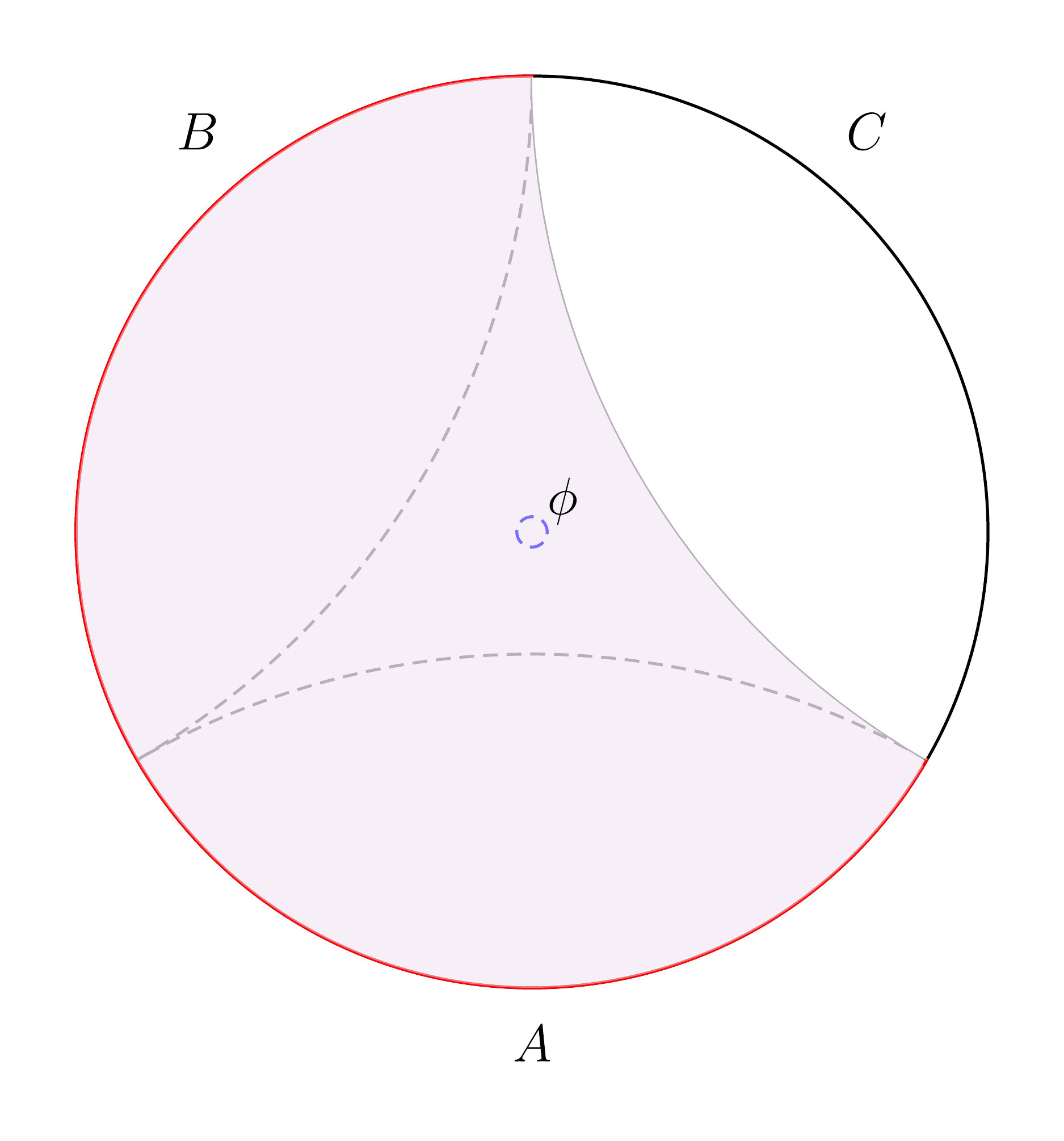}
     \end{subfigure}
     \hfill
     \begin{subfigure}[t]{0.4\textwidth}
        \centering
         \includegraphics[width=\textwidth]{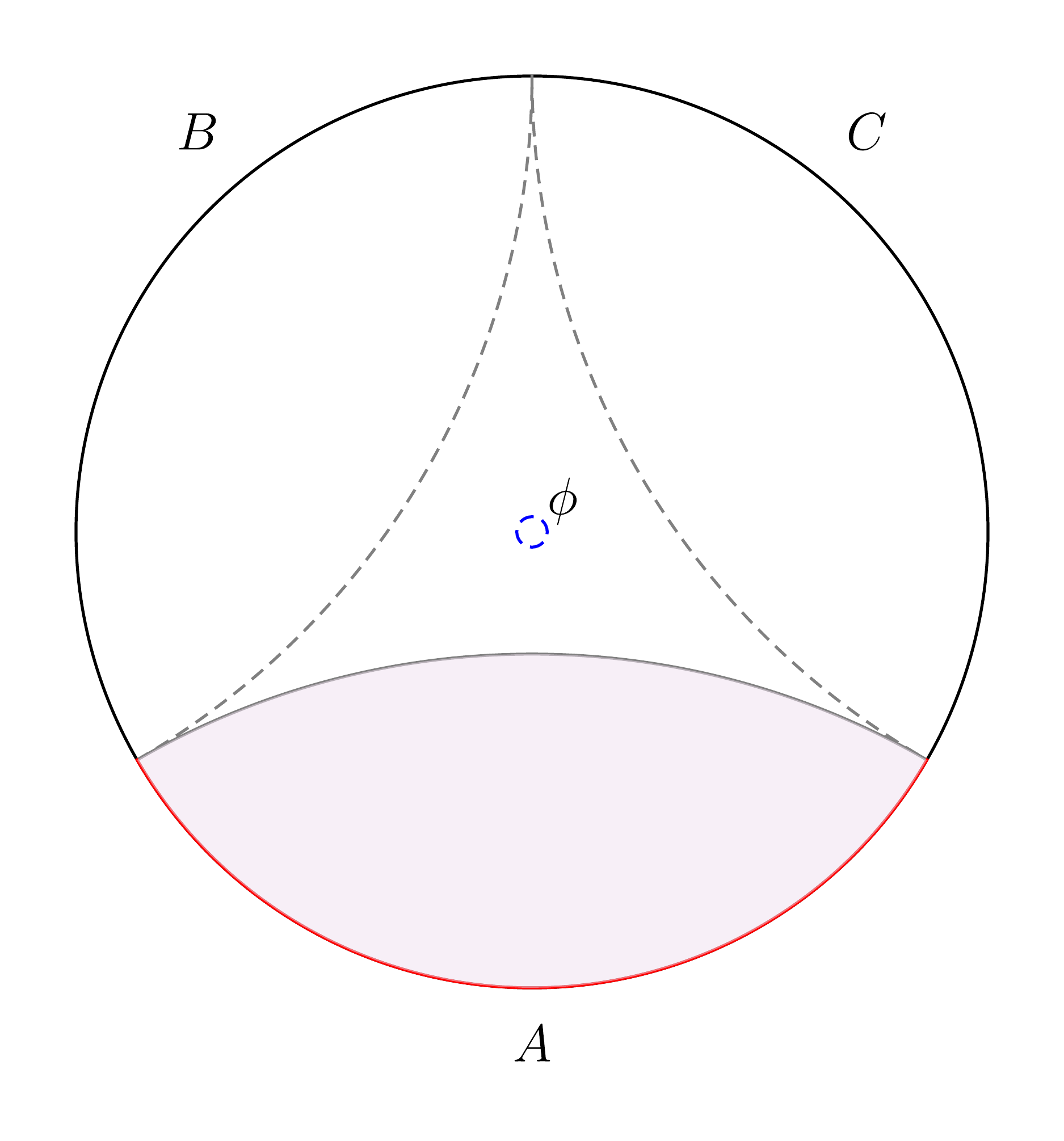}
     \end{subfigure}
        \caption{The paradox of reconstruction, which illustrates the error-correcting property of the bulk. The bulk field $\phi$ in the center can be reconstructed from $\rho_{AB}$ (left) and likewise from $\rho_{BC}$ or $\rho_{CA}$, but not from $\rho_A$ (right) nor from $\rho_B$ nor $\rho_C$.}
        \label{QEC}
\end{figure}

Consider the setup in Figure~\ref{QEC}. We have divided a Cauchy slice of the boundary CFT into three disjoint, connected regions $A, B, C$. We identify the three entanglement wedges\footnote{We use the notation $AB = A \cup B$, which is standard in the holographic literature.} $\mathcal{W}(AB),~\mathcal{W}(BC),~\mathcal{W}(CA)$ and observe that they all intersect in the middle of the bulk spacetime. This is very intriguing. Subregion duality says that a bulk field theory operator $\phi(x)$ supported in that region should be reconstructible (in the sense of equation~\ref{hkllreconstr}) in terms of CFT operators, which act in $AB$ alone, or in $BC$ alone, or in $CA$ alone. Therefore $\phi(x)$ should act trivially on $C$ and $A$ and $B$, but the only such operator is the identity!

The insight of \cite{errorref, ewrproof}---and a way out of this paradox---is that bulk reconstruction (\ref{hkllreconstr}) should not be understood as a fundamental equality of operators, but as equality of correlation functions in a certain subspace of the Hilbert space called code subspace $\mathcal{H}_{\rm code}$. For perturbative fields such as $\phi(x)$, a natural code subspace is the space of states where bulk effective field theory is valid, in which case $\mathcal{H}_{\rm code}^{\perp}$ comprises stringy bulk states as well as states whose quantum numbers are parametric in $N$. In a certain precise sense, the choice of code space can be understood as renormalizing the bulk Newton's constant \cite{rterror}. 

All preceding assertions involving bulk fields---from (\ref{hkllreconstr}) to the JLMS relation~(\ref{jlms})---must be understood as valid only within $\mathcal{H}_{\rm code}$. For example, the proof of subregion duality \cite{ewrproof} technically establishes that any operator $\phi_{\rm code}$ which acts on $\mathcal{H}_{\rm code}$ and localizes in $\mathcal{W}(B)$ has at least one reconstruction in the $B$-factor of the full Hilbert space. 

\paragraph{A toy model} The $\phi(x)$ in the example above must be reconstructible in the CFT even when we lose access to any one of regions $A, B, C$---though not if we lose access to two of them. The task is thus to reconstruct $\phi(x)$ in the presence of a single-site error (if the data in one region get corrupted) or a single-site erasure. A standard quantum computing protocol \cite{Shor:1995, Gottesman:1997zz} achieves error/erasure correction. (For an earlier application of quantum error correction to gravity see \cite{verlinderserrorcorr}.) The insight of \cite{errorref} is that local bulk fields in AdS/CFT work the same way. 

The simplest model comprises three physical qutrits: systems, which can be in one of three states $|0\rangle, |1\rangle, |2\rangle$. The three physical qutrits stand for the three regions $A, B, C$. Our code subspace $\mathcal{H}_{\rm code}$ will contain one so-called logical qutrit, which represents the Hilbert space of bulk perturbation theory. If we embed a copy of $\mathcal{H}_{\rm code}$ in $\mathcal{H}_A \otimes \mathcal{H}_B \otimes \mathcal{H}_C$ via
\begin{align}
    \ket{\tilde{0}} & = \frac{1}{\sqrt{3}}\big( \ket{000} + \ket{111} + \ket{222}\big) \nonumber \\
    \ket{\tilde{1}} & = \frac{1}{\sqrt{3}}\big( \ket{012} + \ket{120} + \ket{201}\big) \label{codeembed} \\
    \ket{\tilde{2}} & = \frac{1}{\sqrt{3}}\big( \ket{021} + \ket{102} + \ket{210}\big) \nonumber\,
\end{align}
then the logical qutrit will be protected against error/erasure of any one physical qutrit. Here protection means that any superposition $|\psi\rangle = q_0 |0\rangle + q_1 |1\rangle + q_2 |2\rangle \in \mathcal{H}_{\rm code}$ can be recovered from $|\tilde{\psi}\rangle = q_0 |\tilde{0}\rangle + q_1 |\tilde{1}\rangle + q_2 |\tilde{2}\rangle$ by a unitary transformation, which acts non-trivially only on the remaining two qutrits. Specifically, for the erasure of the third qutrit we use
\begin{align}
& \ket{00} \rightarrow \ket{00}\;\;\ket{11} \rightarrow \ket{01}\;\;\ket{22} \rightarrow \ket{02} 
\nonumber \\
U_{AB}\!: \qquad     
& \ket{01} \rightarrow \ket{12}\;\;\ket{12} \rightarrow \ket{10}\;\;\ket{20} \rightarrow \ket{11} \\
& \ket{02} \rightarrow \ket{21}\;\;\ket{10} \rightarrow \ket{22}\;\;\ket{21} \rightarrow \ket{20}
\nonumber
\end{align}
and find:
\begin{equation}
    \left(U_{12} \otimes I_{3}\right) \ket{\tilde{\psi}} = \ket{\psi} \otimes \frac{1}{\sqrt{3}}\left( \ket{00} + \ket{11} + \ket{22}\right)
    \label{explcorr}
\end{equation}
Converting this into a transformation of operators, we recognize that an operator $\mathcal{O}_{\rm code}$ can be emulated by a physical operator $\mathcal{O}_{AB} = U_{AB} \mathcal{O}_{\rm code} U_{AB}^\dagger$ in a way that is independent of the $C$-qutrit. The operator $\mathcal{O}_{AB}$ symbolizes the reconstruction of a bulk operator $\mathcal{O}_{\rm code}$ in the entanglement wedge of $AB$. 

Correlation functions of operators of the form $U_{AB} \mathcal{O}_{\rm code} U_{AB}^\dagger$ are, by construction, the same as those of $\mathcal{O}_{\rm code}$. But operators, which cannot be written as $U_{AB} \mathcal{O}_{\rm code} U_{AB}^\dagger$ (or similarly with $U_{BC}$ or $U_{CA}$), have no meaning in $\mathcal{H}_{\rm code}$ and neither do three-qutrit states outside the span of (\ref{codeembed}). Those are analogous to operators and states, which go outside bulk perturbation theory and which can exhibit non-local behavior in the bulk. 

\paragraph{Explicit recovery channel for bulk reconstruction}
A key piece of data that characterizes an error correcting scheme is the way the logical degrees of freedom (the tilded qutrit in the toy model) is embedded in the larger Hilbert space that offers protection from error. An example is equation~(\ref{codeembed}). In a context where the protagonists are density operators $\rho$ rather than pure states $|\psi\rangle \in \mathcal{H}$, we express this piece of data as a \emph{quantum channel}---a trace-preserving positive map, which sends density matrices to density matrices:
\begin{equation}
\textrm{quantum channel}~\mathcal{N}: \quad  \rho \to \mathcal{N}[\rho] 
\end{equation}
A quantum channel is just like a quantum mechanical operator $\mathcal{O}: |\psi\rangle \to \mathcal{O}|\psi\rangle$, except that it acts on mixed states $\rho$ rather than pure states $|\psi\rangle$. 

As a generalization of~(\ref{codeembed}), we consider a quantum channel $\mathcal{N}_0$, which embeds a density operator on $\mathcal{H}_{\rm code}$ in some larger space of density operators for error protection. The resulting density operator $\mathcal{N}_0[\rho]$ is then subject to an error or erasure, which is also conceptualized as a quantum channel $\mathcal{E}: \mathcal{N}_0[\rho] \to \mathcal{E}[\mathcal{N}_0[\rho]]$. Let us call the combined effect of encoding and erasure $\mathcal{N}[\rho]$:
\begin{equation}
\mathcal{N}: \quad \rho \xrightarrow{~{\rm encode}~} \mathcal{N}_0[\rho] \xrightarrow{~{\rm erasure}~} \mathcal{E}\circ\mathcal{N}_0[\rho] \equiv \mathcal{N}[\rho]
\label{quantumchanneln}
\end{equation}
Suppose a class of states $\rho$ obeys the following equality of relative entropies
\begin{equation}
S(\rho | \sigma) = S(\mathcal{N}[\rho]\, | \, \mathcal{N}[\sigma])
\label{relequal}
\end{equation}
for some reference state $\sigma$. Then and only then \cite{petz}, there exists an explicit \emph{recovery channel} $\mathcal{R}$ such that $\mathcal{R} \circ \mathcal{N} [\rho] = \rho$. The recovery channel, called the Petz map, works for all $\rho$ obeying (\ref{relequal}). It is given explicitly by
\begin{equation}
\mathcal{P}_{\sigma, \mathcal{N}} =  
\sigma^{1/2} \mathcal{N}^*[\mathcal{N}[\sigma]^{-1/2} (\ldots) \mathcal{N}[\sigma]^{-1/2}] \sigma^{1/2}\,,
\label{petzmap}
\end{equation}
where $\mathcal{N}^*$ is the adjoint of $\mathcal{N}$. Note that the recovery channel---as well as its domain of applicability~(\ref{relequal})---depends on a reference state $\sigma$. But, thankfully, it does not depend on $\rho$, so it can be used to recover to $\rho$: $\mathcal{P}_{\sigma, \mathcal{N}} \circ \mathcal{N}[\rho] = \rho$.

The Petz map allows an explicit realization of subregion duality \cite{subregionchannel}. The error correcting property of the holographic map defines a globally defined quantum channel $\mathcal{N}_0$, which maps states in bulk effective field theory to states in the CFT:
\begin{equation}
\mathcal{N}_0: \quad \rho^{\textrm{bulk effective field theory}} \to \rho^{\rm CFT}
\end{equation}
Further, from equation~(\ref{equalrel}), we know that the holographic map has the following property:
\begin{equation}
S(\rho^{\rm bulk}_{\mathcal{W}(A)} | \sigma^{\rm bulk}_{\mathcal{W}(A)}) 
= S(\rho^{\rm CFT}_A | \sigma^{\rm CFT}_A) + \mathcal{O}(1/N)
\label{ecjlms}
\end{equation}
The $1/N$ corrections are due to shifts in the extremal surface $\Xi(A)$ and redefinitions of entanglement wedge $\mathcal{W}(A)$, which are occasioned by the entanglement of perturbative bulk fields, and which we discuss in Section~\ref{sec:genentropy}. Equation~(\ref{ecjlms}) looks like the necessary and sufficient condition for the Petz map to work, though it is subject to $\mathcal{O}(1/N)$ corrections.

To apply the Petz map for bulk reconstruction, Reference~\cite{subregionchannel} had to tackle several problems. To begin, bulk states $\rho^{\rm bulk}_{\mathcal{W}(A)}$ and $\sigma^{\rm bulk}_{\mathcal{W}(A)}$ are not even in the domain of $\mathcal{N}_0$, which takes global bulk states to global CFT states. The authors of \cite{subregionchannel} first replace $\rho^{\rm bulk}_{\mathcal{W}(A)} \to \rho^{\rm bulk}_{\mathcal{W}(A)} \otimes \tau_{\,\overline{\mathcal{W}(A)}}$ for some fiducial, maximal rank $\tau$, and show that this only adds small, controllable error in all subsequent formulas. In terms of $\mathcal{N}_0$ and $\tau$, the CFT state on $A$ is:\footnote{This formula assumes that the bulk and boundary Hilbert spaces factorize, but as shown in \cite{subregionchannel}, the assumption can be removed. We sketched the problem of assuming that Hilbert spaces admit tensor product decomposition around equation~(\ref{hmod2sided}).}
\begin{equation}
\rho^{\rm CFT}_A = {\rm Tr}_{\bar{A}}\, \mathcal{N}_0 [\rho^{\rm bulk}_{\mathcal{W}(A)} \otimes \tau_{\,\overline{\mathcal{W}(A)}}]
\label{eccftstate}
\end{equation}
We think of the quantum channel $\rho^{\rm bulk}_{\mathcal{W}(A)} \to \rho^{\rm CFT}_A$ defined by~(\ref{eccftstate}) as $\mathcal{N}$ in equation~(\ref{quantumchanneln}); the tracing out of $\bar{A}$ is like the erasure step. Equation~(\ref{ecjlms}) is now in the general form~(\ref{relequal}), but it is subject to $\mathcal{O}(1/N)$ corrections. In this circumstance, a \emph{twirled Petz map} \cite{twirledpetz}
\begin{equation}
\mathcal{R}_{\sigma, \mathcal{N}} = \frac{\pi}{2} 
\int_{-\infty}^\infty \frac{dt}{\cosh \pi t +1}\, \sigma^{-it/2}\, 
\mathcal{P}_{\sigma, \mathcal{N}} [\mathcal{N}[\sigma]^{it/2} (\ldots) \mathcal{N}[\sigma]^{-it/2}] \,\sigma^{it/2}
\end{equation}
allows an approximate recovery, in the sense that $|| \rho - \mathcal{R}_{\sigma, \mathcal{N}} \circ \mathcal{N}[\rho] ||$ is also of order $\mathcal{O}(1/N)$ under reasonable operator norms. 

The adjoint channel $\mathcal{R}_{\sigma, \mathcal{N}}^*$ then maps bulk operators $\phi(x)$ to CFT operators:
\begin{equation}
\phi(x) \to \mathcal{R}_{\sigma, \mathcal{N}}^*[\phi(x)]
\end{equation}
This is an upgraded version of HKLL reconstruction (\ref{hkllreconstr}), in the sense that the CFT correlation functions of operators $\mathcal{R}_{\sigma, \mathcal{N}}^*[\phi(x)]$ reproduce bulk correlation functions of $\phi(x)$ up to $\mathcal{O}(1/N)$ corrections. Unlike the original equation~(\ref{hkllreconstr}), however, the encoded operators $\mathcal{R}_{\sigma, \mathcal{N}}^*[\phi(x)]$ may not be smearings of local CFT operators. 

\section{Tensor networks}
\label{sec:tns}

It would be nice to have a visualization tool for the facts we learned in Sections~\ref{sec:rt}-\ref{sec:error}. Condensed matter theorists often study quantum states of multi-partite systems, frequently with peculiar entanglement structures, and have long felt a similar need. This is why they invented tensor networks. 

\begin{figure}[tbp]
\centering 
\includegraphics[width=\textwidth]{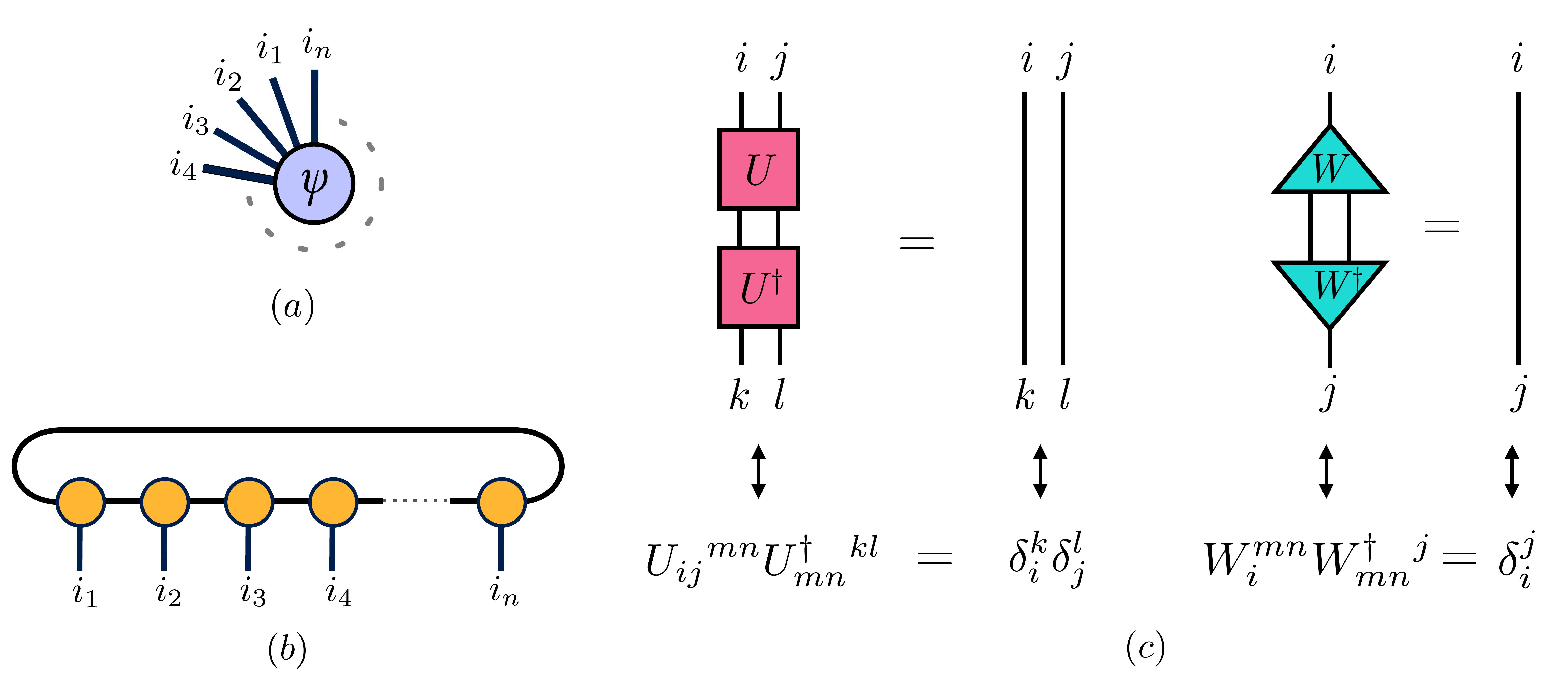}
\hfill
\caption{\label{fig:tns} 
Examples of tensor networks. (a) A featureless tensor network composed of a single
tensor prepares a generic $n$-body wavefunction. (b) A tensor network composed
of a chain of tensors contracted together called a matrix product state. (c) The unitary (resp.
isometric) character of the disentanglers and isometries in the MERA network (Figure~\ref{fig:meraflat} below) means that these tensors cancel out when contracted with their hermitian conjugates. Figure is reproduced with permission from \cite{Czech:2015kbp}.
}
\end{figure}

To represent exactly the quantum state of an $n$-body system, one needs a number of parameters that grows exponentially in $n$. A method to circumvent this difficulty is to consider special classes of states, which require fewer parameters but which hopefully encompass (or closely approximate) the quantum state of interest. We design such classes using a graphical notation, in which nodes with $k$ legs correspond to $k$-index tensors and individual legs represent vectors in Hilbert spaces. Joining two legs to make an internal bond means replacing two vectors $|\psi\rangle \in \mathcal{H}_1$ and $|\phi \rangle \in \mathcal{H}_2$ with a number---that is, applying a bilinear form. More physically, the joining of legs projects $|\psi\rangle \otimes |\phi \rangle$ onto entangled pairs of $\mathcal{H}_1 \otimes \mathcal{H}_2$ whose entanglement structure is determined by the bilinear form. Annotated examples of simple tensor networks are shown in Figure~\ref{fig:tns}. 

Designing a tensor network for a chosen class of states is an art as much as it is a science. Certain architectures are attractive because (i) they describe a many-body state using few parameters or (ii) they graphically encapsulate an interesting feature of a quantum state. One example is the Multi-scale Entanglement Renormalization Ansatz (MERA) network \cite{Vidal:2007hda, Vidal:2008zz, Evenbly:2007hxg}, which achieves (i): it captures ground states of 1+1-dimensional CFTs very efficiently. In Reference~\cite{brian}, Swingle made an explosively impactful observation: MERA also achieves (ii) in that it {\it looks like} the holographically dual anti-de Sitter space!

That observation ushered the ingress of tensor networks into gravity and holography. 
Below we describe in some detail three important classes of tensor networks useful to holographers, though other architectures \cite{Yang:2015uoa, tnsfortime, Czech:2016nxc, Evenbly:2017hyg, RTN2, VanRaamsdonk:2018zws, Bao:2019fpq, Caputa:2020fbc} are also useful. Tensor networks are used primarily as visualization tools and toy models, and have more than once motivated novel conjectures about the role of quantum information theory in gravity, for example in \cite{hartmanmaldacena, firstcomplexity, Miyaji:2015yva, Miyaji:2015fia, Takayanagi:2017knl}. 

\subsection{MERA}
\begin{figure}[tbp]
\centering 
\includegraphics[width=.9\textwidth]{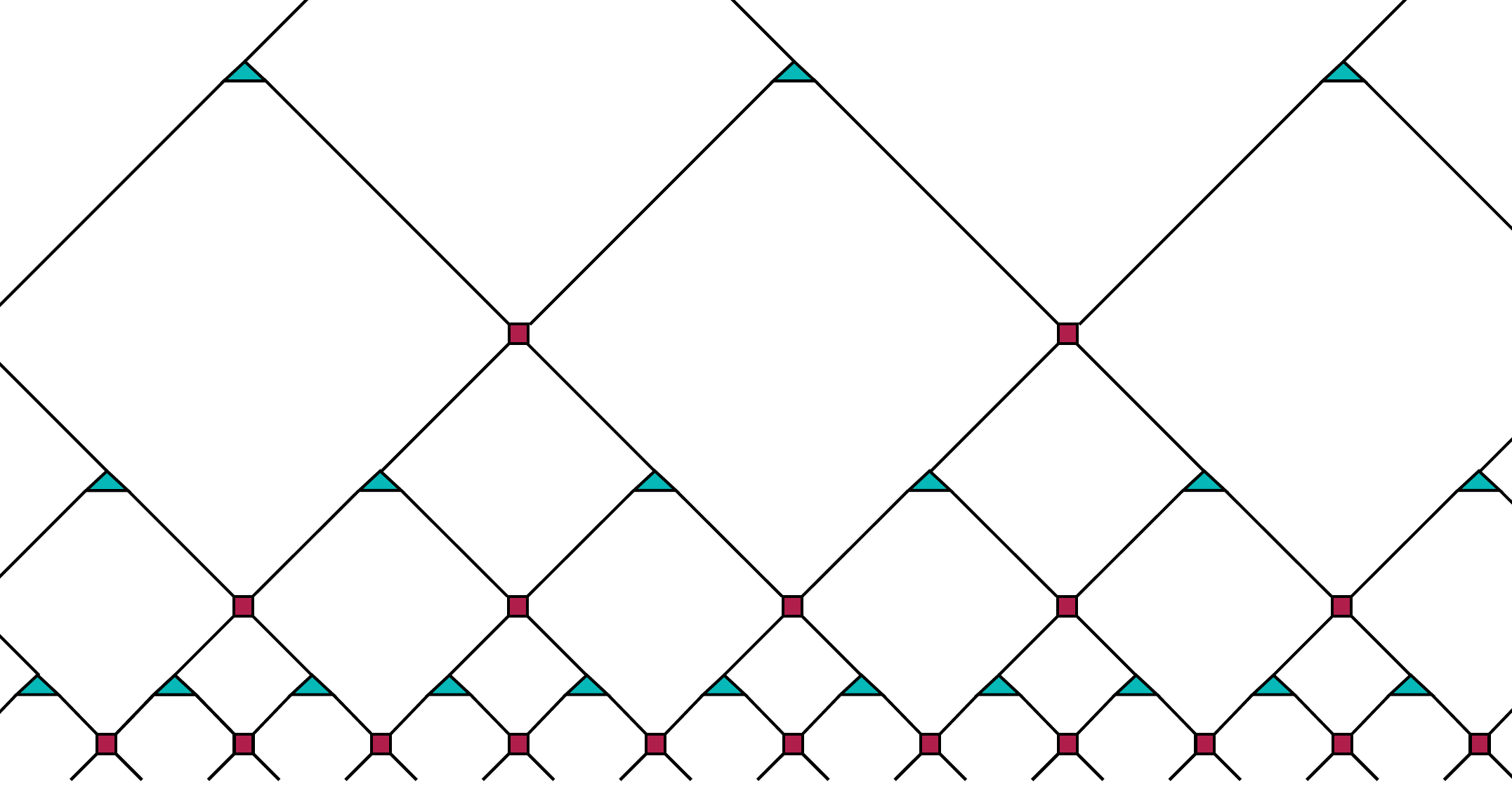}
\hfill
\caption{\label{fig:meraflat} 
The MERA tensor network. The blue triangles are isometries and the red squares are called disentanglers. Figure is reproduced with permission from \cite{Czech:2015kbp}.}
\end{figure}

\begin{figure}[tbp]
\centering 
\includegraphics[width=.8\textwidth]{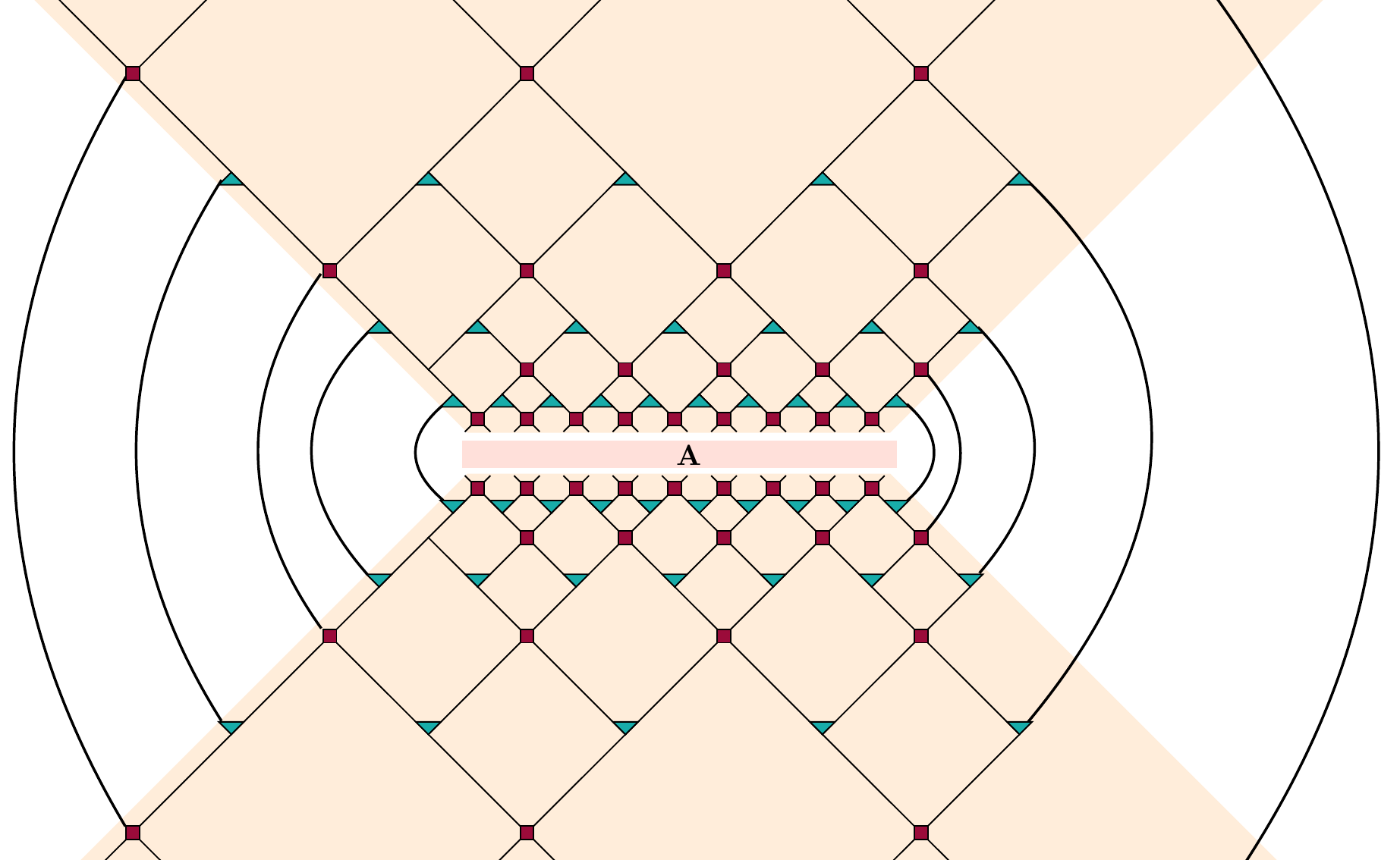}
\hfill
\caption{\label{fig:merarho} 
The density matrix of a region $A$ in the MERA representation. The two upside-down copies of the MERA network prepare $|\Psi\rangle$ and $\langle \Psi|$ in $\rho_A = {\rm Tr}_{\bar{A}} |\Psi\rangle \langle \Psi|$. The partial trace Tr$_{\bar{A}}$ produces cancellations because the component tensors are isometric; see panel~(c) of Figure~\ref{fig:tns}. Figure is reproduced with permission from \cite{Czech:2015kbp}.}
\end{figure}

The MERA network \cite{Vidal:2007hda, Vidal:2008zz, Evenbly:2007hxg} is shown in Figure~\ref{fig:meraflat}. The four-legged tensors are called disentanglers and the three-legged tensors are called isometries.  Disentanglers are unitary when read top to bottom. Isometries are called isometries because they can be extended to unitary transformations by the addition of an extra leg. This means that both unitaries and isometries cancel out when contracted with Hermitian conjugates, as shown in panel~(c) of Figure~\ref{fig:tns}. This in turn implies that the reduced density matrix of a region $\rho_A = {\rm Tr}_{\bar{A}} |\Psi\rangle \langle \Psi|$ takes the simplified graphical form shown in Figure~\ref{fig:merarho}.

\begin{figure}[tbp]
\centering 
\includegraphics[width=.9\textwidth]{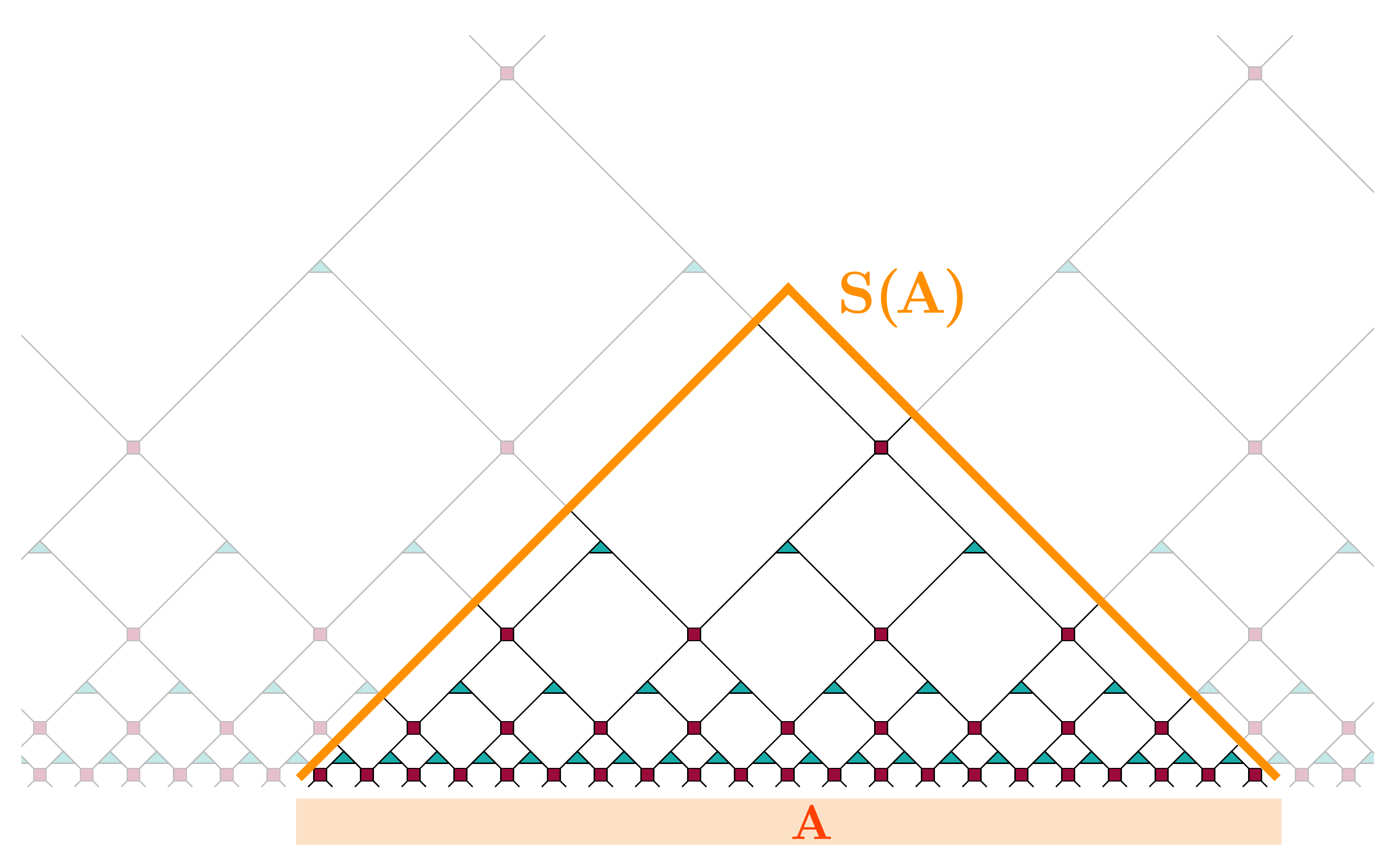}
\hfill
\caption{\label{fig:meracut} 
A cut through a tensor network exposes a Hilbert space $\mathcal{H}_{L,R}^{\rm cut}$ on either side. For any cut that subsumes $A$, log(dimension) of that Hilbert space is an upper bound on $S(A)$; see equation~(\ref{tnbound}). In practice, $S(A)$ is often proportional to $\log \dim \mathcal{H}_{L(i)}^{\rm cut}$ of the shortest cut. Figure is reproduced with permission from \cite{Czech:2015kbp}.}
\end{figure}

\begin{figure}[tbp]
\centering 
\includegraphics[width=.30\textwidth]{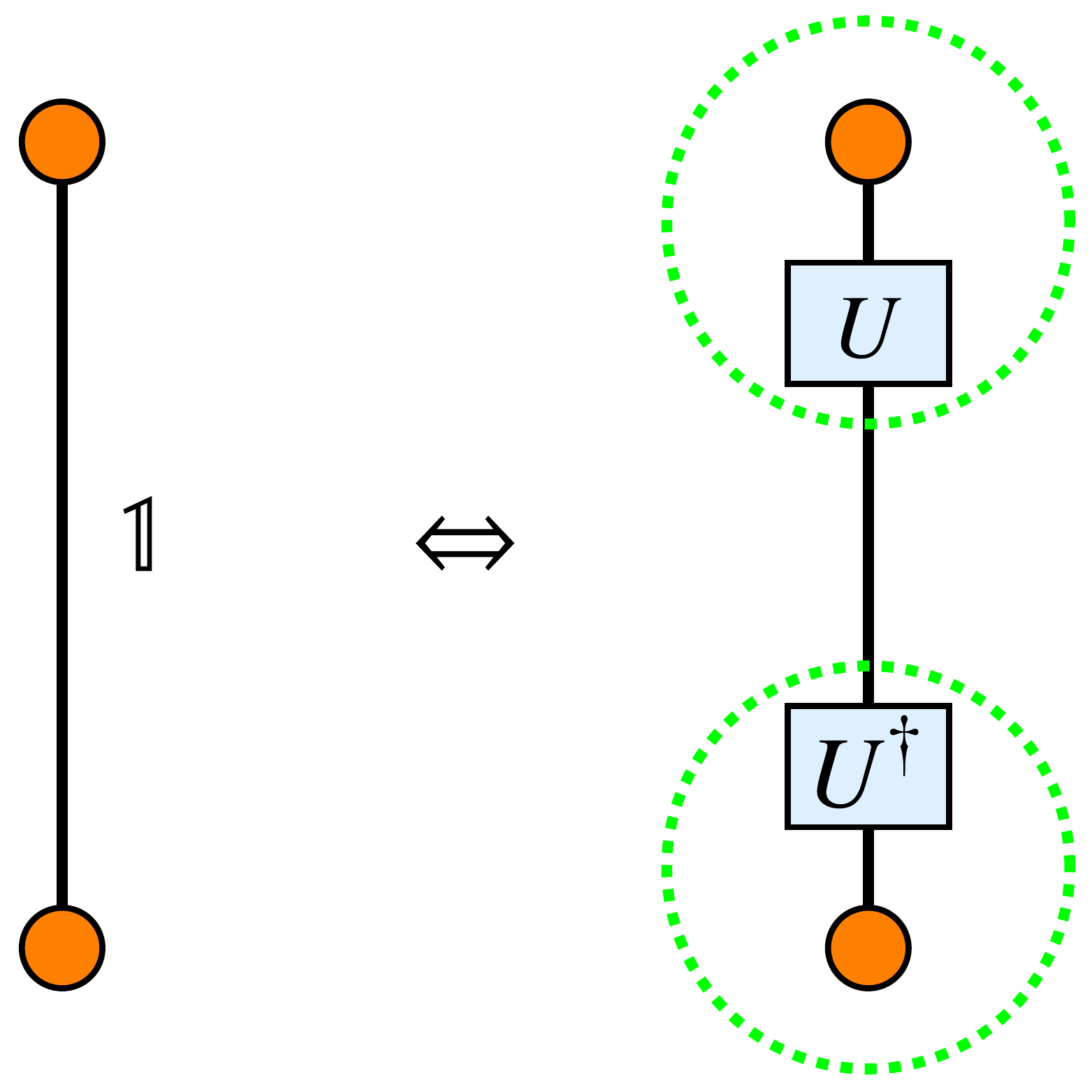}
\hfill
\caption{Every pair of contracted tensors is subject to a gauge symmetry because a pair of canceling unitaries can be reabsorbed into the definition of the tensors. Here, the big green circles can replace the original orange circles.}
\label{fig:uu}
\end{figure}

It is easy to see that the MERA network looks like the spatial slice of metric~(\ref{poincareh2}). The coordinate density of tensors dies off exponentially in successive layers, which captures the $g_{xx} = e^{2\rho}$ metric component of
\begin{equation}
\frac{dx^2 + dz^2}{z^2} = d\rho^2 + e^{2\rho} dx^2 \qquad {\rm where}~z = e^{-\rho}.
\label{h2again}
\end{equation}
Beyond this, there are several more striking similarities:
\begin{itemize}
\item Consider the entanglement entropy of an interval $A$. When we cut a tensor network into two pieces---one containing $A$ and the other containing $\bar{A}$---we expose on each side of the cut an additional Hilbert space $\mathcal{H}_{L,R}^{\rm cut}$; see Figure~\ref{fig:meracut} for illustration. $\mathcal{H}_{L}^{\rm cut}$ is the tensor product of Hilbert spaces $\mathcal{H}_{L(i)}^{\rm cut}$ associated to individual ruptured bonds $i$. The severed halves of the network now prepare some $|\phi\rangle_A \otimes |\phi\rangle_L^{\rm cut} \in \mathcal{H}_A \otimes \mathcal{H}_{L}^{\rm cut}$ and $|\psi\rangle_{\bar{A}} \otimes |\psi\rangle_R^{\rm cut} \in \mathcal{H}_{\bar{A}} \otimes \mathcal{H}_{R}^{\rm cut}$. Because the original state is obtained from these two by projecting onto entangled states of $\mathcal{H}_{L}^{\rm cut} \otimes \mathcal{H}_{R}^{\rm cut}$, the entanglement entropy $S(A)$ is bounded by:
\begin{equation}
S(A) \leq \log \dim \mathcal{H}_{L}^{\rm cut} 
= \sum_{i \in {\rm cut}} \log \dim \mathcal{H}_{L(i)}^{\rm cut} \equiv \sum_{i \in {\rm cut}} \log b_i
\label{tnbound}
\end{equation}
The quantities $b_i = \dim \mathcal{H}_{L(i)}^{\rm cut}$ on the right hand side are called bond dimensions. If all internal Hilbert spaces have equal dimensions, bound~(\ref{tnbound}) simply counts the number of severed bonds, so the tightest bound is supplied by the shortest cut. This is reminiscent of the Ryu-Takayanagi proposal~(\ref{rtsimplest}), which identifies entanglement entropies with the shortest cuts through bulk space. This argument applies to all tensor networks. 
\item The similarity is semi-quantitative, because vacuum entanglement entropies of intervals of size $L$ in two-dimensional CFTs evaluate to \cite{holzheyetal, cardycalabrese}
\begin{equation}
S(L) \propto \log (L/\mu),
\label{sinterval}
\end{equation}
where $\mu$ is an ultraviolet cutoff. In the MERA network, $\log (L/\mu)$ is also proportional to the number of legs in the shortest cut, as can be gauged by inspecting Figure~\ref{fig:meracut}. (We interpret $\mu$ as the distance between neighboring tensors in the bottom-most layer.) Even the shape of the cut shown in Figure~\ref{fig:meracut} shows a vague similarity with the Ryu-Takayanagi geodesic~(\ref{poincaregeod}). Most persuasively, in realistic networks the bound~(\ref{tnbound}) is either saturated or falls short of saturation by a fixed relative margin \cite{scutproportional}, so we may treat (\ref{tnbound}) with the shortest cut as an estimate of (\ref{sinterval}). 
\item Figure~\ref{fig:merarho} shows a MERA representation of the reduced density matrix $\rho_A$. A macroscopic part of the network cancels out of this object. We interpret this as a manifestation of subregion duality, with the canceled part representing the complementary entanglement wedge $\mathcal{W}(\bar{A})$, which is independent of $\rho_A$.
\item We may insert $\mathbb{1} = U U^\dagger$ on any internal bond of the tensor network, and then reabsorb $U$ and $U^\dagger$ separately into the nodes at each end; see Figure~\ref{fig:uu}. 
This is a local symmetry in the nodes of a tensor network, which applies independently of the details of the quantum state and theory. We interpret this symmetry as a gauge symmetry in the bulk of a holographic geometry. A standard argument in holography establishes that every global symmetry of a holographic CFT gives rise to a gauge symmetry in the dual AdS; see e.g.~\cite{beninilectures, Harlow:2018tng}. We stress that this similarity between tensor networks and bulk geometries is universal and applies equally well to other tensor networks. 
\item Even when a CFT has no interesting global symmetries, the global conformal symmetry can fall under the purview of the preceding argument. In the bulk, a part of this gauge symmetry corresponds to the freedom of choosing a vielbein \cite{Witten:1988hc} because global conformal symmetries are AdS isometries.
\end{itemize}


\paragraph{A wrinkle:} The MERA network also shows some characteristics, which are not present in the bulk AdS$_3$. The cuts that minimize~(\ref{tnbound}) always travel at 45$^\circ$ angles to the horizontal and bend at right angles at their deepest points; see Fig.~\ref{fig:meracut}. While neither feature has a counterpart in AdS$_3$, both are strongly reminiscent of lines of causality in Lorentzian spacetimes. This and related considerations motivated a proposal that MERA may be better associated with two-dimensional de Sitter space \cite{ds2beny}, which was later formalized and interpreted holographically using material of Section~\ref{sec:ksks} in \cite{Czech:2015kbp}. The final word on holographic interpretations of MERA came in \cite{guifrenull}, which argued that we should actually envision MERA as living on the causal cone of a bulk point in AdS$_3$. Meanwhile, Reference~\cite{Evenbly:2017hyg} devised a clever modification of the MERA network, which evades the concerns of \cite{Czech:2015kbp} and appears maximally faithful to geometry~(\ref{h2again}).

But the minutiae of how MERA relates to AdS$_3$ are of secondary importance. The power of Swingle's observation resides primarily in highlighting tensor networks as visualization tools and toy models to guide our intuitions about information theory in holography. 

\subsection{HaPPY code}
A prerequisite for this subsection is the notion of a {\it perfect tensor}. The defining property of a perfect tensors with $k$ legs is that it is isometric (can be extended to unitary transformations) for any division of its legs into $n \geq k/2$ covariant and $k-n$ contravariant indices. The existence of perfect tensors was proved in \cite{ptexist1, ptexist2, ptexist3}. Their properties depending on the number of legs and bond dimension are a rich area of research. 

\begin{figure}
     \centering
     \begin{subfigure}[t]{0.4\textwidth}
         \centering
         \includegraphics[width=\textwidth]{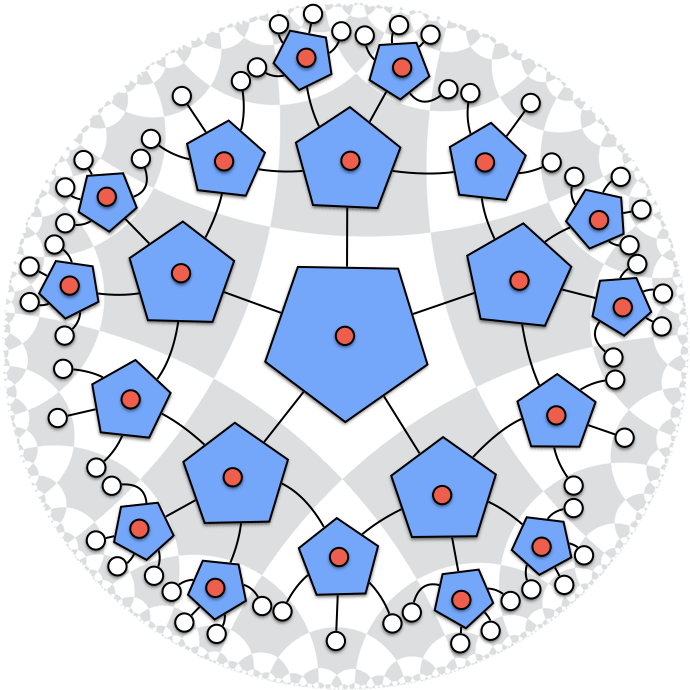}
         \label{pentagon}
     \end{subfigure}
     \hfill
     \begin{subfigure}[t]{0.4\textwidth}
        \centering
         \includegraphics[width=\textwidth]{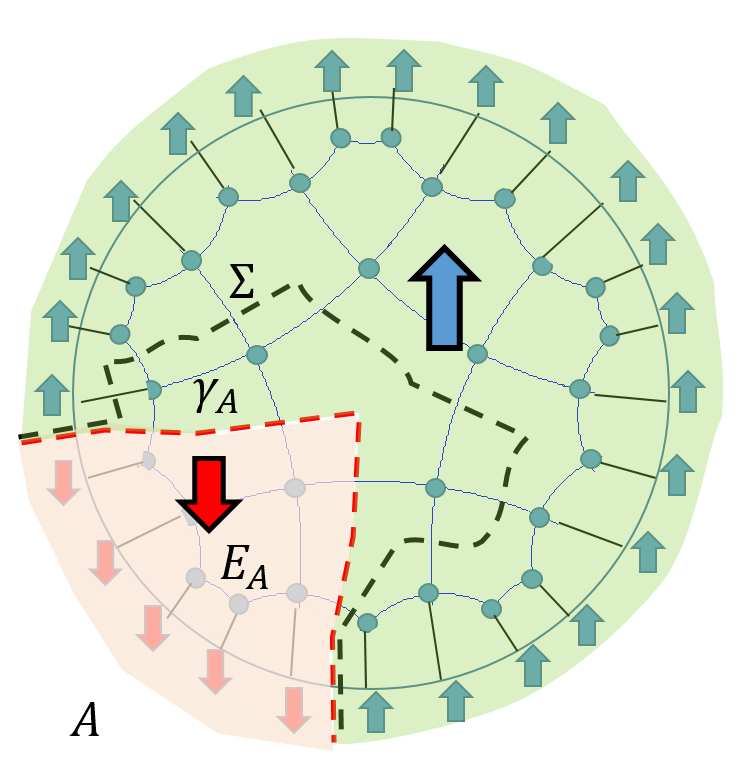}
         \label{domain}
     \end{subfigure}
        \caption{Left: The pentagonal HaPPY code. Every pentagon stands for a six-leg perfect tensor: one leg (shown as a red dot) supporting bulk fields, and five legs that compose the discrete geometry. Right: Two cuts in the network homologous to $A$. The tensors between $\Sigma$ and $A$ do not make up an isometry $\mathcal{H}_\Sigma \to \mathcal{H}_A$ because some tensors are used in the direction from more to fewer legs. \textchg{(Each of the three tensors between $\Sigma$ and $\gamma_A$ has three legs cut by $\Sigma$ and one leg cut by $\gamma_A$, so going from $\Sigma$ to $A$ via $\gamma_A$ cannot be an isometry.)} The tensors between $\gamma_A$ and $A$ do compose into an isometry. The up and down arrows refer to domains of the ferromagnetic Ising model from Section~\ref{sec:rtn}. The figures are reproduced with permission from \cite{Pastawski:2015qua} (left) and \cite{Hayden:2016cfa} (right).}
        \label{fig:happy}
\end{figure}

Consider a uniform tiling of two-dimensional hyperbolic space (equal time slice of AdS$_3$) by pentagons in Figure~\ref{fig:happy}. (Other tessellations of hyperbolic space work as well.) On each pentagon we place a {perfect tensor} with six legs, and associate five legs of the tensor with edges of the tessellation. We then contract legs of neighboring tensors across edges of the tessellation. This leaves one uncontracted leg per pentagon. For a holographic interpretation, we now associate the Hilbert space on the open bulk legs with the Hilbert space $\mathcal{H}_{\rm bulk}$ of perturbative bulk fields in AdS$_3$. (A similar identification for the MERA network was previously made in \cite{xiaoliangehm}.) The network also has open legs on the (asymptotic) boundary of hyperbolic space; their Hilbert spaces symbolize $\mathcal{H}_{\rm CFT}$.

This tensor network carries the optimistic acronym HaPPY after \cite{Pastawski:2015qua}. It is a toy model for how error correction and subregion duality work in the AdS/CFT correspondence. Here is why:
\begin{itemize}
\item The network inherits from the tessellation a large discrete subgroup of the isometry group of the spatial slice of AdS$_3$, though part of it may be broken by non-isotropies in the perfect tensors. 
\item Let us view the network as a map from $\mathcal{H}_{\rm bulk}$ to $\mathcal{H}_{\rm CFT}$ (or rather, their discrete analogues). Referring to Section~\ref{sec:error}, this map is analogous to sending logical qutrits $|0\rangle,~|1\rangle,~|2\rangle$ to their tilded counterparts in equation~(\ref{codeembed}). In this sense, the network realizes the erasure-correcting property of holographic duality. 
\item Consider a bulk state in the Hilbert space $\mathcal{H}_{\rm bulk}^i$ carried by some open bulk leg $i$. (We phrase the discussion in terms of states, but we could easily rephrase it in terms of operators, as we did in going from equation~(\ref{explcorr}) to $\mathcal{O}_{AB} = U_{AB} \mathcal{O}_{\rm code} U_{AB}^\dagger$.) Analogous to the statement that (\ref{explcorr}) protects against the erasure of one but not two physical qutrits, we may ask which boundary subregions $A$ afford a reconstruction of a state in $\mathcal{H}_{\rm bulk}^i$. The answer (qualified by certain discrete artefacts) is those regions $A$, whose minimal cuts through the network encircle the leg $i$. We interpret this as reflecting subregion duality: a state of a local bulk Hilbert space $\mathcal{H}_{x}$ can be reconstructed from $A \subset {\rm CFT}$ if $x \in \mathcal{W}(A)$.  
\item In the previous point, the minimal cut that divides $A$ from $\bar{A}$ defines a HaPPY network analogue of an entanglement wedge. The minimal cut itself is therefore a discrete version of the Ryu-Takayanagi surface. Using bound~(\ref{tnbound}) and assuming it is saturated (or off saturation by a fixed factor), this gives an estimate of entanglement entropies of boundary regions. The minimal cuts follow AdS$_3$ geodesics as closely as the tessellation allows, which gives qualitative agreement with proposal~(\ref{rtsimplest}). 
\end{itemize}

The argument backing the assertion about subregion duality is illuminating. After an erasure of $\mathcal{H}_{\bar{A}} \subset {\rm CFT}$, reconstructing a bulk state of $\mathcal{H}_{\rm bulk}^i$ requires an isometric (extendible to unitary) embedding $\mathcal{H}_{\rm bulk}^i \to \mathcal{H}_A$. 
Perfect tensors provide isometric maps so long as one goes from a smaller to a large Hilbert space. Compositions (contractions) of perfect tensors likewise define isometric embeddings, but again---only if one never composes a perfect tensor in the direction from fewer to more legs. To determine the range of reconstruction afforded by $A$, we may therefore start from $\mathcal{H}_A$---the image of the desired isometry---and compose successive layers of perfect tensors backwards toward $\mathcal{H}_{\rm bulk}$. In this process of undoing the isometry, we must never go in the direction \emph{from fewer to more} indices. A minimal cut in the network is an obstruction to this process and therefore defines a boundary of the tensor network avatar of $\mathcal{W}(A)$. This reasoning is illustrated in the right panel of Figure~\ref{fig:happy}.

\subsection{Random tensor network}
\label{sec:rtn}
One slightly dissatisfying feature of the HaPPY network is that the AdS geometry was hard-wired from the outset. For an understanding of holographic spacetimes as `entanglement maps,' one would like a procedure that adapts the architecture of the network to the quantum state at hand. Networks that do this were proposed in \cite{Bao:2019fpq, Caputa:2020fbc}, but those constructions are outside the scope of this review. 

Instead, we review another tensor network \cite{Hayden:2016cfa}, which made a more modest step in a similar direction: it disposed of the perfect property of perfect tensors. Choosing not to hard-wire a symmetry brings us closer to extrapolating the lessons of holography outside AdS settings. It will also be welcome by readers who hold the aesthetic belief that all symmetries are emergent. It is a beautiful fact that the resulting Random Tensor Network (RTN) exhibits many of the same gratifying features that the HaPPY network does, and more.

\paragraph{Definition} We again take a hyperbolic tessellation with pentagons but now place on every plaquette $x$ a five-leg tensor $V_x$, which is {\it random} with respect to a suitable Haar measure and independently chosen for each $x$. (This version does not have bulk degrees of freedom, but an appropriate extension that mirrors \cite{Pastawski:2015qua} is straightforward.) Interpreting the tensor $V_x$ as a pure state of a five-partite system, and recalling that joining two legs in a bond means projecting onto a fixed entangled pair, the resulting state can be written as
\begin{equation}
    \ket{\Psi}=\left( \bigotimes_{x \sim y}\,\bra{xy} \right) \left( \bigotimes_{x} \ket{V_{x}} \right)
\end{equation}
where $x \sim y$ means that pentagons $x$ and $y$ share an edge. 

A rough reason why RTN behaves similarly to the HaPPY network is that random tensors, in the limit of large bond dimension, are approximate isometries. We shall go beyond this heuristic and compute explicitly the spectrum of $\rho_A$ averaged over choices of tensors. The calculation will be handy in Section~\ref{sec:renyi}  where we sketch the proof of the Ryu-Takayanagi proposal~(\ref{rtsimplest}). 

\paragraph{A replica trick calculation} A common method to compute the von Neumann entropy~(\ref{vonneumann}) of a density matrix $\rho_A$ is to consider an auxiliary quantity called the R\'{e}nyi entropy:
\begin{equation}
S^n_A = \frac{1}{1-n} \log {\rm Tr}\, \rho_A^n
\label{defrenyi} 
\end{equation}
When the limit $\lim_{n \to 1} S^n_A$ is smooth and sensible, it equals (\ref{vonneumann}). This method of computing entanglement entropy is called the replica trick \cite{holzheyetal} because (\ref{defrenyi}) involves $n$ copies of $\rho_A$. 

We focus on $S_A^2$ and follow the presentation in \cite{michaeltalk}. First, draw two copies of $|\Psi\rangle \langle \Psi|$ and consider how the free legs 
on the boundary are contracted to effect ${\rm Tr}\rho_A^{\, 2}$:
\begin{equation}
\left( 
\bigotimes_{\substack{\textrm{boundary} \\ \textrm{contractions}}}  
\left(\delta_{u^1}^{x^1} \delta_{u^2}^{x^2}~{\rm or}~\delta_{u^2}^{x^1} \delta_{u^1}^{x^2} \right) \right) 
\begin{array}{c}
\left( \bigotimes_{x^1\sim y^1}\bra{x^1 y^1} \right) 
\left( \bigotimes_{x} \ket{V_{x}} \bra{V_x} \right)
\left( \bigotimes_{u^1\sim v^1} \ket{u^1 v^1} \right) \\
\\
\left( \bigotimes_{x^2\sim y^2}\bra{x^2 y^2} \right) 
\left( \bigotimes_{x} \ket{V_{x}} \bra{V_x} \right)
\left( \bigotimes_{u^2\sim v^2} \ket{u^2 v^2} \right)
\end{array}
\end{equation}
(Note that these contractions are effected by Kronecker $\delta$s instead of $\langle xy|$ because we are contracting bra states with ket states. For example, contracting interior legs in $\langle \Psi|$ is done with $|u v\rangle$ while corresponding contractions in $|\Psi \rangle$ are done with $\langle xy|$.) The first option contracts indices from the same copy of $|\Psi\rangle \langle \Psi|$; applied on $\bar{A}$, it turns $|\Psi\rangle \langle \Psi|$ into $\rho_A$. The second option mixes the copies; applied on $A$, it makes $\rho_A^2$. Thus, the correct expression is:
\begin{equation}
\left( \bigotimes_{x,u \in {\bar A}}  
\left(\delta_{u^1}^{x^1} \delta_{u^2}^{x^2}\right) \right) 
\left( \bigotimes_{x,u \in {A}}  
\left(\delta_{u^2}^{x^1} \delta_{u^1}^{x^2}\right) \right) 
\begin{array}{c}
\left( \bigotimes_{x^1\sim y^1}\bra{x^1 y^1} \right) 
\left( \bigotimes_{x} \ket{V_{x}} \bra{V_x} \right)
\left( \bigotimes_{u^1\sim v^1} \ket{u^1 v^1} \right) \\
\\
\left( \bigotimes_{x^2\sim y^2}\bra{x^2 y^2} \right) 
\left( \bigotimes_{x} \ket{V_{x}} \bra{V_x} \right)
\left( \bigotimes_{u^2\sim v^2} \ket{u^2 v^2} \right)
\end{array}
\label{tracerho2}
\end{equation}
We now average~(\ref{tracerho2}) over the choices of tensors. Their randomness implies \cite{avrandom} that
\begin{equation}
\overline{\ket{V_{x}} \bra{V_x} \otimes \ket{V_{x}} \bra{V_x}} \quad \propto \quad \mathbb{1}_x + \sigma_x\,,
\label{averageswapping}
\end{equation}
where $\sigma_x$ swaps each pair of identical legs incident to $V_x$ between the two copies of $|\Psi\rangle \langle \Psi|$. This swap is exactly the difference between the two types of contractions on the boundary of the network:
\begin{equation}
\delta_{u^2}^{x^1} \delta_{u^1}^{x^2} = \sigma_x (\delta_{u^1}^{x^1} \delta_{u^2}^{x^2})
\end{equation}
Using~(\ref{averageswapping}), equation~(\ref{tracerho2}) turns into a sum of products, each of which is generated by an independent choice of applying or not applying the swaps $\sigma_x$ at all sites $x$. 

We can encode the choices with a binary variable $h_x = \pm 1$ such that applying $\sigma_x$ corresponds to $h_x = -1$. Every edge $x \sim y$ contributes a single factor of its bond dimension when $h_x \neq h_y$ but two factors of bond dimension when $h_x = h_y$; it is the difference between ${\rm Tr}\, (\mathbb{1}_{\mathcal{H}_{x \sim y}})^2$ versus $({\rm Tr}\, \mathbb{1}_{\mathcal{H}_{x \sim y}})^2$. If we call the bond dimension $D$, we see that the average of ${\rm Tr}\rho_A^2$ is proportional to:
\begin{equation}
\sum_{h_x = \pm 1}\, \prod_{x \sim y}\, D^{h_x h_y/2}
\end{equation}
This is the partition function of a classical ferromagnetic Ising model on the lattice, on which the RTN lives. The inverse temperature is $\beta = \log D / 2$. The system is non-trivial because it is subject to frustrated boundary conditions: $h_x = 1$ for $x \in \bar{A}$ and $h_x = -1$ for $x \in A$. 

The partition function is dominated by a solution with two ordered phases ($h_x = -1$ in a region adjacent to $A$ and $h_x = +1$ in a region adjacent to $\bar{A}$) separated by a domain wall. \textchg{We mark the domains with large up/down arrows in Figure~\ref{fig:happy}.} Of course the optimal course for the domain wall is one that minimizes its area. This is an analogue of the Ryu-Takayanagi formula. We recognize the long range ordered phases as avatars of entanglement wedges.

\paragraph{Comments:}
\begin{itemize}
\item The second R\'{e}nyi entropy is essentially the free energy. After properly accounting for normalization, it becomes the length of the minimal domain wall times $\log D$. 
\item A similar but more complicated calculation for the higher $n$ R\'{e}nyi entropies also gives the minimal area times $\log D$. From the definition~(\ref{defrenyi}) it is clear that all the R\'{e}nyi's can be equal only if $\rho_A$ is a normalized multiple of the identity matrix. We say that the RTN has a flat entanglement spectrum.
\item Consequently, the von Neumann entropy of $\rho_A$ is also $\textrm{(minimal area)} \times \log D$.
\item We can introduce bulk degrees of freedom as in the HaPPY network. This makes RTN into an error-correcting code. Incorporating bulk degrees of freedom can shift the domain wall due to possible bulk entanglement. This is a harbinger of quantum extremal surfaces, which we discuss in Section~\ref{sec:genentropy}.
\item It was not really important in the calculation that the RTN lived on hyperbolic space. We do need a boundary on which to set boundary conditions, and the existence of minimal surfaces to have a stable, non-trivial dominant solution. The latter is a rough analogue of negative curvature in the bulk.
\item There is a generalization of RTN, which also encodes the passage of time \cite{RTN2}. 
\end{itemize}

\subsection{Surface/state correspondence}
\label{sec:ssc}
In practice, all the above networks are drawn on finite pieces of paper or displayed on finite monitors. This is different from anti-de Sitter space whose spatial size is infinite. As we explained in Section~\ref{sec:adscft}, this infinity derives from the arbitrarily small scales in the holographic CFT. 

Are we missing an essential ingredient of holography by considering finite tensor networks? References~\cite{Miyaji:2015yva, Miyaji:2015fia} conjecture that the answer is no. They propose a {\it surface/state correspondence}, wherein any convex, topologically trivial, spacelike, \textdel{curve}\textchg{codimension-2 surface} in a gravitational system defines a pure state in some Hilbert space. Topologically non-trivial \textdel{curves}\textchg{codimension-2 surfaces} may define mixed states or entangled states of multiple Hilbert spaces. Hilbert spaces associated to topologically trivial \textdel{curves}\textchg{codimension-2 surfaces} embed in one another when one \textdel{curves}\textchg{such surface} encircles the other. These are the same properties as those enjoyed by Hilbert spaces $\mathcal{H}^{\rm cut}$ on cuts in tensor networks, which is why we include this proposal in the present section.

The proposal is not trivial because it presumes that a gravitational system can be adequately studied inside a small box. Equivalently, it seems to presume that it makes sense to study a holographic CFT with an {\it ultraviolet} cutoff of order the system size. These are highly non-trivial assumptions because a proper implementation of a finite IR cutoff in gravity is a tricky subject with a long history; see e.g.~\cite{vijayrg, janverlindesrg, janrg, tomhongmukundrg, ttbar, ttbarssc, Grado-White:2020wlb, nimarg}. At the same time, the surface/state correspondence is a very fruitful assumption: adopting it motivates many interesting conjectures, some of which are discussed in the next section.

\section{Information-theoretic concepts in holography}

\subsection{R{\'e}nyi entropy and a proof of the Ryu-Takayanagi proposal}
\label{sec:renyi}
In Section~\ref{sec:rtn} we encountered the R{\'e}nyi entropies, which allowed an efficient calculation of the von Neumann entropy via the replica trick:
\begin{equation}
\lim_{n \to 1^+} S_A^n = S(A)
\qquad {\rm with}~~S^n_A = \frac{1}{1-n} \log {\rm Tr}\, \rho_A^n
\label{defrenyi-h}
\end{equation}
Just like their limit $S(A)$ in proposal~(\ref{rtfull}), the R{\'e}nyi entropies---or rather, a slight generalization of them defined below---also have a nice bulk interpretation. In fact, identifying it affords an efficient proof \cite{Dong:2016fnf} of the Ryu-Takayanagi proposal, which we still owe to the reader. (The same proof, couched in a slightly different language, appeared earlier in \cite{Lewkowycz:2013nqa}, but the underlying strategy goes back to the 1970s when it was used to prove the Bekenstein-Hawking entropy formula \cite{Gibbons:1976ue}.) Other references on the bulk dual of CFT R{\'e}nyi entropies include \cite{Headrick:2010zt, Akers:2018fow, Bao:2019aol, DHoker:2020bcv}.

In replica trick calculations, the exponent $n$ in $\rho_A^n = e^{-n H_{\rm mod}}$ acts like the inverse temperature $\beta \leftrightarrow n$. Using standard thermodynamic relations, we interpret
\begin{equation}
-\frac{\partial F_n}{\partial T} = \beta^2 \frac{\partial F_n}{\partial \beta}
\quad \leftrightarrow \quad 
\tilde{S}^n_A = n^2 \partial_n \left( -\frac{1}{n} \log {\rm Tr}\, e^{-n H_{\rm mod}^A}\right) = S^n_A+n(n-1)\partial_n S^n_A \,\,\,\,
\label{identifyrenyi}
\end{equation}
as a new type of $n$-dependent entropy whose $n\to 1$ limit is also $S(A)$ and from which $S^n_A$ can be easily extracted. Quantities $\tilde{S}^n_A$, $S^n_A$ and $S(A)$ can be straightforwardly computed provided we interpret the free energies $F_n$ on the left hand side of (\ref{identifyrenyi}).

\begin{figure}[tbp]
\centering 
\includegraphics[width=0.5\textwidth]{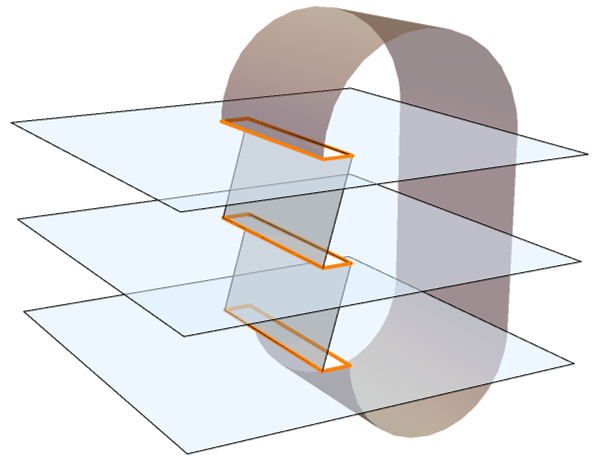}
\hfill
\caption{\label{fig:replica} 
The $n$-fold branched cover $M_n$ of the CFT with cuts at $A$ (orange), here shown for $n=3$. This is the asymptotic boundary of the gravitational solution, which is relevant for computing $F_{n=3}$ in equation~(\ref{identifyrenyi}).}
\end{figure}

The AdS/CFT correspondence posits an equivalence between its two sides, i.e. an equality of the AdS and CFT partition functions. In particular, the CFT free energy per temperature $\beta F = -\log \mathcal{Z}_{\rm CFT}[M]$ on a manifold $M$ equals, up to quantum effects suppressed by large $N$, the action of the classical gravitational solution $I_{\rm bulk}[M]$ whose asymptotic boundary is $M$. In (\ref{identifyrenyi}), the free energy per temperature $\beta F_n$ is computed by a Euclidean gravitational solution whose boundary is dictated by the CFT computation of $\rho_A^n$. This is an $n$-fold branched cover $M_n$ of the Euclidean CFT, in which $n$ copies of the CFT geometry are glued along cuts at $A$ such that crossing $A$ takes us from the $k^{\rm th}$ to the $(k+1)^{\rm st}$ copy; see Figure~\ref{fig:replica}. For computing $F_n$ itself, we take a $\mathbb{Z}_n$ orbifold of this solution because $F_n = (1/n) \beta F_n$. Therefore, we are interested in gravitational solutions, which are $\mathbb{Z}_n$ orbifolds of the solutions for $M_n$. 

In the bulk, it is tempting to take the gravitational solution with boundary $M_n / \mathbb{Z}_n$ to be the same as the old solution on $M$ because the replacement $M \to M_n / \mathbb{Z}_n$ is a global and not a local operation \cite{fursaev}. This reason is valid everywhere except where the $n$ copies of the $M$-solution are glued together in the $M_n$-solution \cite{Headrick:2010zt} . On that locus, the quotient by $\mathbb{Z}_n$ introduces a conical deficit angle $\Delta \phi = 2\pi (n-1)/n$. Therefore, Einstein's equations require that we source the $M_n / \mathbb{Z}_n$ geometry inherited from $M$ with a membrane of tension $(n-1)/4nG_N$ at the fixed set of the $\mathbb{Z}_n$ symmetry \cite{Dong:2016fnf}. With the tensionful membrane in place, the action of the full solution has the following $n$-dependence:
\begin{equation}
I_{\rm bulk}[M_n / \mathbb{Z}_n] = \frac{n-1}{4Gn} (\textrm{Area of membrane})_n 
+ (\textrm{$n$-independent terms})
\end{equation}
We recognize that $\tilde{S}^n_A$ is simply the area of the membrane in the classical solution. Because the membrane carries tension, classically it always settles on an extremal area.

We found that the tension of the brane reflects the R{\'e}nyi index $n$. We may now take the limit $n \to 1$, in which the tension $(n-1)/4nG \to 0$ but the brane retains its position on the extremal surface. This shows that the extremal area also equals $\lim_{n \to 1} \tilde{S}_A^n = S(A)$ and proves the proposal (\ref{rtfull}). An obvious caveat of this proof is that it presumes that the replica symmetry $\mathbb{Z}_n$ is not broken in the orbifolded solution on $M_n / \mathbb{Z}_n$ and can be sensibly extended to non-integer $n$. 

\subsection{Differential entropy}
\label{sec:ks}

Thus far we have discussed extremal surfaces or---restricting to equal time slices of static and time reversal-symmetric spacetimes---minimal surfaces. If they represent entanglement entropies as in proposal~(\ref{rtsimplest}), can we likewise interpret non-extremal surfaces in terms of quantum information theory? This question turns out to be messy in higher dimensions \cite{Balasubramanian:2018uus}, but in $d=2$ it has a surprisingly sharp answer rich in implications \cite{diffentpaper}.

\begin{figure}[tbp]
\centering 
\includegraphics[width=.5\textwidth]{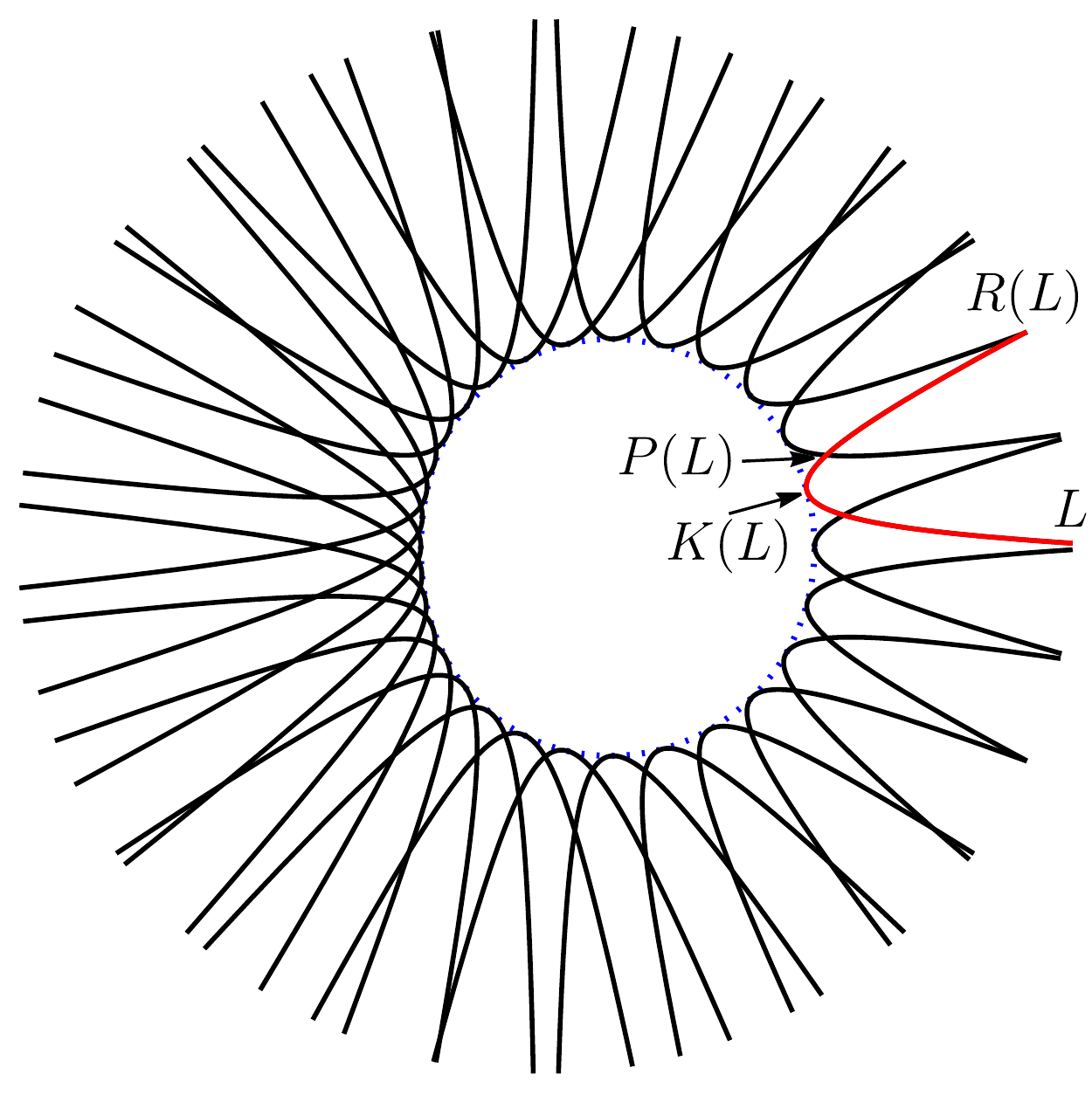}
\hfill
\caption{\label{fig:tangentcurves} 
A family of geodesics tangent to a bulk curve. Each geodesic is identified by its left $L$ and right $R(L)$ endpoints. The notation~$P(L)$ and $K(L)$ is used in equation~(\ref{diffentproof}).}
\end{figure}

Consider a differentiable curve $\mathcal{C}$ on an equal time slice of a static or time reversal-symmetric, asymptotically AdS$_3$ geometry. For simplicity, we assume that $\mathcal{C}$ is closed and convex, i.e. its extrinsic curvature on the equal time slice never changes sign. We take the holographically dual CFT to live on $S^1 \times \mathbb{R}_{\rm time}$, so that the metric near the asymptotic boundary approaches
\begin{equation}
ds^2 \to d\rho^2 + \sinh^2 \rho\, d\theta
\end{equation}
with $\theta \equiv \theta + 2\pi$ parameterizing the asymptotic boundary. We can capture full data about the curve $\mathcal{C}$ through its set of tangent geodesics. Tangency to $\mathcal{C}$ is a single condition, which sets a functional dependence between the left and right endpoints of the tangent geodesics. We will express this condition as one function of one variable $L(R)$, which tells us that a geodesic with boundary endpoints at $\theta_L=L(R)$ and $\theta_R = R$ is tangent to the curve $\mathcal{C}$; see Figure~\ref{fig:tangentcurves}. Of course, our focus on geodesics is motivated by the Ryu-Takayanagi proposal~(\ref{rtsimplest}). A rewriting of $\mathcal{C}$ using geodesics is automatically a rewriting in information-theoretic terms.

The circumference of $\mathcal{C}$ is computed by a CFT quantity called {\it differential entropy} \cite{diffentpaper}:
\begin{align}
\frac{\textrm{length of $\mathcal{C}$}}{4G\hbar} & = 
\oint dR \,
\frac{\partial S(L,R)}{\partial R} \Big|_{L = L(R)}
= \oint \!\Big(S\big(L(R), R+dR\big) - S\big(L(R), R\big)\Big) 
\nonumber \\
& = - \oint dL\, \frac{\partial S(L,R)}{\partial L} \Big|_{R = R(L)}
\equiv S_{\rm diff}[L(R)]
\label{diffent}
\end{align}
Here $S(L,R)$ is the entanglement entropy of a boundary interval $(L,R) \subset {\rm CFT}$, which by (\ref{rtsimplest}) equals the length of the geodesic with the same endpoints in units of $4G\hbar$. The equivalent expression in the second line is obtained via integration by parts; $R(L)$ is the inverse function of $L(R)$, which also identifies geodesics tangent to $\mathcal{C}$.

\paragraph{Proof} We need to label two types of bulk points; see Figure~\ref{fig:tangentcurves}. We will call $K(L)$ the point where the geodesic subtending $(L, R(L))$ touches the curve $\mathcal{C}$, and call $P(L)$ the point where that geodesic intersects its infinitesimally shifted neighbor $(L+dL, R(L+dL))$. We will also use expressions like $S\big(L, K(L)\big)$ and $S\big(P(L), K(L)\big)$ to refer to lengths of geodesic segments connecting those points, ignoring the factor $4G\hbar$. This extension of the notation $S(L,R)$ will only be used in the next paragraph.

Working to first order in $dL$, we have:
\begin{align}
- dL\, \frac{\partial S(L,R)}{\partial L}\Big|_{R = R(L)} 
= &\, S\big(L, R(L)\big) - \Big[S\big(L+dL, R(L)\big)\Big]
\nonumber \\
= &\, S\big(L, R(L)\big) - \Big[S\big(L+dL, P(L)\big) + S\big(P(L), R(L)\big)\Big] \nonumber \\ 
= &\, S\big(L, P(L)\big) - S\big(L+dL, P(L)\big) \nonumber\\
= &\, S\big(L, K(L)\big) - S\big(L+dL, K(L+dL)\big) \nonumber \\
&\,\,\, + \Big[S\big(K(L), P(L)\big) + S\big(P(L), K(L+dL)\big) \Big]  \nonumber  \\
= &\, - dL\, \frac{dS\big(L, K(L)\big)}{dL} + \Big[ S\big(K(L), K(L+dL)\big) \Big] 
\label{diffentproof}
\end{align}
Both pairs of terms in $[\ldots]$ are equal because first order perturbations do not change extremal quantities such as lengths of geodesic segments. Rewriting~(\ref{diffentproof}) proves (\ref{diffent}) because $S\big(K(L), K(L+dL)\big)$ is the length element along the curve and the other term is a total derivative. This proof is adapted from \cite{hholes, xidongdiffent}.

\paragraph{Comments:}
\begin{itemize}
\item Formula~(\ref{diffent}) can be generalized to open curves \cite{Czech:2014ppa} and 
curves with time dependence. There is also a generalization, which relaxes the tangency condition into so-called {\it null vector alignment} \cite{hholes}. 
\item The proof in~(\ref{diffentproof}) only uses the extremality of the function $S(\cdot,\cdot)$, so it automatically extends to theories where entropies are computed by extremizing functionals other than length. This includes non-Einstein gravity theories and settings with quantum corrections, which we discuss in Section~\ref{sec:genentropy}. 
\item In spacetimes with matter or horizons, curves that lie too deep (too close to a matter source or horizon) have tangent geodesics, which are not minimal and therefore cannot be interpreted using the Ryu-Takayanagi proposal. In some cases such geodesics also compute an information theoretic quantity called {\it entwinement} \cite{entwinement}. Generally, one must be careful in applying and interpreting equation~(\ref{diffent}) whenever a spacetime admits more than one geodesic with the same endpoints.
\item Although individual entanglement entropies $S(L,R)$ are cutoff-dependent, differential entropy is independent of the choice of ultraviolet cutoff in the CFT. You can see from equation~(\ref{sinterval}) that the cutoff in $d=2$ is additive and does not survive the derivative---a feature that is shared by all CFT$_2$ states because all states are similar sufficiently deep in the ultraviolet. Cutoff independence is the first plausibility check of (\ref{diffent}) because a compact curve does not go near the asymptotic boundary. The non-additive dependence on cutoffs in $d>2$ is part of the reason why generalizing (\ref{diffent}) to higher dimensions is difficult \cite{Balasubramanian:2018uus}.
\item Differential entropy isolates an appropriate infrared part of the quantum entanglement of the CFT state. It is unclear whether it computes the von Neumann entropy of some density operator in the CFT \cite{veronika, brianisaac}.\textdel{, though in the surface/state correspondence of Section~\ref{sec:ssc} this would be automatic}
\item A timelike analogue of (\ref{diffent}) should be the proper time along a timelike curve (worldline) in the bulk. But the timelike generalization must involve some radically novel ingredients because timelike geodesics in AdS (which may be tangent to a worldline) never reach the boundary. Only recently was the proper time along a worldline computed using CFT ingredients \cite{Jafferis:2020ora}. The methodology bears some resemblance to (\ref{diffent}), but it employs many new ingredients, which go beyond what we can cover in this review. Handling time is more challenging than space. 
\end{itemize}

\subsubsection{Bulk curves encode quantum streaming protocols} 
An interesting and important way to interpret the von Neumann entropy of a state is that it quantifies resources necessary for quantum teleportation \cite{qteleport}. We will find that formula~(\ref{diffent}) has a similar, {\it operational} interpretation \cite{streaming}. We start with a lightning review of:

\paragraph{Quantum teleportation} Consider a four-partite setup: a reference system $R$, a memory system $M$, and systems $A$ and $B$ controlled by Alice and Bob, all assumed to comprise qubits for simplicity. We will conceptualize messages to be stored in a memory system as `knowing how to purify the reduced state $\rho_R$,' i.e. holding a purification of $\rho_R$. In other words, we will say that the memory $M$ stores the message if the joint state of $R \cup M$ is pure. For systems $A$ and $B$, we assume that they are initialized in a combination of maximally entangled pairs (also called EPR pairs). Thus, assuming the memory $M$ stores the message, the full state of this four-partite system can be written in some basis as:
\begin{equation}
\left( \frac{|\!+\!+\rangle_{RM} + |\!-\!-\rangle_{RM}}{\sqrt{2}}\right)^{\otimes m}
\otimes
\left(\frac{|\!+\!+\rangle_{AB} + |\!-\!-\rangle_{AB}}{\sqrt{2}}\right)^{\otimes n} 
\label{qtinitial}
\end{equation}

Assume that the memory system $M$ is under Alice's control. This might mean that it physically sits on Alice's desk, but in our analysis it signifies that Alice can apply all operators, which act on $\mathcal{H}_M \otimes \mathcal{H}_A$. We further assume that Bob and his system are far away, so that operators acting on $\mathcal{H}_M \otimes \mathcal{H}_B$ or $\mathcal{H}_A \otimes \mathcal{H}_B$ are not at our disposal. The goal of quantum teleportation is to use the allowed tools so that, at the very end, it is Bob's system $B$ that stores the message, i.e. the state of $B \cup R$ is pure.

Alice can achieve this by measuring successive qubits of the composite system $M \cup A$ under her control in a basis of EPR pairs. From the rewriting
\begin{align}
& \left( \frac{|\!+\!+\rangle_{RM} + |\!-\!-\rangle_{RM}}{\sqrt{2}}\right)
\otimes
\left(\frac{|\!+\!+\rangle_{AB} + |\!-\!-\rangle_{AB}}{\sqrt{2}}\right)
\\
= &
\,\frac{1}{2}\frac{|\!+\!+\rangle_{AM} + |\!-\!-\rangle_{AM}}{\sqrt{2}}
\!\otimes\!
\frac{|\!+\!+\rangle_{RB} + |\!-\!-\rangle_{RB}}{\sqrt{2}}
+ 
\frac{1}{2}\frac{|\!+\!+\rangle_{AM} - |\!-\!-\rangle_{AM}}{\sqrt{2}}
\otimes
\frac{|\!+\!+\rangle_{RB} - |\!-\!-\rangle_{RB}}{\sqrt{2}}
\nonumber \\
+ &
\,\frac{1}{2}\frac{|\!+\!-\rangle_{AM} + |\!-\!+\rangle_{AM}}{\sqrt{2}}
\otimes
\frac{|\!+\!-\rangle_{RB} + |\!-\!+\rangle_{RB}}{\sqrt{2}}
-
\frac{1}{2}\frac{|\!+\!-\rangle_{AM} - |\!-\!+\rangle_{AM}}{\sqrt{2}}
\otimes
\frac{|\!+\!-\rangle_{RB} - |\!-\!+\rangle_{RB}}{\sqrt{2}}
\nonumber
\end{align}
we see that any such measurement projects the corresponding qubits of $R$ and $B$ into a pure state, as desired. Moreover, Bob can bring this pure state of $RB$ qubits to the form \mbox{$\big(|\!+\!+\rangle + |\!-\!-\rangle\big)_{RB}/\sqrt{2}$}---the same form in which $M$ initially `knew' $R$---by a unitary transformation applied only to his own qubits $B$. For example, if Alice's local measurement projects onto \mbox{$\big(|\!+\!-\rangle - |\!-\!+\rangle\big)_{AM}/\sqrt{2}$}, Bob would map: 
\begin{equation}
|-\rangle_B \to |+\rangle_B \qquad {\rm and} \qquad |+\rangle_B \to -|-\rangle_B
\label{bobsunitary}
\end{equation}
To know which transformation to apply on his end, Bob has to wait for Alice to send him classical information about the outcome of her measurement. The necessity to wait for this classical information to arrive from Alice saves quantum teleportation from the types of paradoxes that spooked Einstein in his famous letter \cite{spooky}.

To fully teleport the message held in~(\ref{qtinitial}), Alice must perform $m$ measurements in maximally entangled bases. Each of them uses one of the $n$ prearranged entangled states shared by Alice and Bob. If $n < m$ in (\ref{qtinitial}), it is impossible to fully teleport the message.

\paragraph{Salient features of quantum teleportation:}
\begin{itemize}
\item Quantum teleportation consumes prearranged entanglement shared by the sender and the recipient and cannot succeed if there is not enough of it. Therefore, quantum entanglement between sender and recipient is a resource for quantum teleportation.
\item  When the state to be teleported and the resource do not come in the form of qubits, it is more cumbersome to quantify their relation. But the complications go away when we go to an asymptotic setup, wherein we have $Q \gg 1$ copies of a pure state $|\psi\rangle_{RM}$ and $Q$ copies of the entanglement resource $|\phi\rangle_{AB}$, and ask about the average entanglement cost of teleportation per copy. In that case, the necessary condition for teleportation is that $S(A) = S(B) \geq S(R) = S(M)$. Proving this proceeds by converting the problem to one involving qubits using a so-called compression protocol \cite{compression}, which works well when dealing with many copies of a system.
\item This leads to an operational interpretation of entanglement entropy (\ref{vonneumann}) as a resource for quantum teleportation. This means two things: that $S(A) = S(B)$ bounds what states can be teleported using $|\phi\rangle_{AB}$ as a resource, and that $S(R) = S(M)$ defines the minimal entanglement resources necessary to teleport the state of $M$. 
\item Typical textbook presentations focus on individual pure states such as $|+\rangle_M$ instead of full purifications of $R$. In fact, the two tasks---teleporting a pure state or teleporting a purification of a reference $R$---are equally difficult.
\item As soon as Alice performs a measurement, Bob comes to hold the message but does not yet know how to decode it. The decoding happens when he applies a unitary transformation such as (\ref{bobsunitary}). In order to choose the right unitary transformation that decodes the message, Bob needs to receive from Alice classical information about the outcome of her measurement. 
\end{itemize}

\paragraph{Operational interpretation of differential entropy} 
A generalization of quantum teleportation concerns a setup wherein $R \cup M$ is not initially in a pure state but $R\cup M \cup B$ is. Bob now starts out with some partial `knowledge' of the message, but the full message is delocalized between his system $B$ and the memory $M$ held by Alice. The goal, as before, is to use local unitary transformations so that $B$ ends up with a complete purification of $R$. This task is called {\it state merging}. The minimal entanglement cost of state merging is smaller than the cost of teleportation $S(R)$ because Bob has a head start.

In that case, the necessary entanglement (in many-copy settings) is the {\it conditional entropy} \cite{mergingcost1, mergingcost2}:
\begin{equation}
S(M|B) \equiv S(MB) - S(B) = S(R) - S(B) \leq S(A)
\label{condent}
\end{equation}
Note that now $S(A) \neq S(B)$ because in addition to purifying $A$, system $B$ contributes to purifying $RM$. More interestingly, observe that the conditional entropy can be negative. This occurs when merging can be completed without any measurements, by a mere change of basis on Bob's side. In such situations, $-S(M|B)>0$ EPR pairs can be isolated for use in future merging, which is interpreted as a negative cost. The headline `Scientist knows less than nothing' \cite{lessthan0}, which we mentioned in Section~\ref{sec:unruhhawking}, referred to this situation.

Differential entropy~(\ref{diffent}) is a sum of conditional entropies $S(L(R), R+dR) - S(L(R),R)$, which is (\ref{condent}) with $B = (L(R), R)$ and $M = (R, R+dR)$. In this light, the special case of differential entropy where $L(R) \equiv L_i$ has a straightforward interpretation. The expression
\begin{equation}
\int_{L_i}^{R_f} dR \,
\frac{\partial S(L_i,R)}{\partial R} 
= \int_{L_i}^{R_f} S\big((R,R+dR)\, | \, (L_i, R)\big)
= S(L_i, R_f)
\label{sentincr}
\end{equation}
computes the cost of teleporting the interval $(L_i, R_f)$ in incremental, infinitesimal merging steps, one $dR$ at a time. 

The general equation~(\ref{diffent}) differs from (\ref{sentincr}) in that we seem to be merging incoming information not to the cumulative result of previous merging steps---the state on interval $(L_i, R)$---but to a subset of it. In other words, equation~(\ref{diffent}) describes an incremental merging protocol, in which we are not reaping the full benefit of the previous merging steps. What limits this benefit is Bob's ability to apply unitary transformations such as (\ref{bobsunitary}) to decode the incoming information. Equation~(\ref{condent}) presumes that Bob is able to apply arbitrary decoding unitaries to his system $B$. If Bob's experimental setup allows only unitaries acting on $(L(R), R)$ then the cost of merging the state of $(R, R+dR)$ is:
\begin{equation}
S\big((R, R+dR) \, | \, (L(R), R)\big)
\end{equation}
Differential entropy is a sum of exactly such terms.

We conclude \cite{streaming} that differential entropy computes the entanglement cost of quantum teleporting a state in incremental merging steps, subject to the constraint that the most delocalized unitaries at the recipient's disposal act on intervals $(L(R), R)$. Curve $\mathcal{C}$ is a graphical representation of this constraint. If the bulk spacetime geometrizes data about the entanglement of a state, a curve on the spacetime displays a division between different manipulations of it.

A neat way to conceptualize the incremental constrained merging protocol is to think of it as a quantum version of streaming files. Watching a one-hour movie on a streaming service like Netflix\textsuperscript{\textregistered} always takes more bandwidth than downloading the same movie in compressed form. There is a reason why Netflix\textsuperscript{\textregistered} does not send you compressed movies: it is that compression uses infrared data about the movie, which can only be decoded after file transfer is complete. In contrast, a file in streaming format can be decoded on the go using local operations---which is analogous to merging incoming data subject to a locality constraint. The length difference between a convex curve $\mathcal{C}$ connecting boundary points $\theta = L_i, R_f$ and the geodesic length (\ref{sentincr}) is like the extra bandwidth used in streaming a film. Different shapes of curves define different streaming tactics.

\subsubsection{Kinematic space}
\label{sec:ksks}
In a global pure state, the differential entropy formula admits an interesting rewriting \cite{intgeo}:
\begin{equation}
S_{\rm diff}[L(R)] = \frac{1}{2} \int_{(L(R), R)}^{(R,L(R))} 
dL\!\wedge\! dR \,\frac{\partial^2 S(L,R)}{\partial L\, \partial R}
\label{crofton}
\end{equation}
The limits of integration comprise geodesics tangent to curve $\mathcal{C}$, but with different orientations. If the entanglement wedge of interval $(L(R),R)$ meets curve $\mathcal{C}$ only at the tangency point, the entanglement wedge of interval $(R, L(R))$ covers the entire curve but is also tangent to it. When we evaluate~(\ref{crofton}) using Stokes's theorem, each boundary evaluates to one copy of (\ref{diffent}). Note that the region of integration in~(\ref{crofton}) has a very nice characterization: it comprises those geodesics $(L,R)$, which intersect curve $\mathcal{C}$. This is an instance of a Crofton formula \cite{intgeo, Czech:2019hdd}, which expresses a length in terms of intersecting geodesics. 

It is fruitful to interpret the integrand of~(\ref{crofton}) as a Lorentzian metric 
\begin{equation}
ds^2 =  \frac{1}{2} \frac{\partial^2 S(L,R)}{\partial L\, \partial R} dL\, dR
\label{ksmetric}
\end{equation}
and differential entropy as a 1+1-dimensional volume. The space~(\ref{ksmetric}) is called {\it kinematic space}. This metric space is a novel object, different from the CFT and from the bulk. Points in kinematic space are in one-to-one correspondence with CFT intervals or, equivalently, oriented geodesics. The Lorentzian signature of kinematic space has a natural interpretation:
\begin{itemize}
\item Two CFT intervals are timelike-separated in kinematic space when one includes another, with the larger interval living in the future of the other. 
\item Two CFT intervals are spacelike-separated in kinematic space if neither interval contains the other.
\item Two CFT intervals are lightlike-separated in the marginal scenario that separates the preceding two cases, namely when both intervals have the same left or right endpoint. 
\end{itemize}

The integrand of (\ref{crofton}), i.e. the metric of kinematic space, carries deep significance. It is an instance of {\it conditional mutual information}
\begin{equation}
I(A: C|B) = S(AB) + S(BC) - S(B) - S(ABC)\,,
\label{condmutual}
\end{equation}
a quantity which is non-negative in all states by the famous strong subadditivity inequality \cite{Lieb:1973cp}. Because of the positivity of the integrand, we can interpret (\ref{crofton}) as integrating a `density of geodesics.' In the AdS/CFT correspondence, spacelike geodesics represent WKB approximations to CFT two-point functions \cite{vijaysimon, geodesicprop}, so equation~(\ref{crofton}) can be said to collect all bipartite correlations in the CFT state above a sliding ultraviolet scale set by the intervals $(L(R), R)$. In the graphical language of the bulk spacetime, these are the correlations which probe the interior of curve $\mathcal{C}$ at the semiclassical level. 

For the vacuum of a two-dimensional CFT on a plane, metric~(\ref{ksmetric}) becomes
\begin{equation}
ds^2 \propto dL\, dR / (R-L)^2\,,
\label{ksdesitter}
\end{equation}
which is the two-dimensional de Sitter space. The alternative holographic interpretation of the MERA tensor network advocated in \cite{Czech:2015kbp} referred to this (kinematic) space. It is interesting to observe that (\ref{ksmetric}) is the simplest expression that is cutoff-independent even for non-uniform cutoffs. When we allow the cutoff to vary with space, the vacuum entanglement entropy~(\ref{sinterval}) becomes proportional to $(1/2) \log (R-L)^2/\mu(L) \mu(R)$, but this change does not affect~(\ref{ksdesitter}). This property of conditional mutual information will be important in Sections~\ref{sec:berry} and \ref{sec:hec}.

Finally, we comment on kinematic spaces for $d>2$. The simplest (though not only) way to define them is as the space of spherical regions in the CFT \cite{Czech:2016xec}; this generalization also includes intervals off a time slice in CFT$_{d=2}$. In the CFT vacuum, conformal symmetry picks out a unique metric for such kinematic spaces, which turns out to have signature $(d,d)$. Thus, kinematic spaces of spherical regions in higher dimensions are symplectic manifolds---a fact whose full significance is still not fully understood (though see \cite{Penna:2018xqq}).

\subsubsection{Entanglement Berry phases}
\label{sec:berry}
Presentations of Berry phases typically consider a time-dependent and slowly varying Hamiltonian $H(\lambda(t))$, where $\lambda$ coordinatizes a parameter space of possible Hamiltonians and $t$ is physical time. If the system starts out in its ground state, it remains in the ground state of the instantaneous Hamiltonian at all times by the adiabatic theorem. Assuming the Hamiltonian returns to its initial form (that is, $\lambda(t_{\rm final}) = \lambda(t_{\rm initial})$), the relative phase between the initial and final state of the system is unambiguously defined. This is the famous {\it Berry phase} \cite{Berry:1984jv}:
\begin{equation}
\gamma_{\rm Berry} = -i \oint d\lambda\,  \langle \psi(\lambda) | \dot{\psi}(\lambda) \rangle
\equiv -i \oint  \langle \psi | d {\psi} \rangle
\label{berryphase}
\end{equation}
Here $|\psi(\lambda)\rangle$ is a ground state of $H(\lambda)$ with an arbitrarily defined phase and $|\dot{\psi}(\lambda)\rangle = d |\psi(\lambda) \rangle / d\lambda$. Equation~(\ref{berryphase}) assumes for simplicity that $H(\lambda) |\psi(\lambda)\rangle = 0$; otherwise the final state carries an additional dynamical phase, which can be unambiguously isolated from (\ref{berryphase}). 

Equation~(\ref{berryphase}) carries an obvious similarity with (\ref{diffent}). This is not a coincidence. As explained in \cite{Czech:2017zfq}, in quantum field theory it is possible to vary subregions $A$ and construct generalized {\it modular Berry phases} out of the corresponding family of modular Hamiltonians~(\ref{boostgen}). In the AdS/CFT correspondence, modular Berry phases (and generalizations) of the CFT state encode multiple geometric features of the dual bulk geometry, including (transforms of) the Riemann curvature \cite{Czech:2019vih, modularchaos} and lengths of curves such as (\ref{diffent}).

To explain this connection, we give a novel derivation of (\ref{berryphase}), which differs from the usual textbook treatment of Berry phases but which automatically subsumes equation~(\ref{diffent}). 

\paragraph{A new derivation of Berry phases}
Consider a one-parameter family of density matrices $\rho(\lambda)$, with $\lambda = \lambda(t)$ a function of time. We will be studying solutions of $\rho = W W^\dagger$ as a function of time $t$. When $\rho(\lambda)$ is a mixed state, this means studying purifications of $\rho(\lambda)$ (see comments below equation~\ref{traceout}). When $\rho(\lambda) = |\psi(\lambda)\rangle \langle \psi(\lambda)|$, however, we are simply examining $U(1)$ phases that may be attached to $|\psi(\lambda)\rangle$. In this case, referring to $W$ as a `purification' abuses standard terminology, but we wilfully commit this offence to emphasize that ambiguous purifications and Berry phases have a common origin. 

Operators that commute with $\rho(\lambda)$ are called modular zero modes. In holography, subregion duality means that a zero mode of $\rho(\lambda)$ is a symmetry of the dual entanglement wedge. To define (modular) Berry phases, we solve the following parallel transport problem for all $\lambda$:
\begin{align}
& d\lambda\, [V(\lambda), \rho(\lambda)] = \rho(\lambda + d\lambda) - \rho(\lambda) + \textrm{zero modes of $\rho(\lambda)$} 
\label{defv} \\
{\rm and}~ & \textrm{$V(\lambda)$ contains no zero modes of $\rho(\lambda)$}
\label{nozeromodes}
\end{align}
Note that without condition~(\ref{nozeromodes}) the operator-valued one-form $V(\lambda)d\lambda$ would be ambiguous. We also observe that conditions~(\ref{defv}-\ref{nozeromodes}) presume that a projector onto modular zero modes can be defined on operator space. This assumption is subtle but we will proceed undeterred.

We choose an arbitrary initial condition $W(t_{\rm initial})$ and define $W(t)$ recurrently by:
\begin{equation}
W(t + dt) = \big(\mathbb{1} + V(\lambda(t))dt \big) W(t)
\label{berryeom}
\end{equation}
Assuming the trajectory in the space of density matrices is closed, i.e. $\lambda(t_{\rm final}) = \lambda(t_{\rm initial})$, we say that a Berry operator $\Gamma_{\rm Berry}$ relates the initial and final $W$:
\begin{equation}
W(t_{\rm final}) \equiv \Gamma_{\rm Berry} W(t_{\rm initial})
= \left({\rm P}\exp \oint V(\lambda) d\lambda \right) W(t_{\rm initial})
\label{pickberry}
\end{equation}
By construction, $\Gamma_{\rm Berry}$ is valued in zero modes of $\rho(\lambda(t_{\rm initial})) = \rho(\lambda(t_{\rm final}))$. In holography, it will be a symmetry of the relevant entanglement wedge.

\paragraph{Ordinary Berry phases from modular parallel transport} To see how~(\ref{defv}-\ref{nozeromodes}) make contact with usual Berry phases, we apply these equations to pure state density operators $\rho(\lambda) = |\psi(\lambda)\rangle \langle \psi(\lambda)|$. Operator
\begin{equation}
V(\lambda)_{\rm candidate} 
= |\dot{\psi}(\lambda)\rangle \langle {\psi}(\lambda) | 
- |{\psi}(\lambda)\rangle \langle \dot{\psi}(\lambda) |
\label{simplev}
\end{equation}
solves~(\ref{defv}), but generally contains a $2 \langle {\psi}(\lambda) | \dot{\psi}(\lambda) \rangle$-multiple of $\rho(\lambda)$ so fails to satisfy (\ref{nozeromodes}). But if we choose 
\begin{equation}
|\phi(\lambda)\rangle \equiv |\psi(\lambda)\rangle \times
\exp{\Big(-\int_{\lambda(t_{\rm initial})}^\lambda d\lambda' \,  \langle {\psi}(\lambda') | \dot{\psi}(\lambda')\rangle \Big)}
\label{defphi}
\end{equation}
then $\langle \phi(\lambda) | \dot{\phi} (\lambda) \rangle = 0$ and 
\begin{equation}
V(\lambda) = |\dot{\phi}(\lambda)\rangle \langle {\phi}(\lambda) | - |{\phi}(\lambda)\rangle \langle \dot{\phi}(\lambda) |
\end{equation}
solves both (\ref{defv}) and (\ref{nozeromodes}). Moreover, for the initial condition
\begin{equation}
W(t_{\rm initial}) = |\phi(\lambda(t_{\rm initial}))\rangle = |\psi(\lambda(t_{\rm initial}))\rangle
\end{equation}
the solution of~(\ref{berryeom}) is $W(t) = |\phi(\lambda(t))\rangle$. Although definition~(\ref{defphi}) fails to be single-valued over $\lambda$ when the trajectory $\lambda(t)$ closes, the operator $V(\lambda)$ is single-valued and $W(t)$ is a well-defined function of time. Comparing $W(t_{\rm initial})$ to $W(t_{\rm final})$, we recognize that 
\begin{equation}
\Gamma_{\rm Berry} = \exp{\Big(-\oint d\lambda' \,  \langle {\psi}(\lambda') | \dot{\psi}(\lambda')\rangle\Big)},
\end{equation}
which is a pure phase because $|\psi(\lambda')\rangle$ is normalized. That phase is~(\ref{berryphase}).

To understand why conditions~(\ref{defv}-\ref{nozeromodes}) reproduce the Berry phase, observe that:
\begin{equation}
-idt\, H(\lambda) |\phi(\lambda)\rangle = 0 
\qquad {\rm but} \quad 
-i dt \Big(H(\lambda) + i V(\lambda)\Big) |\phi(\lambda)\rangle = \frac{dt}{d\lambda} d\lambda\, |\dot{\phi}(\lambda)\rangle
\end{equation}
If the state is to vary with time, the former contradicts the Schr{\"o}dinger equation. But $i V(\lambda)$ is a Hermitian operator, which---when added to the Hamiltonian---repairs the Schr{\"o}dinger equation and allows us to use $\lambda$ as reparameterized time. We see that condition~(\ref{nozeromodes}) effectively enforces the Schr{\"o}dinger equation, which is the usual starting point for deriving the Berry phase.

\paragraph{Modular parallel transport in holography}
We first rewrite equations~(\ref{defv}-\ref{nozeromodes}) in a way, which is better suited for field theory.  Rather than varying the states $\rho(\lambda)$, we will consider $\lambda$-dependent modular Hamiltonians $H_{\rm mod}(\lambda) = - \log \rho(\lambda)$: 
\begin{align}
& d\lambda\, [V(\lambda), H_{\rm mod}(\lambda)] = H_{\rm mod}(\lambda + d\lambda) - H_{\rm mod}(\lambda) + \textrm{zero modes of $H_{\rm mod}(\lambda)$} 
\label{defvft} \\
{\rm and}~ & \textrm{$V(\lambda)$ contains no zero modes of $H_{\rm mod}(\lambda)$}
\label{nozeromodesft}
\end{align}
When both (\ref{defv}) and (\ref{defvft}) make sense, they have the same solutions $V(\lambda)$. The conditions~(\ref{nozeromodes}) and (\ref{nozeromodesft}) are also the same because zero modes of $H_{\rm mod}$ are simultaneously zero modes of $e^{-H_{\rm mod}}$. So, indeed, equations~(\ref{defvft}-\ref{nozeromodesft}) agree with (\ref{defv}-\ref{nozeromodes}) when both make sense. In our previous example of transporting pure states $\rho(\lambda) = |\psi(\lambda)\rangle \langle \psi(\lambda)|$, however, equation~(\ref{defvft}) could not be used for a technical reason. The corresponding $H_{\rm mod}(\lambda)$s have formally infinite eigenvalues $\sim -\log(0)$ coming from states orthogonal to $|\psi(\lambda)\rangle$, so there are no `small' variations in modular Hamiltonians of pure states. 

We apply (\ref{defvft}-\ref{nozeromodesft}) to a family $\rho(\lambda) = e^{-H_{\rm mod}(\lambda)}$ of reduced states drawn from a common global state dual to a semiclassical geometry, where $\lambda$ parameterizes CFT subregions. In our primary example, we take $\lambda = (L, R)$, which describes intervals on an equal time slice of a two-dimensional CFT. Our goal is to recognize equation~(\ref{diffent}) as characterizing $\Gamma_{\rm Berry}$ for the family of intervals $(L(R), R)$. 

For this, we must interpret condition~(\ref{nozeromodesft}) holographically. As mentioned above, zero modes of $H_{\rm mod}(\lambda)$ are from the bulk point of view symmetries of the dual entanglement wedges $\mathcal{W}(\lambda)$. Thus, equation~(\ref{defvft}) asks for a map from entanglement wedge $\mathcal{W}(\lambda)$ to an infinitesimally different $\mathcal{W}(\lambda+d\lambda)$ while condition~(\ref{nozeromodesft}) demands that the map use no symmetries of $\mathcal{W}(\lambda)$. One such symmetry (equation~\ref{modularshift}) is generated by the modular Hamiltonians; sufficiently near the RT surface it acts as an orthogonal boost. Note that excluding the modular Hamiltonian from $V(\lambda)$ is directly analogous to excluding $\rho(\lambda)$ from $V(\lambda)$ in the construction of the Berry phase. In addition to modular boosts, other symmetries of $H_{\rm mod}(\lambda)$ that act geometrically near the Ryu-Takayanagi surface are RT surface-preserving diffeomorphisms. Those too must be excluded from $V(\lambda)$. 

\begin{figure}[tbp]
\centering 
\includegraphics[width=\textwidth]{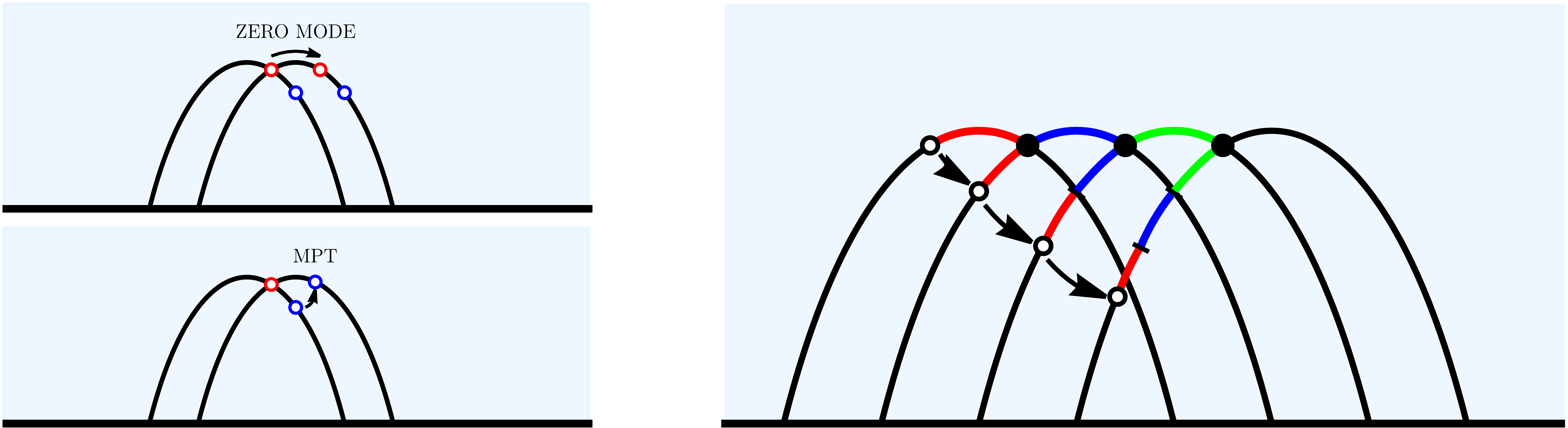}
\hfill
\caption{\label{fig:modularmode} 
Left: A rigid mapping of one geodesic to another is ambiguous by how far their intersection point slides along the target geodesic. Modular parallel transport fixes this ambiguity in equation~(\ref{nozeromodesft}) by setting the amount of sliding to zero. Right: Applying modular parallel transport on a sequence of intersecting geodesics causes a fiducial point to recede from their common envelope $\mathcal{C}$ by an amount equal to the length of $\mathcal{C}$. }
\end{figure}

Let us apply this prescription to intervals $(L(R), R)$; see Figure~\ref{fig:modularmode}. Since we must not apply modular boosts~(\ref{boostgen}), the analysis is confined to an equal time slice of the bulk. Consider the intersection point of geodesics $(L(R), R)$ and $(L(R+dR), R+dR)$, and think of it as a point on $(L(R),R)$. A candidate map $V(\lambda)_{\rm candidate}$ will take it either to itself or to another point on the geodesic $(L(R+dR), R+dR)$ because solutions of (\ref{defvft}) are ambiguous by surface-preserving diffeomorphisms. But sending the intersection point anywhere else would employ the said diffeomorphisms and violate condition~(\ref{nozeromodesft}). We conclude that the solution of (\ref{defvft}-\ref{nozeromodesft}) keeps the intersection point in place. In other words, it acts in the bulk as a rigid rotation about the intersection point.

To find $\Gamma_{\rm Berry}$, we compound the effect of a family of infinitesimal rigid rotations $V(\lambda)d\lambda$ as in (\ref{pickberry}). This is illustrated in Figure~\ref{fig:modularmode}. Note that for a pair of $d\lambda$-separated geodesics the pivot of rotation falls $\mathcal{O}(d\lambda)$ away from the curve $\mathcal{C}$; working at order $\mathcal{O}(d\lambda)$ we may as well take it to live on the curve. Thus, $P\exp \int V(\lambda) d\lambda$ describes the motion of a string, which is being unwound from a $\mathcal{C}$-shaped spool. We may also take a passive view---fix a geodesic and rotate the bulk geometry appropriately---in which case $\Gamma_{\rm Berry}$ describes how a $\mathcal{C}$-shaped wheel rolls on a road (geodesic) without slipping. In either analogy, completing the full trajectory produces a translation along the geodesic by a displacement equal to the circumference of $\mathcal{C}$, i.e. equation~(\ref{diffent}).

\begin{figure}
        \centering
        \includegraphics[width=.96\textwidth]{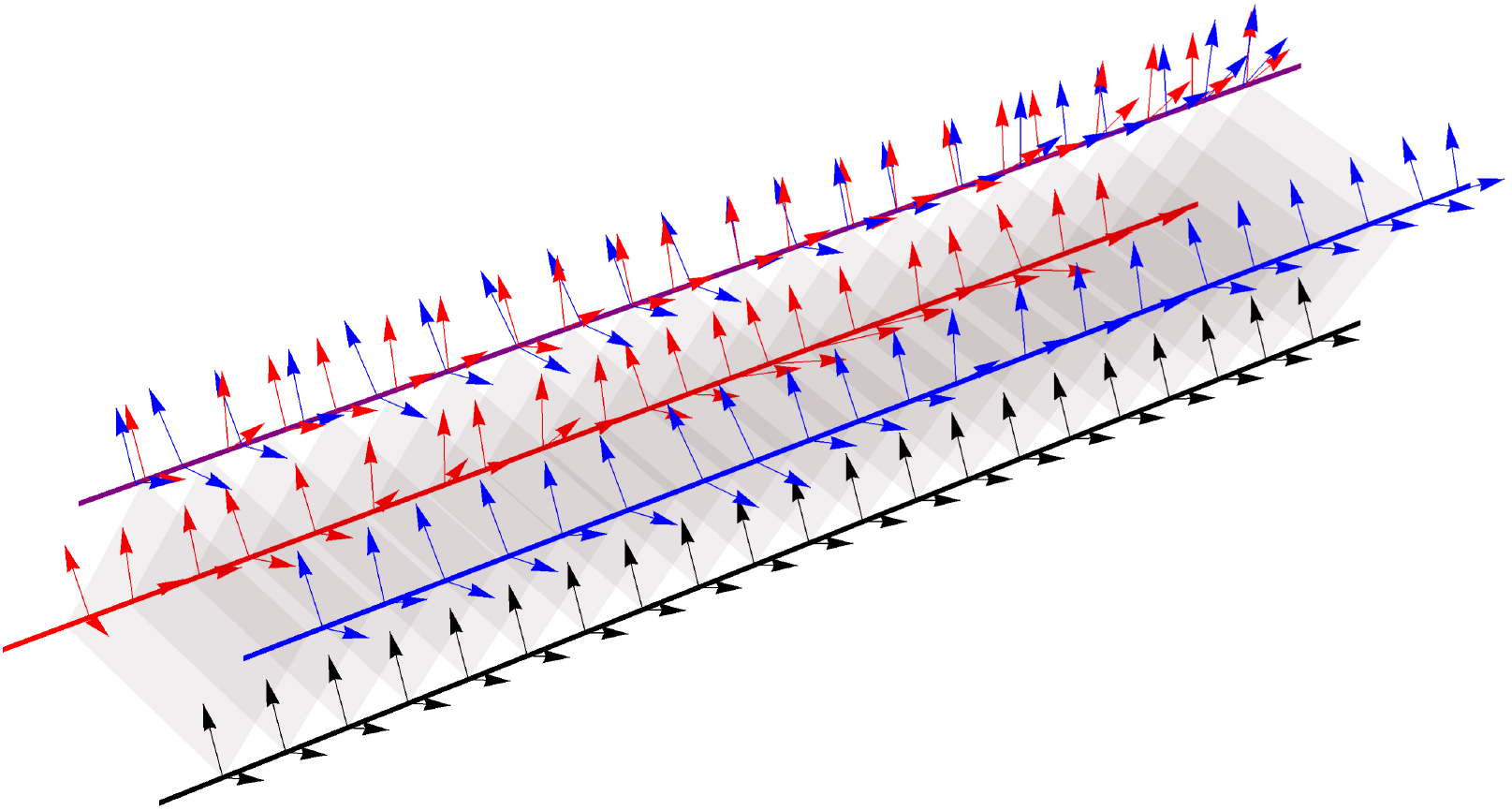}
       \caption{A closed motion of a framed surface in the bulk. If this motion is infinitesimal, the mismatch of the blue and red arrows at top is the modular Berry curvature. Figure is reproduced with permission from \cite{Czech:2019vih}.}
        \label{fig:holonomy}
\end{figure}

\paragraph{Comments:}
\begin{itemize}
\item When we move CFT regions off an equal time slice, $\Gamma_{\rm Berry}$ will generally include both surface-preserving diffeomorphisms and modular boosts fibered over the RT surface.
\item A generic example is shown in Figure~\ref{fig:holonomy}. This is the modular parallel transport of a {\it framed surface}. Note that the Riemann curvature tensor is defined as a similar type of transport, only for {\it framed points} or vielbeins of coordinate vector fields. Thus, modular Berry phases contain much of the same information as the bulk Riemann curvature tensor, only collected surface-wise. For details on how and when the Riemann curvature tensor enters $\Gamma_{\rm Berry}$, see \cite{Czech:2019vih, modularchaos}.
\item The Berry transformation defined by a closed trajectory over an infinitesimal `parallelogram' $d\lambda_1 \wedge d\lambda_2$ is the curvature two-form of the Berry connection. In the case of intervals on a time slice, the conditional mutual information~(\ref{condmutual}) defines a parallelogram-like sequence of motions $AB \to ABC \to BC \to B \to AB$ in kinematic space, which produces a net translation by $I(A,C|B)$. We recognize that the integrand in (\ref{crofton}) is the Berry curvature, which is why we emphasized that it is independent of the ultraviolet cutoff. 
\item Though our presentation focused on $d=2$, the construction of $\Gamma_{\rm Berry}$ works in all $d$. 
\item Applications of (generalized) Berry phases to the geometry of quantum entanglement \cite{geombook} and to AdS/CFT \cite{oblak, kirklin1, kirklin2, nytstory} are a rich subject. It is fascinating that entanglement Berry phases encapsulate such a wealth of data about the bulk geometry: from lengths of curves to integral transforms of Riemann curvature to many others. 
\end{itemize}

\subsection{Modular minimal entropy}
\label{sec:mme}
\begin{figure}
     \centering
     \begin{subfigure}[t]{0.4\textwidth}
         \centering
         \includegraphics[width=\textwidth]{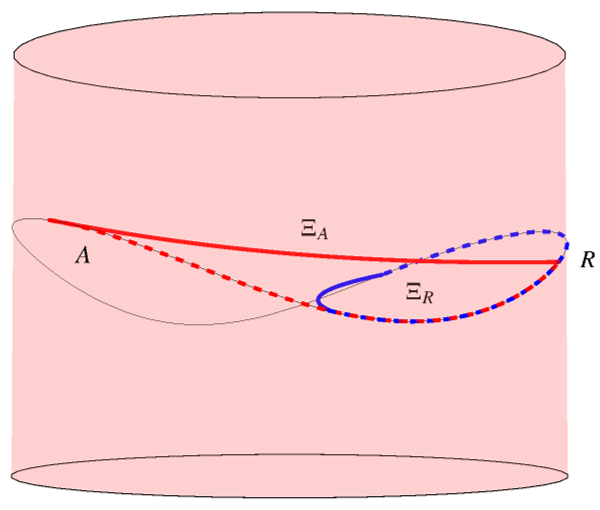}
     \end{subfigure}
     \hfill
     \begin{subfigure}[t]{0.4\textwidth}
        \centering
         \includegraphics[width=\textwidth]{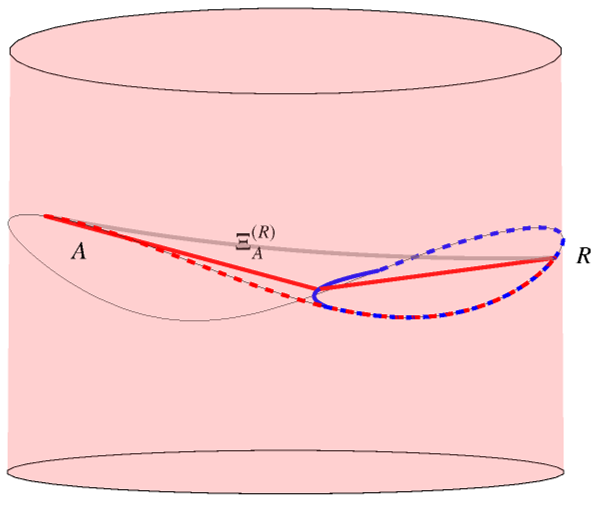}
     \end{subfigure}
        \caption{Left: the generic situation, where the extremal surfaces $\Xi_A$ and $\Xi_R$ of overlapping regions do not cross, but pass by one another separated in time. Right: A surface homologous to $A$ which is constrained to meet $\Xi_R$ (equation~\ref{mme}), proposed to be dual to~(\ref{cftmme}). We included the unconstrained surface $\Xi_A$ (grayed out) for comparison.}
        \label{fig:mme}
\end{figure}

Consider two partially overlapping regions $A$ and $R$ in a holographic CFT. According to the maximin (HRT) proposal~(\ref{hrt}), their entanglement entropies correspond to surfaces $\Xi_{A, R}$, which are minimal on the bulk spatial slices that contain them but maximal with respect to the choice of spatial slice. We will refer to the maximizing spatial slices for $S(A)$ and $S(R)$ as $\Sigma_{A, R}$. The motivating observation of this subsection is that $\Xi_A$ generically lives off $\Sigma_R$ and $\Xi_R$ lives off $\Sigma_A$; see Figure~\ref{fig:mme}. In other words, there is generally no single spatial slice that maximizes (\ref{hrt}) for all choices of regions and two extremal surfaces $\Xi_{A,R}$ typically pass by one another, separated in time. This is a problem if we wish to identify the bulk geometry with a tensor network. It is also an opportunity to study another information theoretic quantity, which diagnoses and quantifies this phenomenon.

For a global state $|\Psi \rangle \in \mathcal{H}_{\rm CFT}$, reference~\cite{Chen:2018rgz} defined the {\it modular minimal entropy}:
\begin{equation}
\bar{S}_R(A) = \min_{s_R} S\big(\rho_A(s_R)\big) 
\qquad {\rm where}~~\rho_A(s_R) = {\rm Tr}_{\bar{A}}\, e^{-is_R H_{\rm mod}^R} |\Psi \rangle \langle \Psi | e^{is_R H_{\rm mod}^R}
\label{cftmme}
\end{equation}
The unitary evolution by the modular Hamiltonian $e^{-i s_R H_{\rm mod}^R} |\Psi\rangle$ is (up to unimportant differences) called Connes cocycle flow and its holographic representation was studied in \cite{connescocycle}. Since $S(A) = S(\rho_A(0))$, we have $\bar{S}_R(A) \leq S(A)$. 

The proposal of \cite{Chen:2018rgz} is that $\bar{S}_R(A)$ is given by the area of an HRT-like surface, which is constructed in a way similar to (\ref{hrt}) but more restricted; see the right panel of Figure~\ref{fig:mme}. In maximizing over the spatial slices $\Sigma$, one restricts to only those slices which contain the HRT surface $\Xi_R$ of region $R$. When regions $A$ and $R$ have a partial overlap, this forces every surface $\Xi \subset \Sigma$ homologous to $A$ to meet $\Xi_R$. One then minimizes over surfaces $\Xi$. Last but not least, one must check whether the extremal surface found in this way intersects with $\Xi_R$ by a constant boost angle $K_\perp$. In AdS$_3$ this is trivial since the extremal surfaces there are one-dimensional. But in higher dimensions a maximin surface satisfying this condition may not exist; when it does, Reference~\cite{Chen:2018rgz} proposes that its area is dual to $\bar{S}_R(A)$:
\begin{equation}
\bar{S}_R(A) =
\max_{
\mathrlap{\!\!\!\!\!\!\!\!
\Sigma\, \Big| \, \subalign{
& {A \subset \partial \Sigma} \\ 
& {\Xi_R \subset \Sigma}
}}} 
\,\,\,\,\,\,\,\,
\left(\min_{\Xi | \Xi \subset \Sigma\,\,{\&}\,\,\Xi \sim A}
\frac{{\rm Area}(\Xi)}{4 G \hbar}\right)
\qquad {\rm if}~K_\perp(\Xi\, \cap\, \Xi_R) = {\rm const.}
\label{mme}
\end{equation}
This proposal is consistent with $\bar{S}_R(A) \leq S(A)$ because (\ref{mme}) is maximized over a subset of the maximization domain in (\ref{hrt}):
\begin{equation}
\bar{S}_R(A) = 
\max_{
\mathrlap{\!\!\!\!\!\!\!\!
\Sigma\, \Big| \, \subalign{
& {A \subset \partial \Sigma} \\ 
& {\Xi_R \subset \Sigma}
}}} 
\,\,\,\,\,\,\,\,
\left(\min_{\Xi | \Xi \subset \Sigma\,\,{\&}\,\,\Xi \sim A}
\frac{{\rm Area}(\Xi)}{4 G \hbar}\right)
\leq
\max_{\Sigma | A \subset \partial \Sigma}
\left(\min_{\Xi | \Xi \subset \Sigma\,\,{\&}\,\,\Xi \sim A}
\frac{{\rm Area}(\Xi)}{4 G \hbar}\right) = S(A)
\end{equation}
The vanishing of the difference
\begin{equation}
S(A) - \bar{S}_R(A)
\end{equation}
is proposed to be a necessary condition for the HRT surfaces $\Xi_{R, A}$ to be drawn from a common Cauchy slice $\Sigma$. Note that when $A \subset R$ or $R \subset A$, $\Xi_A$ and $\Xi_R$ can always be embedded in a common bulk Cauchy slice \cite{Wall:2012uf} and their overlap is an empty set. In this circumstance, the maximin surface in (\ref{mme}) is $\Xi_A$ and $\bar{S}_R(A) = S(A)$ accordingly. 

The motivation for (\ref{mme}) comes from the identification (\ref{boostgen}) of $H_{\rm mod}^R$ with the boost symmetry in the directions orthogonal to $R$. A surface with a constant hyperbolic kink $K_\perp = \Delta s_R$ at the intersection with $\Xi_R$ will become an ordinary, kink-free minimal surface if the cocycle flow $e^{-i \Delta s_R H_{\rm mod}^R}$ is applied to its background geometry. The minimization in (\ref{cftmme}) identifies just the amount of $R$-modular time when this happens.

\subsection{Entanglement wedge cross section and entanglement of purification}
How wide is an entanglement wedge at its narrowest? The von Neumann entropy (\ref{vonneumann}) and modular minimal entropy (\ref{mme}) are both represented by minimal surfaces on certain spatial slices. We may anticipate that the minimal cross section of an entanglement wedge also encodes some information theoretic CFT quantity. 

\begin{figure}
     \centering
     \begin{subfigure}[t]{0.4\textwidth}
         \centering
         \includegraphics[width=\textwidth]{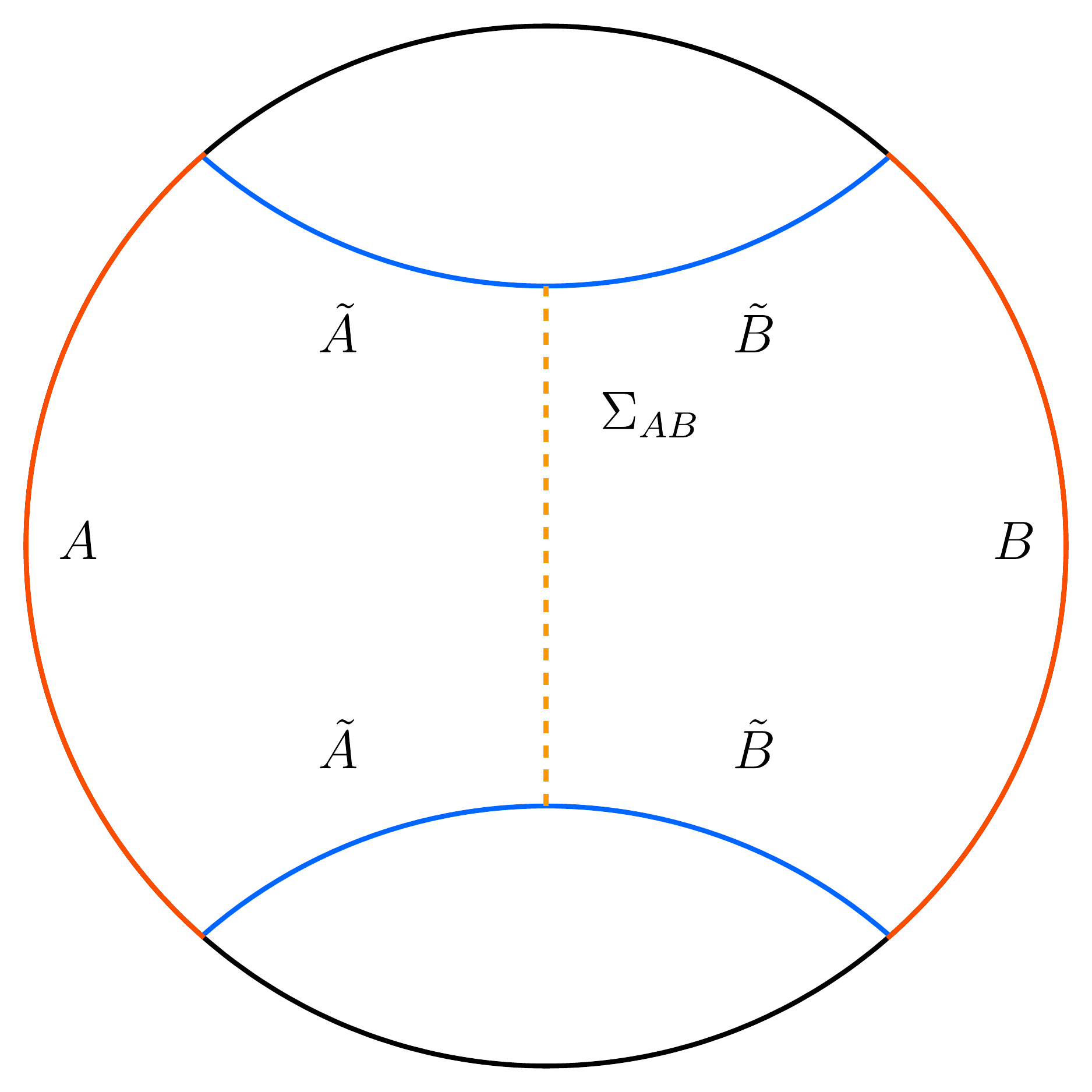}
         \label{ewcs1}
     \end{subfigure}
     \hfill
     \begin{subfigure}[t]{0.4\textwidth}
        \centering
         \includegraphics[width=\textwidth]{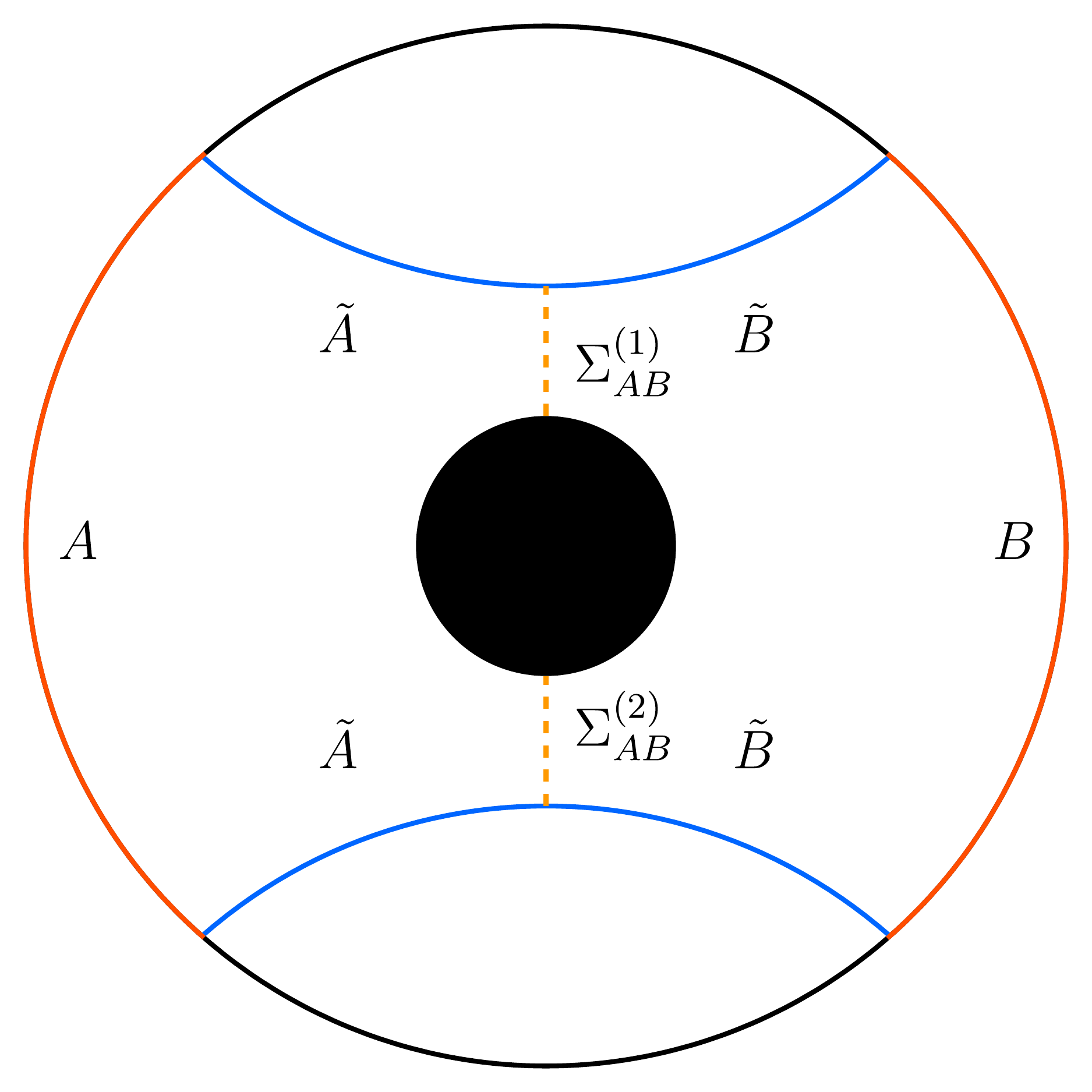}
         \label{ewcs2}
     \end{subfigure}
        \caption{Examples of entanglement wedge cross sections (orange dashed lines) in pure AdS and in a black hole geometry. The optimal purification $\tilde{A}\tilde{B}$ in equation~(\ref{EOP}) is taken to live on the blue surface $\Xi_{AB}$ that binds $\mathcal{W}(AB)$. The entanglement wedge cross section divides this purifier into $\tilde{A}$ and $\tilde{B}$ as shown.}
        \label{fig:ewcs}
\end{figure}

To formalize the notion of the cross-section of an entanglement wedge $\mathcal{W}(R)$ dual to $\rho_R$, we divide the CFT region into two components: $R = A \cup B$. The bulk quantity we are looking for is the area of a minimal cut across some spatial slice of the entanglement wedge, which leaves $A$ and $B$ on its two sides. Reference~\cite{Takayanagi:2017knl} proposed that this minimal cross sectional area (in units of $4G\hbar$) is the so-called entanglement of purification \cite{spurification}:
\begin{equation}
\label{EOP}
E_P(\rho_{AB})
=\min_{WW^\dagger = \rho_{AB}} S_{A\tilde{A}}\big(|W\rangle\langle W|\big)
=\min_{WW^\dagger = \rho_{AB}} S_{B\tilde{B}}\big(|W\rangle\langle W|\big)
\end{equation}
Here, as in equation~(\ref{entangledstate}), $|W\rangle$ is a purification of $\rho_R = \rho_{AB}$, but we now imagine subdividing the purifying Hilbert space into $\mathcal{H}_{\tilde{A}} \otimes \mathcal{H}_{\tilde{B}}$ and attaching $\tilde{A}$ to $A$ and $\tilde{B}$ to $B$. Equation~(\ref{EOP}) asks for the most efficient purification of $\rho_{AB}$ so the resulting systems $A\tilde{A}$ and $B\tilde{B}$ are minimally entangled. This is the most natural way of quantifying the quantum entanglement shared by $A$ and $B$ in $\rho_{AB}$, because it wipes out (extremizes over) artefacts of quantum entanglement between $AB$ and $\tilde{A}\tilde{B}$ in the parent pure state $|W\rangle$. 

The clearest motivation for proposal~(\ref{EOP}) comes from the surface/state correspondence, which associates Hilbert spaces to cuts through the bulk geometry. After we associate $\mathcal{H}_{\tilde{A}} \otimes \mathcal{H}_{\tilde{B}}$ to the HRT surface that defines $\mathcal{W}(AB)$, the division between $\mathcal{H}_{\tilde{A}}$ and $\mathcal{H}_{\tilde{B}}$ is decided by the minimization in (\ref{EOP}). The optimal division has $\tilde{A}$ extending between the boundary region $A$ and the cross-section, and similarly for $\tilde{B}$ extending between $B \subset {\rm CFT}$ and the cross-section; see Figure~\ref{fig:ewcs}. 

A simple sanity check concerns product states $\rho_A \otimes \rho_B$. In this case $\rho_A$ and $\rho_B$ can be individually purified by some $|\Psi\rangle_{A\tilde{A}} \in \mathcal{H}_A \otimes \mathcal{H}_{\tilde{A}}$ and $|\Psi\rangle_{B\tilde{B}} \in \mathcal{H}_B \otimes \mathcal{H}_{\tilde{B}}$ so $E_P(\rho_A \otimes \rho_B) = 0$. Holographically, neglecting $\mathcal{O}(N^{-1})$ corrections, this corresponds to an entanglement wedge being a disjoint union of component entanglement wedges: $\mathcal{W}(AB) = \mathcal{W}(A) \cup \mathcal{W}(B)$. Further consistency checks of proposal~(\ref{EOP}) against various entropic inequalities were considered in \cite{Takayanagi:2017knl}. There are also alternative proposals for how dual CFTs may compute entanglement wedge cross-sections \cite{Tamaoka:2018ned, Dutta:2019gen, shinseysproposal, shinsey2, xiaoliangmichael}. 

\section{Intriguing aspects of the Ryu-Takayanagi formula}
The Ryu-Takayanagi proposal is intriguing because it relates gravity as a dynamical theory of geometry to quantum entanglement. It is all the more intriguing because \emph{not all} geometries represent the entanglement of quantum states, and \emph{not all} quantum states have geometrizable entanglement. Even when a bulk geometry consistently represents the entanglement of a quantum state, its global properties often harbor further mysteries. 

We describe these puzzling aspects of the Ryu-Takayanagi formula below.

\subsection{Not all geometries represent quantum entanglement of boundary states}
Equivalently, this section could be titled `not all bulk stress tensors can arise in holographic duality.' This is because Einstein's equations, or other equations of motion in non-Einstein theories of gravity, relate the spacetime geometry to the stress tensor sourcing it. Any local condition, which a bulk geometry must satisfy for consistency with the Ryu-Takayanagi proposal, is an energy condition when applied on-shell.

Thus far, we have encountered one such condition: the null energy condition. It was assumed in the derivation of black hole thermodynamics (above equation~\ref{2ndlaw}) and in proving \cite{Wall:2012uf} entanglement wedge nesting~(\ref{nesting}) and containment of causal wedges~(\ref{notcausalwedge}), without which subregion duality would make no sense. The null energy condition is a benign assumption on realistic, semiclassical stress tensors \cite{mauliknec} and we have solid reason to believe that a suitable quantum predecessor of it holds in all consistent quantum field theories \cite{qnec, qnec2}. Interestingly, Reference~\cite{ethms} (see also \cite{Lin:2014hva, inviolable}) found that the Ryu-Takayanagi proposal also implies other, generally stronger conditions, which too must hold in the bulk spacetime. 

The CFT quantity that seeds this analysis is the relative entropy~(\ref{relativeentropy}), which we used in Section~\ref{sec:jlms} to justify subregion duality. Here we focus on entropy relative to the vacuum
\begin{equation}
S(\rho|\sigma_{\rm vac}) = {\rm Tr}\,(\rho \log \rho - \rho \log \sigma_{\rm vac}) 
\qquad {\rm where}~\sigma_{\rm vac} = {\rm Tr}_{\bar{B}} |0\rangle\langle 0|
\label{relvacuum}
\end{equation}
for ball-shaped regions $B \subset {\rm CFT}$. This quantity is non-negative, as is evident from expanding in small $d\sigma = \rho - \sigma_{\rm vac}$. (Recall that $S(\rho|\sigma)$ is a measure of distinguishability of states so $S(\rho | \sigma)$ as a function of $\rho$ is minimized by $\rho = \sigma$.) Reference~\cite{ethms} used the covariant proposal~(\ref{hrt}) to argue that $S(\rho|\sigma_{\rm vac})$ is a certain energy measure on the entanglement wedge dual to $\rho$. In any geometry in which the covariant (HRT) proposal~(\ref{hrt}) holds, this energy must not be negative because $S(\rho|\sigma_{\rm vac}) \geq 0$. 

To identify the holographic dual to $S(\rho|\sigma_{\rm vac})$, rewrite it as a difference of free energies:
\begin{align}
S(\rho|\sigma_{\rm vac}) & = 
{\rm Tr}\,(\sigma_{\rm vac} \log \sigma_{\rm vac} - \rho \log \sigma_{\rm vac})) + 
{\rm Tr}\,(\rho \log \rho - \sigma_{\rm vac} \log \sigma_{\rm vac})
\nonumber \\
& = \big(H_{\rm vac}(\rho) - H_{\rm vac}(\sigma_{\rm vac}) \big) - \big(S(\rho)-S(\sigma_{\rm vac})\big) 
\label{relfreeE}
\end{align}
The modular Hamiltonian in any state is a CFT operator localized in the region $B$. In the vacuum, conformal symmetry fixes $H_{\rm vac} = -\log \sigma_{\rm vac}$ to a particular smearing of the CFT stress tensor \cite{chm}, which can be viewed as the generator of a diffeomorphism acting on the boundary. The other term in (\ref{relfreeE}) is a difference of entanglement entropies (or areas of extremal surfaces). In the Iyer-Wald formalism \cite{waldformalism, danieljieqiang}, the area of a surface is a N{\"o}ther charge that generates boosts orthogonal to the surface \cite{entisnother}; applied on $\Xi_B$, it generates evolution in modular time. Thus, both terms in (\ref{relfreeE}) are differences (between $\rho$ and $\sigma_{\rm vac}$) of diffeomorphism generators, which act respectively on $B$ and the associated HRT surface $\Xi_B$. From the homology condition $\Xi_B \sim B$, $B \cup \Xi_B$ is the boundary of a bulk time slice $\tilde{\Sigma}$. If we extend the action of the diffeomorphisms acting on $B$ and $\Xi_B$ to the interior of $\tilde{\Sigma}$ by choosing some vector field $\xi$, equation~(\ref{relfreeE}) will become the difference of Hamiltonian generators of $\xi$, i.e. energies defined with respect to $\xi$. This is how $S(\rho|\sigma_{\rm vac}) \geq 0$ is a bulk energy theorem. 

Every ball $B$ on the boundary imposes one positive energy condition, which must be satisfied if the HRT proposal~(\ref{hrt}) holds. Interestingly, their union does not reduce to any (currently known) set of local criteria. In this way, the HRT proposal reveals previously unsuspected conditions on consistent spacetime geometries. Given how tightly the HRT proposal is tied to the material of Section~\ref{sec:intro}, it is fascinating to ponder that a spacetime geometry must obey these exotic constraints, lest it somehow contaminate seemingly unrelated aspects of general relativity such as black hole thermodynamics. 

\subsection{Not all boundary states have geometrizable entanglement}
\label{sec:hec}
It is obvious that not every state of a holographic CFT can be dual to a semiclassical bulk geometry. For example, a linear superposition of two states, which individually represent wildly different bulk geometries, cannot be expected to describe a single semiclassical bulk \cite{paperwithdon}. Even when two CFT states differ at order $\mathcal{O}(N^0)$---that is, in bulk language, at the level of perturbative fields which live on top of a common background geometry---the metric will not solve Einstein's equations \cite{donfunnyprl}. The application of the Ryu-Takayanagi proposal to such cases was studied in \cite{ahmedrt}.

Interestingly, the Ryu-Takayanagi proposal reveals other, more intriguing limitations on states that may be dual to semiclassical bulk geometries. Such constraints are best understood in the context of time reversal-symmetric states, in which entanglement entropies of subregions are computed by minimal homologous surfaces on the bulk slice fixed by time reversal; see~(\ref{rtfull}). It turns out that the minimality and homology properties of the requisite surfaces automatically imply non-trivial inequalities between entanglement entropies of CFT subregions. The inequalities are non-trivial in that not all quantum states satisfy them.

Below we discuss in detail two inequalities that follow from proposal~(\ref{rtfull}), then sketch the complete set of such inequalities. We initially assume time reversal symmetry and return to time-dependent geometries and states at the end of the subsection. Regions $A, B, \ldots$ are assumed disjoint, but each region may have many connected components.

\paragraph{Subadditivity and monogamy of mutual information} 
The subadditivity inequality reads:
\begin{equation}
S(A) + S(B) \geq S(AB)
\label{subadditivity}
\end{equation}
It is convenient to define a quantity called mutual information
\begin{equation}
I(A:B) \equiv S(A) + S(B) - S(AB),
\label{defmi}
\end{equation}
which rephrases subadditivity as $I(A:B) \geq 0$. We will prove~(\ref{subadditivity}) in holography, following~\cite{matttadashi}.

\begin{figure}
        \centering
        \includegraphics[width=.5\textwidth]{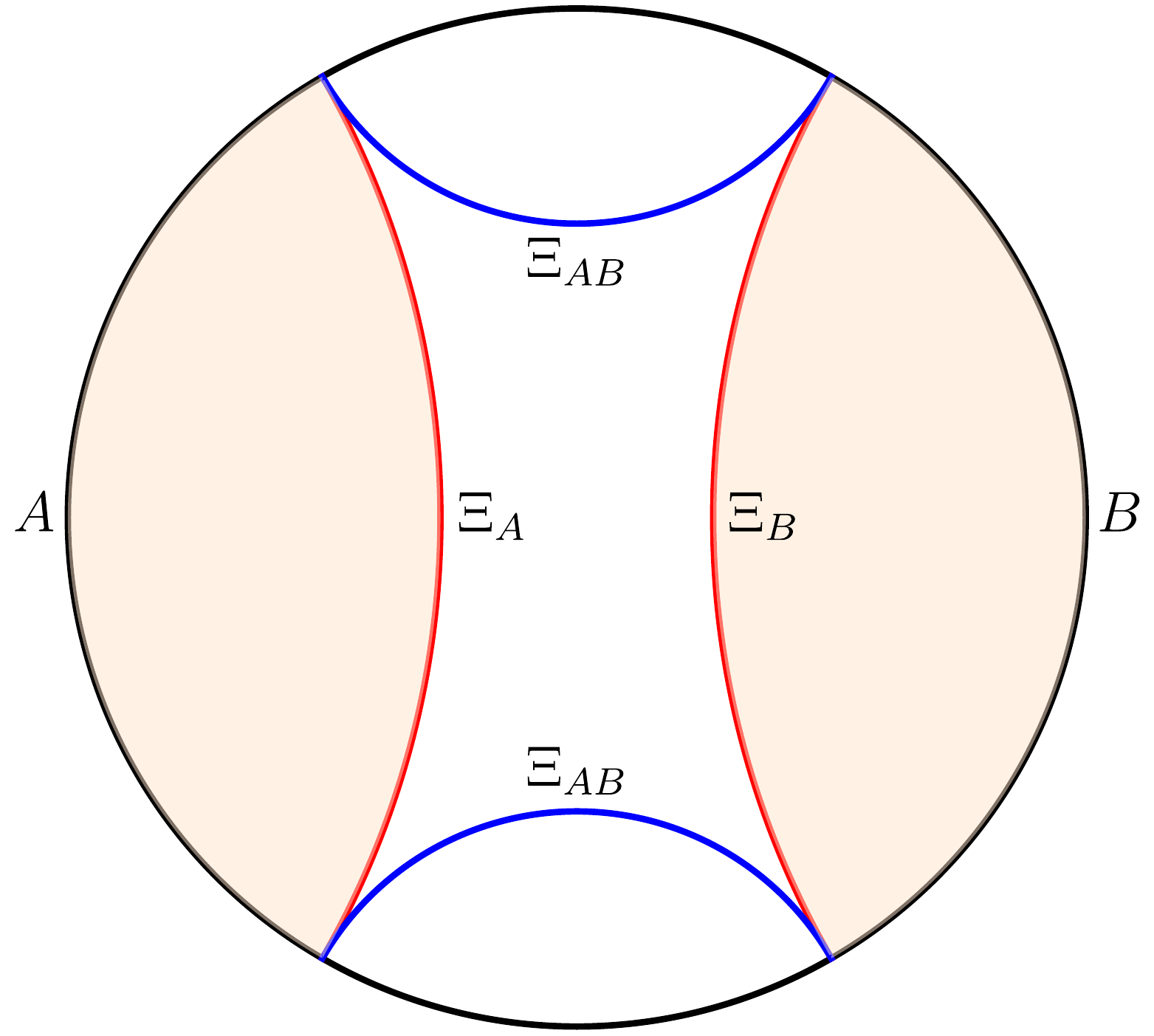}
       \caption{The surface $\Xi_A \cup \Xi_B$ is homologous to $AB$, so its area in units of $4G\hbar$ is no smaller than $S(AB)$. The mutual information $I(A:B)$ is the difference of the two areas.}
        \label{fig:MI}
\end{figure}

Assume that the CFT state is dual to a semiclassical bulk geometry, which is time reversal-symmetric. According to the Ryu-Takayanagi proposal~(\ref{rtfull}), each term in inequality~(\ref{subadditivity}) is the area of a minimal surface $\Xi_A$ (respectively $\Xi_B$, $\Xi_{AB}$) on the privileged bulk slice fixed by time reversal, which is homologous to $A$ (respectively $B$, $AB$). We display the surfaces in Figure~\ref{fig:MI}.  

Note that the surface $\Xi_A \cup \Xi_B$ satisfies the homology condition $\Xi_A \cup \Xi_B \sim AB$. As such, it is part of the domain of minimization in proposal~(\ref{rtfull}) applied to region $AB$. Therefore
\begin{equation}
S(A) + S(B) = \frac{{\rm Area}(\Xi_A \cup \Xi_B)}{4 G \hbar} \geq 
\min_{\Xi \sim AB} \frac{{\rm Area}(\Xi)}{4 G \hbar} = S(AB),
\label{subaddproof}
\end{equation}
so the Ryu-Takayanagi proposal directly implies~(\ref{subadditivity}). 

Now consider the monogamy of mutual information \cite{monogamyref}:
\begin{equation}
S(AB) + S(BC) + S(CA) \geq S(A) + S(B) + S(C) + S(ABC)
\label{monogamy}
\end{equation}
It is called that because of the asymmetric rewriting
\begin{equation}
I(A:BC) \geq I(A:B) + I(A:C),
\end{equation}
which says that partners $B$ and $C$ do not have common mutual information with $A$. 
Proving (\ref{monogamy}) proceeds similarly to (\ref{subaddproof}), but there is a new element \cite{monogamyref}. In~(\ref{subaddproof}), we used the surfaces on the left hand (greater than) side of~(\ref{subadditivity}) wholesale in order to assemble a surface homologous to the region on the right hand (less than) side of the inequality. Now we will need to divide the surfaces $\Xi_{AB}$, $\Xi_{BC}$, $\Xi_{CA}$ into pieces, and then re-glue those pieces to assemble surfaces homologous to $A, B, C$, and $ABC$.

To describe the requisite cutting and re-gluing, we must go back to the homology condition in proposal~(\ref{rtfull}). Homology $\Xi_{AB} \sim AB$ means that the minimal surface $\Xi_{AB}$ that computes $S(AB)$, taken together with the CFT region $AB$, forms the boundary of a submanifold $\tilde{\Sigma}_{AB}$ of the privileged bulk slice: $\Xi_{AB} \cup AB = \partial \tilde{\Sigma}_{AB}$. A holographic representation of the left hand side of (\ref{monogamy}) therefore provides three submanifolds of the bulk slice: $\tilde{\Sigma}_{AB}$, $\tilde{\Sigma}_{BC}$, and $\tilde{\Sigma}_{CA}$. We now form
\begin{align}
& \Sigma_A = \tilde{\Sigma}_{AB} \cap \overline{\tilde{\Sigma}_{BC}} \cap \tilde{\Sigma}_{CA}
\qquad
\Sigma_B = \tilde{\Sigma}_{AB} \cap {\tilde{\Sigma}_{BC}} \cap \overline{\tilde{\Sigma}_{CA}}
\qquad
\Sigma_C = \overline{\tilde{\Sigma}_{AB}} \cap {\tilde{\Sigma}_{BC}} \cap \tilde{\Sigma}_{CA}
\nonumber \\
& \Sigma_{ABC} = \tilde{\Sigma}_{AB} \cup \tilde{\Sigma}_{BC} \cup \tilde{\Sigma}_{CA}\,,
\label{choosesigma}
\end{align}
where overlines stand for taking the complement on the privileged bulk slice.
By construction, these submanifolds meet the asymptotic boundary at $A, B, C, ABC$, respectively. From the definition of homology, this is equivalent to saying that:
\begin{align}
& \partial \Sigma_A \setminus A \equiv \Xi'_A \sim A
\qquad \qquad
\partial \Sigma_B \setminus B \equiv \Xi'_B \sim B
\qquad \qquad
\partial \Sigma_C \setminus C \equiv \Xi'_C \sim C
\nonumber \\
& \partial \Sigma_{ABC} \setminus ABC \equiv \Xi'_{ABC} \sim ABC
\label{constrhomology}
\end{align}
In other words, we have defined four surfaces, each of which satisfies a homology condition for one region on the right hand side of (\ref{monogamy}). It is straightforward, if somewhat tedious, to verify that
\begin{equation}
\Xi'_A \cup \Xi'_B \cup \Xi'_C \cup \Xi'_{ABC} \subset \Xi_{AB} \cup \Xi_{BC} \cup \Xi_{CA}\,,
\label{inclusion}
\end{equation}
where overlapping components of different surfaces are counted with multiplicity. In other words, the surfaces $\Xi'_{A,B,C,ABC}$ are not only homologous to $A,B,C,ABC$, but are indeed obtained by cutting and re-gluing the surfaces $\Xi_{AB,BC,CA}$, with some pieces of the latter left to spare. With these provisos in hand, we copycat proof~(\ref{subaddproof}):
\begin{align}
& S(AB) + S(BC) + S(CA) \nonumber \\
=\, &
\frac{{\rm Area}(\Xi_{AB})}{4 G \hbar} 
+ \frac{{\rm Area}(\Xi_{BC})}{4 G \hbar} 
+ \frac{{\rm Area}(\Xi_{CA})}{4 G \hbar}
\nonumber \\
\geq\, & \frac{{\rm Area}(\Xi'_{A})}{4 G \hbar} 
+ \frac{{\rm Area}(\Xi'_{B})}{4 G \hbar} 
+ \frac{{\rm Area}(\Xi'_{C})}{4 G \hbar}
+ \frac{{\rm Area}(\Xi'_{ABC})}{4 G \hbar}
\nonumber \\
\geq\, &
\min_{\Xi_A \sim A} \frac{{\rm Area}(\Xi_A)}{4 G \hbar} +
\min_{\Xi_B \sim B} \frac{{\rm Area}(\Xi_B)}{4 G \hbar} +
\min_{\Xi_C \sim C} \frac{{\rm Area}(\Xi_C)}{4 G \hbar} +
\min_{\Xi_{ABC} \sim ABC} \frac{{\rm Area}(\Xi_{ABC})}{4 G \hbar}
\nonumber \\
= \,& S(A) + S(B) + S(C) + S(ABC)
\label{monoproof}
\end{align}
The first inequality follows from the inclusion~(\ref{inclusion}) of surfaces $\Xi'_{A,B,C,ABC}$ in $\Xi_{AB,BC,CA}$, and the second one from the fact that the surfaces $\Xi'_{A,B,C,ABC}$ each belong to the relevant domains of minimization because of the homology condition~(\ref{constrhomology}).

\paragraph{Comments on subadditivity and monogamy:}
\begin{itemize}
\item Subadditivity~(\ref{subadditivity}) is an identity, which holds in all quantum systems, so it is not special to holography. However, the general proof of subadditivity is difficult \cite{arakilieb}, in marked contrast to the one-liner~(\ref{subaddproof}) afforded by assuming the Ryu-Takayanagi proposal. 
\item Monogamy of mutual information~(\ref{monogamy}) is not an identity; there are states in quantum mechanics, which violate it. The simplest example is the GHZ state on four qubits:
\begin{equation}
|GHZ\rangle_4 = \frac{1}{\sqrt{2}} |0000\rangle + \frac{1}{\sqrt{2}} |1111\rangle
\label{ghz4}
\end{equation}
Letting $A,B,C$ be the first, second and third qubit, we see that:
\begin{equation}
S(AB) = S(BC) = S(CA) = S(A) = S(B) = S(C) = S(ABC) = \log 2
\label{sghz}
\end{equation} 
As such, the monogamy of mutual information is a non-trivial, necessary condition for a quantum state, if the latter is to be holographically dual to a semiclassical geometry that obeys the Ryu-Takayanagi proposal~(\ref{rtfull}).
\item The preceding discussion assumed time reversal symmetry, but Reference~\cite{Wall:2012uf} proved both (\ref{subadditivity}) and (\ref{monogamy}) by assuming only the covariant proposal~(\ref{hrt}). We next turn to other inequalities, which are only known to hold assuming time reversal-symmetry.  
\end{itemize} 

\subsubsection{Holographic entropy cone}
\label{sec:hecdetail}
Before discussing individual inequalities, it is useful to define a novel concept: the holographic entropy cone \cite{Bao:2015bfa}. (See also \cite{hecformal} for a more formal presentation of this concept.) When we select $n$ disjoint regions $A, B, C, \ldots$ in the CFT, we have $2^n - 1$ distinct unions of them such as $AB$ and $ABC$ in inequality~(\ref{monogamy}); the $-1$ excludes the empty set. Consider a $(2^n - 1)$-dimensional vector space parameterized by entanglement entropies such as $S(A)$, $S(B), \ldots$, $S(AB)$, $S(AC), \ldots$, $S(ABC), \ldots$ Every inequality such as (\ref{subadditivity}) or (\ref{monogamy}) divides this vector space into two halves: one allowed by the inequality, and one disallowed. The dividing locus is a hyperplane, which describes states that saturate the relevant inequality. In the holographic analysis, such states are especially interesting because they are \emph{marginally consistent} with the Ryu-Takayanagi proposal: an infinitesimal deformation of such a state can produce a violation of (\ref{rtfull}). Note that all hyperplanes of marginality (a.k.a. inequalities) meet at the origin because setting all entropies to vanish is marginally consistent with the Ryu-Takayanagi proposal.

A collection of hyperplanes, which all meet at one point, defines a cone. In this way, the inequalities obeyed by holographic entanglement entropies---whatever they be---define the \emph{holographic entropy cone}. Other types of entropic cones have been extensively studied in the literature; see~\cite{wintertalk} for a pedagogical introduction. The most obvious cones are the quantum entropy cone, which comprises all assignments of entanglement entropies to subsystems allowed by quantum mechanics, as well as the Shannon entropy cone, which comprises all consistent Shannon entropies of joint probability distributions. These cones have tremendously rich and not yet understood structures, which set limits on the types of correlations allowed by quantum and classical statistics. For example, already at $n=4$ random variables, the Shannon cone is known not to be polyhedral \cite{nonpolyhedral}---that is, it is demarcated by infinitely many independent inequalities. The same is conjectured for the quantum cone of $n=4$ named regions. Both those cones include subadditivity~(\ref{subadditivity}) as a binding hyperplane. 

Our focus here is the \emph{holographic} entropy cone. This cone comprises assignments of entanglement entropies consistent with the Ryu-Takayanagi proposal. (The cone comprising entanglement entropies consistent with the quantum-corrected version of the Ryu-Takayanagi proposal---equation~(\ref{qrt})---is studied in \cite{qhec}.) Its boundaries set limits on correlations, which are geometrizable in the sense of Ryu-Takayanagi. This cone is strictly smaller than the quantum entropy cone. For example, monogamy~(\ref{monogamy}) is a facet of the holographic entropy cone but cuts through the interior of the quantum entropy cone. Correspondingly, the entanglement entropies~(\ref{sghz}) of the GHZ state~(\ref{ghz4}) live within the quantum entropy cone but outside the holographic entropy cone. As a final preliminary, we point out that the fundamental property of a cone---convexity---is manifestly satisfied by assignments of entropies consistent with the Ryu-Takayanagi proposal. This is because different such assignments $S(A^{(i)}), S(B^{(i)}), S(C^{(i)}), \ldots$ labeled by $i$ can always be combined by taking independent holographic CFT$^{(i)}$s and setting $A = \cup_i A^{(i)}$ (and similarly for other regions). The corresponding bulk geometry can be a disjoint union of geometries or a wormhole connecting CFT$^{(i)}$s living on distinct asymptotic boundaries. 

\paragraph{Properties of the holographic entropy cone, assuming time reversal symmetry:}
Reference~\cite{Bao:2015bfa} studied the holographic entropy cone assuming time reversal symmetry. They identified many additional inequalities beyond subadditivity and monogamy. Below are some of their salient features. 

N.B.: The ensuing comments assume that the inequalities are presented in the same format as (\ref{subadditivity}) and (\ref{monogamy}), i.e. all coefficients on either side are positive. As in (\ref{subadditivity}, \ref{monogamy}), we place the `greater than' side on the left.

\begin{itemize}
\item All the inequalities are balanced. A balanced inequality is one in which every region $A, B, C \ldots$ appears the same number of times on both sides. For example, $A$ appears in (\ref{monogamy}) two times on the left (in $AB$ and $CA$) and on the right (in $A$ and $ABC$), as does $B$ and $C$. \textchg{When all relevant regions have smooth boundaries,} the balanced property means that the inequalities are only sensitive to UV-finite (cutoff-independent) entanglement. \textchg{(Regions with non-smooth boundaries have a richer structure of UV divergences, which do not generically cancel out of balanced inequalities.)}. An inequality with unbalanced appearances of $A$ (with all other regions balanced) would be trivially true or false because it would be overwhelmed by the UV-divergent entanglement of $A$. \textdel{The balanced condition means that the holographic inequalities discovered in \cite{Bao:2015bfa} do not pit against one another the UV-divergences in the quantum entanglements of different regions.}
\item Reference~\cite{Bao:2015bfa} automated the procedure of proving holographic inequalities. Going back to the proof~(\ref{monoproof}) of monogamy, they defined a mechanical process, which identifies submanifolds $\Sigma_{A, B \ldots}$ and surfaces $\Xi'_{A, B \ldots}$ such as (\ref{choosesigma}) and (\ref{constrhomology}) for more general inequalities. However, they did not demonstrate that every valid inequality can be proved this way, which leaves a theoretical possibility that some other, elusive inequalities may be inaccessible to their technical apparatus. 
\item Details of the automated proof procedure are too involved for this review. However, to aid the reader who consults the original reference, we briefly explain how our presentation relates to the more formal terminology of \cite{Bao:2015bfa}. Its authors reduced the proof of an inequality to finding a map, which sends vertices of a hypercube with axes labeled by terms on the left hand side of the inequality, to vertices of a hypercube with axes labeled by terms on the right hand side. The requisite map, which is then used to define regions and surfaces analogous to (\ref{choosesigma}) and (\ref{constrhomology}), must satisfy a certain `initial condition' and a so-called `contraction property.' The step~(\ref{choosesigma}) in our proof of monogamy is actually derived from this type of map. The `initial conditions' of \cite{Bao:2015bfa} guarantee that the resulting surfaces satisfy the homology conditions, analogous to (\ref{constrhomology}). The `contraction property' ensures that the surfaces can be assembled from pieces of the $\Xi$s which represent terms on the left hand side, analogous to (\ref{inclusion}). 
\item Whenever we discover one valid inequality, we also discover many others because the inequalities must be valid independently of how we label the regions. To exploit this fully, we define an additional region $O$, which purifies the union of the $n$ named regions $A \cup B\cup C \cup\ldots$. With the $(n+1)^{\rm th}$ region $O$ included, we can apply the full symmetric group $S_{n+1}$ to any valid inequality. To see the power of relabelling, apply $A \to B \to O$ to (\ref{subadditivity}) and use $S(O) = S(AB)$ and $S(BO) = S(A)$ to get:
\begin{equation}
S(B) + S(AB) \geq S(A)
\label{permutesub}
\end{equation}
Interestingly, the monogamy of mutual information~(\ref{monogamy}) is invariant under the full $S_4$, which permutes $A$, $B$, $C$, and $O$. In addition to permuting regions, we may also recombine them. For example, by substituting $A \to AB$ and $B \to CDE$ we obtain the following instance of subadditivity:
\begin{equation}
S(AB) + S(CDE) \geq S(ABCDE)
\label{recombsub}
\end{equation} 
\item The combinatorics of the full cone quickly become formidable. For example, the $n=5$ cone has 372 facets / inequalities \cite{cuenca}. To organize this data, we look for a minimal set of inequalities that generate all others by permuting or recombining regions and taking linear combinations.
\item The holographic entropy cone is known for up to $n=5$ regions \cite{cuenca}, but not beyond. The minimal generating inequalities for five regions comprise subadditivity~(\ref{subadditivity}), monogamy of mutual information~(\ref{monogamy}), and six other, highly non-trivial inequalities, listed in Table~1 of \cite{cuenca}. To impart a proper sense of awe, we copy one of them below:
\begin{align}
& 3(ABC\! +\! ABD\! +\! ACE)\! +\! ABE\! +\! ACD\! +\! ADE\! +\! BCD\! +\! BCE\! +\! BDE\! +\! CDE \nonumber \\
\geq\, & 2(AB \! +\!  AC \! +\!  BD \! +\!  CE \! +\!  ABCD \! +\!  ABCE) \! +\!  AD \! +\!  AE \! +\!  BC \! +\!  DE \! +\!  ABDE \! +\!  ACDE
\nonumber \\
&
\label{crazyineq}
\end{align}
In this inequality (and only here), we write $ABC$ instead of $S(ABC)$ to locally ease the notation. \textchg{At present, there is no surefire way to generate this or other holographic inequalities. For the best attempt to do so, see \cite{repackaging}.}
\item There is one \textdel{known} type of inequality (up to relabelling etc.), \textdel{which is valid} \textchg{whose validity has been proven} for an arbitrary number $n$ of regions \cite{Bao:2015bfa}. Assuming $n = 2k + 1$ is odd, it reads:
\begin{align}
& S(A_1 A_2 \ldots A_{k+1}) + S(A_2 A_3 \ldots A_{k+2}) + \ldots + S(A_n A_1 \ldots A_{k}) 
\label{cyclic} \\
\geq\,  & 
S(A_1 A_2 \ldots A_{k}) + S(A_2 A_3 \ldots A_{k+1}) + \ldots + S(A_n A_1 \ldots A_{k-1})
+ S(A_1 A_2 \ldots A_n)
\nonumber
\end{align}
A similar inequality holds for even $n$, but it is implied by (\ref{cyclic}) and subadditivity. As such, only~(\ref{cyclic}) is part of the fundamental generating set of inequalities. 
\item Let the regions $A_i$ in (\ref{cyclic}) be intervals in a CFT$_2$ on a circle. Assume that the regions partition the full circle, and that their labels $i$ are cyclically ordered on the circle. If we send $n \to \infty$ such that the size of every $A_i$ goes to zero, the sum of terms
\begin{equation}
S(A_i A_{i+1} \ldots A_{i+k}) - S(A_{i} A_{i+1} \ldots A_{i+k-1}) = 
S(A_{i+k} | A_{i} \ldots A_{i+k-1})
\end{equation}
becomes the same integral we encountered in equations~(\ref{diffent}) and (\ref{sentincr}), i.e. differential entropy. The shape of the bulk curve computed by the differential entropy is set by the rates at which the sizes of the intervals $A_i$ approach zero in the $n \to \infty$ limit. Because we have assumed that $\cup_{i=1}^n A_i$ is the full CFT, the final term in (\ref{cyclic}) is the overall entropy of the CFT, or equivalently the area of the horizon in the bulk. In this setup, the $n \to \infty$ limit of the cyclic inequality~(\ref{cyclic}) is:
\begin{equation}
S_{\rm diff} \geq S(\rho_\textrm{full CFT}).
\end{equation}
It says that a curve which encircles the horizon must be longer than the horizon. It is fascinating that the cyclic inequalities~(\ref{cyclic}) provide a finite $n$ `resolution' of the differential entropy formula, and that the `resolution' is valid even when the concept of differential entropy does not apply (in higher dimensions, when regions $A_i$ are labeled out of order, or if they have many connected components.)
\item All known \textchg{and yet-to-be discovered, currently unknown fundamental inequalities enjoy a property called superbalance:} they are saturated by collections of EPR pairs \cite{balanced}. For example, an EPR pair shared by $A$ and $B$ contributes $\log 2$ to every term that includes one but not both of $A$ and $B$; there are exactly two such terms on each side of (\ref{monogamy}) and eight such terms on each side of~(\ref{crazyineq}). For EPR pairs shared by named regions and the purifier $O$, \textchg{superbalance reduces to} the balanced condition. 
\item \textchg{Superbalance} means that the mutual information~(\ref{defmi})---the amount by which subadditivity is away from saturation---uniquely isolates from an entropy vector contributions of EPR pairs. If we subtract off EPR pairs, the remaining part of the entropy vector is solely responsible for the non-saturation of other inequalities. For example, the entropy vector
\begin{align}
& S(A) = S(B) = S(AB) = 2\log 2 \nonumber \\
& S(AC) = S(BC) = 3 \log 2 \label{exsvector} \\
& S(C) = S(ABC) = \log 2 \nonumber
\end{align}
saturates all fundamental inequalities except the subadditivity~$I(A:B)\geq 0$ and monogamy:
\begin{align}
& S(A) + S(B) - S(AB) = 2 \log 2 \geq 0 
\label{subadditivityhere} \\
& S(AB) + S(BC) + S(CA) - S(A) - S(B) - S(C) - S(ABC) = \log 2 \geq 0 
\label{monogamyhere}
\end{align}
Inequality~(\ref{subadditivityhere}) tells us that the state must contain (an equivalent of) one $A$-$B$ EPR pair, which contributes to~(\ref{exsvector}):
\begin{align}
& S_{\rm EPR}(A) = 
S_{\rm EPR}(B) = 
S_{\rm EPR}(AC) = 
S_{\rm EPR}(BC) = \log 2 \nonumber \\
& S_{\rm EPR}(C) = S_{\rm EPR}(AB) = S_{\rm EPR}(ABC) = 0 
\label{eprhere}
\end{align}
$S_{\rm EPR}(.)$ is fully responsible for the non-saturation of subadditivity in~(\ref{subadditivityhere}). If we subtract it off, the remainder $S_{\rm PT}(.) = S(.) - S_{\rm EPR}(.)$ reads:
\begin{align}
& S_{\rm PT}(A) = S_{\rm PT}(B) = S_{\rm PT}(C) = S_{\rm PT}(ABC) = \log 2 \nonumber \\
& S_{\rm PT}(AB) = S_{\rm PT}(AC) = S_{\rm PT}(BC) = 2 \log 2 \label{ptvector}
\end{align}
This vector represents a contribution to the entanglement pattern~(\ref{exsvector}), which is qualitatively distinct from EPR pairs. This contribution is invisible to subadditivity~(\ref{subadditivity}), but it is detected by and fully accounts for the monogamy inequality~(\ref{monogamyhere}). We will interpret the entanglement pattern~(\ref{ptvector}) in Section~\ref{sec:flows}.
\end{itemize}

\paragraph{The holographic cone of average entropies}
\textchg{
The last point---that holographic entropy inequalities decompose entropy vectors into atomic constituents---recently found a generalization to arbitrarily many regions \cite{coa} (see also \cite{ coacuenca}). To explain the assertion, we define $S^p$ as the average of all $p$-region entropies, with and without the purifier $O$ included. For example, for $n=3$ named regions, we have:
\begin{align}
S^1 & = \frac{1}{4} \big( S(A) + S(B) + S(C) + S(O) \big) \\
S^2 & = \frac{1}{6} \big( S(AB) + S(AC) + S(AO) + S(BC) + S(BO) + S(CO) \big)
\end{align}
With the purifier in the game, complementary regions automatically have equal entropies (e.g. $S(O) = S(ABC)$ for $n=3$ named regions). This means that at arbitrary $n$ we have $\myfloor{\frac{n+1}{2}}$ independent quantities $S^p$, where $\myfloor{\ldots}$ is the floor function. Referring back to the discussion above~(\ref{permutesub}), we recognize that $S^p$ form the complete set of $S_{n+1}$-invariants of an entropy vector. Instead of asking about detailed values of individual regions' entropies---like inequalities~(\ref{crazyineq}) and (\ref{cyclic}) do---we may now ask what values of $S^p$ are achievable by holographic states. This question defines the `holographic cone of average entropies.'  This cone is conjectured to have the simplest imaginable structure \cite{coa}, which we presently explain.
}

\textchg{
Because we have $\myfloor{\frac{n+1}{2}}$ quantities, the simplest possible cone is bound by $\myfloor{\frac{n+1}{2}}$ hyperplanes (inequalities). It is a simplex whose one face (the one that excludes the origin) is pushed to infinity. Equivalently, any cross-section of the cone is an $(\myfloor{\frac{n+1}{2}}-1)$-dimensional simplex. Such cones, called simplicial, have a special property: any vector in the interior can be \emph{uniquely} written as a linear combination of $\myfloor{\frac{n+1}{2}}$ extremal vectors with non-negative coefficients. (For non-simplicial convex cones, the corresponding statement is weaker because there is no uniqueness.) This fact means that for $n$ named regions, we have $\myfloor{\frac{n+1}{2}}$ atomic constituents, which uniquely make up any holographically achievable set of average $p$-region entropies.
}

\textchg{
The reasoning, which supports this conjecture, is outside the scope of this review. To pique the reader's interest, and to contextualize this claim, we only mention three intriguing aspects of the conjecture:
\begin{itemize}
\item The atomic ingredients, which build up any holographically achievable vector of average $p$-region entropies, are graded by the degree of erasure correction they offer. We discussed the role of erasure correction in holography in Section~\ref{sec:error}.
\item The same atomic ingredients can be identified with stages of unitary evaporation of black holes. We may think of individual Hawking particles as the named regions $A, B, C, \ldots$ and the black hole itself as the purifier $O$. As a black hole evaporates, the resulting vector of average entropies $S^p$ visits all the atomic ingredients of the cone of average entropies one by one. We illustrate that statement in Figure~\ref{fig:islands} in Section~\ref{sec:genentropy} where we discuss the black hole information paradox.
\item When we add up all permutation images of a valid inequality (for example, (\ref{permutesub}) is a permutation image of $S(A) + S(B) \geq S(AB)$), we get another valid inequality, which only involves the permutation invariants $S^p$. With the sole exception of (\ref{cyclic}), every known holographic inequality reduces under this symmetrization procedure to the binding inequalities of the cone of average entropies. In this precise sense, the cone of average entropies `unifies' almost all known holographic inequalities. 
\end{itemize}
For an explanation of these claims, we refer the reader to \cite{coa}. For completeness, we present the bounding inequalities of the cone of average entropies. Except for the symmetrized version of subadditivity, which is $2S^1 \geq S^2$, all of them take a unified, $n$-independent form:
\begin{equation}
2 (p-1) (p+1) S^p \geq p (p+1) S^{p-1} + p (p-1) S^{p+1}
\end{equation}
}

\textdel{
The entropy vector~(\ref{exsvector}) can be decomposed into pieces, which individually account for the non-saturation of inequalities~(\ref{subadditivityhere}-\ref{monogamyhere}), and which represent qualitatively distinct types of entanglement. Can we extend the relation between fundamental inequalities and distinct types of entanglement to the less trivial inequalities, beyond subadditivity and monogamy? An affirmative answer would tell us that holographic geometries are dual to states, which are unitarily equivalent to (combinations of) a discrete set of basic `atoms' of entanglement, which include EPR pairs among others. This is a tantalizing but formidable possibility. But how do we go about constructing a quantum state whose entanglement pattern is solely responsible for the non-saturation of inequality~(\ref{crazyineq})? Partial progress on this question was reported in \cite{repackaging}.}

\paragraph{Time dependence} The above discussion was based on the non-covariant Ryu-Takayanagi proposal~(\ref{rtfull}), which applies when the quantum state and the bulk geometry are time reversal-symmetric. What can we say without assuming time reversal symmetry?

In general dimensions, not much: among the fundamental inequalities uncovered so far, only subadditivity and monogamy have been proven~\cite{Wall:2012uf} starting from the covariant (HRT) proposal~(\ref{hrt}). The situation is markedly different in $d=2$ boundary dimensions, where all inequalities identified in \cite{Bao:2015bfa} were proven \cite{kscone}. The $d=2$ proof is carried out in kinematic space, which we discussed in Section~\ref{sec:ksks}. We sketch it briefly below.

First, a comment on the difference between $d = 2$ and $d>2$. The obstacle to proving inequalities in time-dependent settings is that HRT surfaces $\Xi$ which compute entropies on the left hand side do not in general intersect. We discussed this fact in Section~\ref{sec:mme}. This means that the $\tilde{\Sigma}$s employed in (\ref{choosesigma}) are not submanifolds of a common slice of the bulk geometry. Consequently, their unions, intersections and complements are meaningless and the proof strategy laid out in (\ref{choosesigma}-\ref{monoproof}) is unavailable. 

\paragraph{Time dependence in $d=2$ boundary dimensions} 
Even though surfaces $\Xi$ do not intersect, in $d=2$ it is possible to define a topologically protected generalization of `intersection,' which applies to a pair of intervals. For simplicity, suppose the CFT lives on a line. We say that intervals $(L_1, R_1)$ and $(L_2, R_2)$ `intersect' if and only if their endpoints are ordered as $L_1 < L_2 < R_1 < R_2$ or $L_2 < L_1 < R_2 < R_1$ on a Cauchy slice of the CFT. It turns out that the proofs in \cite{Bao:2015bfa} can be rephrased using this topological notion of `intersection.' When we introduced kinematic space in Section~\ref{sec:ksks}, we observed below equation~(\ref{crofton}) that bulk lengths in AdS$_3$/CFT$_2$ can be understood in terms of counting intersecting geodesics. Ultimately, this is the reason why manipulating bulk lengths---as was done in \cite{Bao:2015bfa}---boils down to manipulating the topological `intersections.' Unfortunately, this technique is unlikely to generalize to $d>2$.

One further comment concerning the time-dependent $d=2$ setup is in place. When we inspect equations~(\ref{crofton}) and (\ref{condmutual}), it is evident that counting `intersections' in kinematic space can only be sensitive to one inequality, namely strong subadditivity:
\begin{equation}
S(AB) + S(BC) - S(B) - S(ABC) \geq 0
\label{ssa}
\end{equation}
If so, how can this technique offer a proof of other inequalities, including~(\ref{monogamy}) and (\ref{crazyineq})? The answer is that in $d=2$ (and only there) all known holographic inequalities are linear combinations of (\ref{ssa}) \emph{phase by phase}. We illustrate this point using the monogamy of mutual information for four intervals $A,B,C,D$, which partition a circle. 

Exploiting its invariance under permutations of constituent intervals, the same monogamy inequality can be written in the following two ways: 
\begin{align}
\big(S(AB) + S(BC) - S(B) - S(ABC)\big) - \big(S(A) + S(C) - S(AC)\big) & 
\geq 0 \\
\big(S(AB) + S(AD) - S(A) - S(ABD)\big) - \big(S(B) + S(D) - S(BD)\big) &
\geq 0
\end{align}
Suppose the intervals $A,B,C,D$ are cyclically ordered on the circle and consider $S(AC) = S(BD)$. As discussed in equation~(\ref{2intoptions}) and shown in the right panel of Figure~\ref{phase}, it equals either $S(A) + S(C)$ or $S(B) + S(D)$, whichever is smaller. This means that the second parenthesis must vanish in one of the two expressions, and monogamy boils down to:
\begin{align}
S(AB) + S(BC) - S(B) - S(ABC) \geq 0 & \qquad {\rm if}~~S(A) + S(C) \leq S(B) + S(D) 
\label{mono1st} \\
S(AB) + S(AD) - S(A) - S(ABD) \geq 0 & \qquad {\rm if}~~S(A) + S(C) \geq S(B) + S(D)
\label{mono2nd}
\end{align} 
We see that in $d=2$ monogamy is independent of strong subadditivity only in the sense that no rewriting like (\ref{mono1st}) or (\ref{mono2nd}) works independently of the phase of $S(AC)$. All inequalities proved in \cite{Bao:2015bfa} can be similarly reduced (phase by phase) to strong subadditivity. Indeed, applied in $d=2$, the proof of \cite{Bao:2015bfa} is basically a list of how the given inequality can be reduced to a linear combination of strong subadditivities, which covers a complete set of alternative phases of holographic entropies \cite{kscone}. It is not known if the extra power wielded by strong subadditivity in $d=2$ carries a deeper significance.

\subsection{Bit threads: Areas versus flows}
\label{sec:flows}
As anyone who tried to patch a leaking faucet knows, the minimal cross-sectional area sets the maximal rate of flow of a fluid. Formally, this is known as the max-flow/min-cut theorem. For our purposes, the theorem means that the Ryu-Takayanagi proposal for $S(A)$ can be reformulated as the maximal flux out of $A$ of a divergenceless flow $v^\mu$ with bounded norm:
\begin{tcolorbox}
{\bf Max-flow version of the RT proposal for time reversal-symmetric states:}
\begin{equation}
S(A) = \max \int_A v \equiv \max \int_A \sqrt{h}\, n_\mu v^\mu
\label{rtflow}
\end{equation}
where $v^\mu$ is a divergenceless vector field satisfying $|v| \leq 1/4G\hbar$.
\end{tcolorbox}
\noindent
The maximal norm is set to the universal coefficient of gravitational entropy formulas, so we do not have to write it explicitly in equations involving flows. Reformulation~(\ref{rtflow}) was proposed in \cite{Freedman:2016zud}, whose presentation we follow closely in the ensuing write-up. (A generalization of bit threads that incorporates quantum corrections of Section~\ref{sec:genentropy} is discussed in \cite{qbitthreads1, qbitthreads2}; see also \cite{lorentzianflows, highercurvflows} for other extensions of the concept.)

Rephrasing~(\ref{rtflow}) enjoys many conceptual advantages. It `explains' the homology condition in~(\ref{rtfull}) when we realize that the maximal flow is only constrained at the bottleneck, which is the minimal surface $\Xi_A$. Indeed, the max-flow/min-cut theorem only directly implies the first equality in:
\begin{equation}
S(A) = \max \int_{\Xi_A} v = \max \int_A v
\label{flowhomology}
\end{equation}
The second equality is a consequence of the homology condition $\Xi_A \cup A = \partial \tilde{\Sigma}$, which for any divergenceless flow (maximal or not) implies:
\begin{equation}
-\int_{\Xi_A} v + \int_A v = \int_{\tilde{\Sigma}} \sqrt{g}\, \nabla_\mu v^\mu = 0
\end{equation}
The minus sign appears because equation~(\ref{flowhomology}) assumes that $A$ and $\Xi_A$ have differently (inward versus outward) pointing normals. 

Secondly, when we cross a phase transition in holographic entanglement entropy (see text around Figure~\ref{phase}), the minimal surface changes abruptly but the maximal flow changes continuously. It is tempting to interpret the original Ryu-Takayanagi proposal~(\ref{rtfull}) as revealing the presence of a Hilbert space on a minimal surface, with degrees of freedom packed with density $(4G\hbar)^{-1}$ per unit area; indeed, the surface/state correspondence discussed in Section~\ref{sec:ssc} posits just that. But this picture is difficult to reconcile with phase transitions, for how can a Hilbert space jump discontinuously? The flow language, instead, emphasizes bipartite correlations between the two sides of the RT surface, and evades discussing discontinuities altogether.

The same advantage of flows comes handy when we consider information theoretic quantities, which are computed in holography by \emph{differences} of minimal surfaces. Conditional entropy, which we discussed around equation~(\ref{condent}), is a case in point. We will also consider the mutual information~(\ref{defmi}) and conditional mutual information~(\ref{condmutual}).

We first set up a modicum of notation. Let $v(A)$ be a flow that maximizes~(\ref{rtflow}). Note that $v(A)$ is highly ambiguous because a maximal flow can be arbitrarily deformed away from the bottleneck $\Xi_A$. This flexibility is in fact powerful enough to allow the same flow to be simultaneously maximizing for a collection of regions, so long as they contain one another in a sequence like Russian dolls. We denote this type of flow $v(A, B\ldots)$. We have
\begin{equation}
\int_A \! v(A,B\ldots) = \max\! \int_A \! v \qquad {\rm and} \qquad
\int_{AB}\! v(A,B\ldots) = \max\! \int_{AB}\! v \qquad {\rm and}\,\ldots,
\end{equation}
which makes sense because $A \subset AB \subset AB \cup \ldots$

\paragraph{Flows for entropies that involve differences of areas} For conditional entropy, we have directly from~(\ref{rtflow}):
\begin{equation}
S(B|A) = S(AB) - S(A) = \int_{AB} v(A, B) - \int_A v(A,B) = \int_B v(A,B) = \min_{v(AB)} \int_B v(AB)
\label{condentflow}
\end{equation}
This expression can be narrated in many ways. 
It is the flux out of $B$ of a flow, which is maximal on $A$ and $AB$, or the amount by which $B$ must supplant an $A$-maximal flow so it becomes $AB$-maximal. It is also the minimum flux through $B$ over all $AB$-maximizing flows: since a flux through $AB$ is the sum of fluxes through $A$ and $B$, achieving the maximum flux though $A$ is equivalent to achieving the minimum flux though $B$.  

That final rewriting offers an interpretation of mutual information and the subadditivity inequality~(\ref{subadditivity}):
\begin{equation}
I(A:B) = S(B) - S(B|A) = 
\max_{v(AB)} \int_B v(AB) - \min_{v(AB)} \int_B v(AB) \geq 0
\label{miflow}
\end{equation}
We can write $S(B) = \max_{v(AB)} \int_B v(AB)$ because $B \subset AB$, so the maximal flux through $B$ is achievable by an $AB$-maximizing flow. This proof of subadditivity is arguably simpler than~(\ref{subaddproof}). We can think of the difference between the maximum and minimum flow through $B$---the mutual information $I(A:B)$---as quantifying the flexibility in distributing flux lines through $\Xi_{AB}$ between $A$ and $B$. We will see shortly that a similar concept---a suitably quantified flexibility or flabbiness of a flow---also underlies the monogamy of mutual information~(\ref{monogamy}). 

For conditional mutual information~(\ref{condmutual}), in turn, we use the rewriting:
\begin{equation}
I(B:C|A) = S(B|A) - S(B|AC)
\label{cmihere}
\end{equation}
In (\ref{condentflow}), we may set $v(A, B)$ to $v(A,B,C)$ because $B \subset AB \subset ABC$ and an $A$- and $AB$-maximizing flow can also maximize flux through $ABC$. Similarly, the flow used in computing $S(B|AC)$ {\`a} la (\ref{condentflow}) can be $v(A, C,B)$. This is because a flow that is maximal on $AC$ and $ABC$ may as well be maximal on $C$ since $C \subset AC \subset ABC$.  This gives:
\begin{equation}
I(B:C|A) = \int_B v(A,B,C) - \int_B v(A, C, B) \geq 0
\label{cmiflux}
\end{equation}
The quantity is positive because the first term maximizes flux through $B$ sooner. Equation~(\ref{cmiflux}) proves the strong subadditivity inequality. 

\paragraph{Perfect tensor entanglement} 
The flow picture is very suggestive of bipartite correlations between degrees of freedom, which live at the endpoints of a flux line---in this context called a \emph{bit thread}. This in turn brings to mind EPR pairs. On the other hand, we cannot simply equate every {bit thread} with an EPR pair. This is because---as we observed around equation~(\ref{ptvector})---states with semiclassical bulk duals also contain qualitatively distinct types of entanglement, which may nevertheless be presented as maximal flows. In tackling this issue, Reference~\cite{holomonogamy} discovered an interpretation of the non-EPR constituent~(\ref{ptvector}) of holographic entropy vectors, which accounts for the non-saturation of the monogamy of mutual information~(\ref{monogamy}). 

We need one additional concept: a \emph{multiflow}. Let $A_i$ be $n$ non-overlapping regions in a CFT. A multiflow is a collection of flows $v_{ij} = -v_{ji}$, which have vanishing flux on all $A_k$ except on $A_i$ and $A_j$. All $v_{ij}$s satisfy the conditions of divergencelessness and norm bound spelled out in (\ref{rtflow}). Colloquially, a multiflow represents traffic circulating among $n$ sites, where $v_{ij}$ is specifically the traffic from $A_i$ to $A_j$. A theorem proved in \cite{holomonogamy} asserts that it is possible to find a multiflow such that:
\begin{equation}
\sum_{j=1}^n \int_{A_i} v_{ij} = S(A_i) \qquad \textrm{for all $i$}
\label{multiflow}
\end{equation} 
In other words, a collection of constituent flows $v_{ij}$ exists such that $\sum_j v_{ij}$ is an $A_i$-maximizing flow for each $i$. This traffic pattern uses up the throughput in and out of every site, and does so simultaneously for all sites. 

\begin{figure}
     \centering
     \begin{subfigure}[t]{0.3\textwidth}
         \centering
         \includegraphics[width=\textwidth]{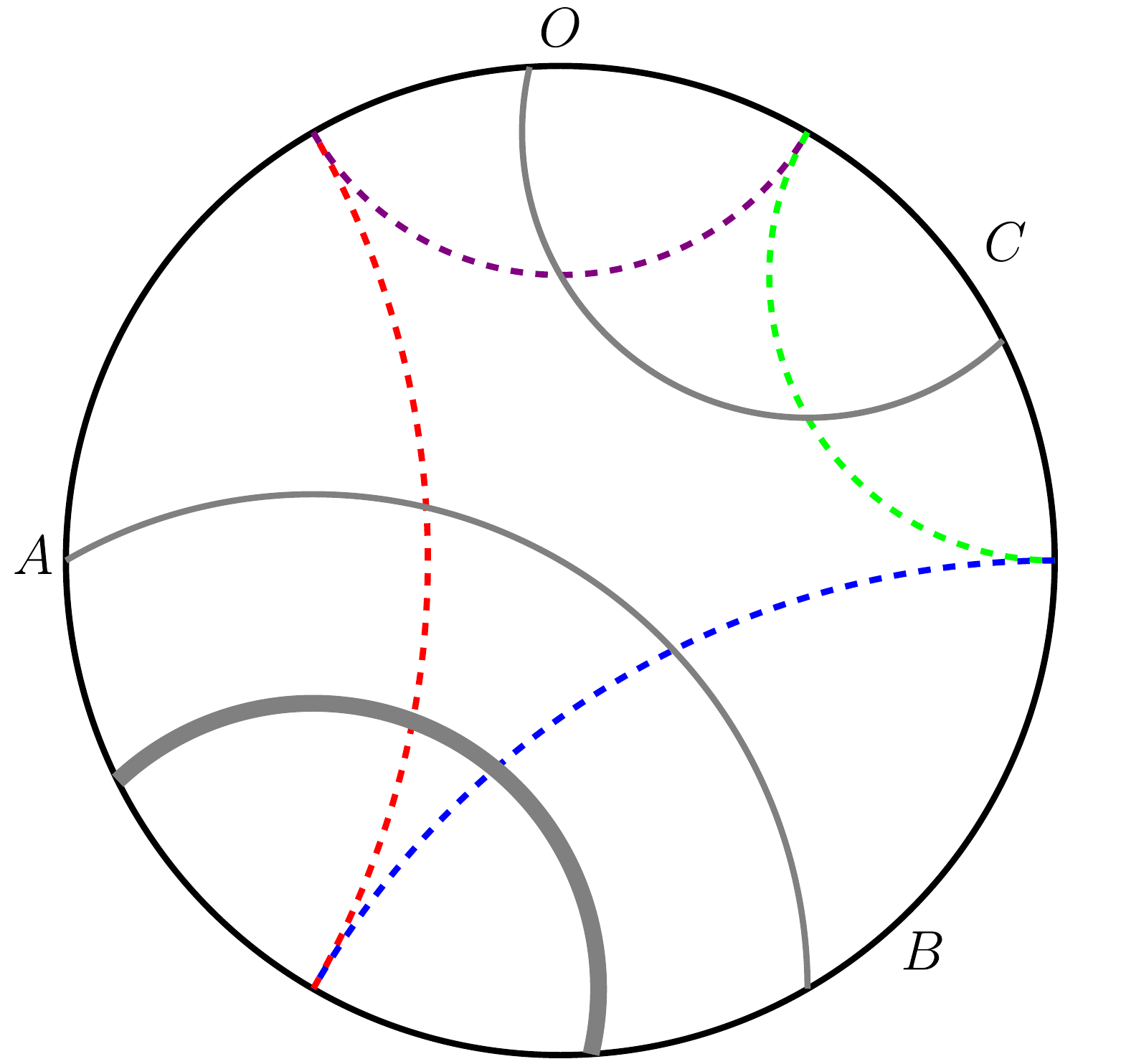}
         \label{flow1}
     \end{subfigure}
     \hfill
     \begin{subfigure}[t]{0.3\textwidth}
        \centering
         \includegraphics[width=\textwidth]{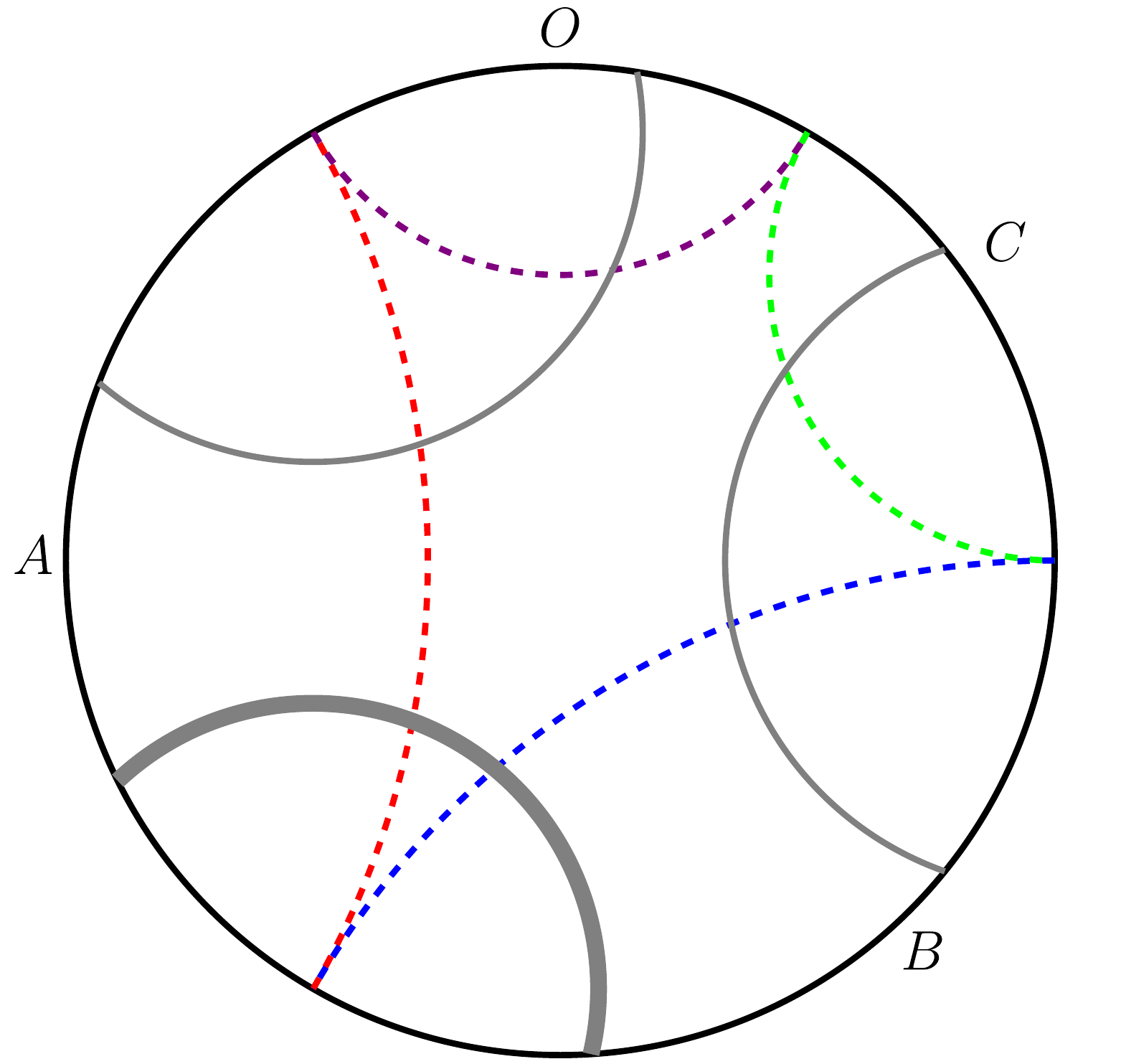}
         \label{flow2}
     \end{subfigure}
     \hfill
     \begin{subfigure}[t]{0.3\textwidth}
        \centering
         \includegraphics[width=\textwidth]{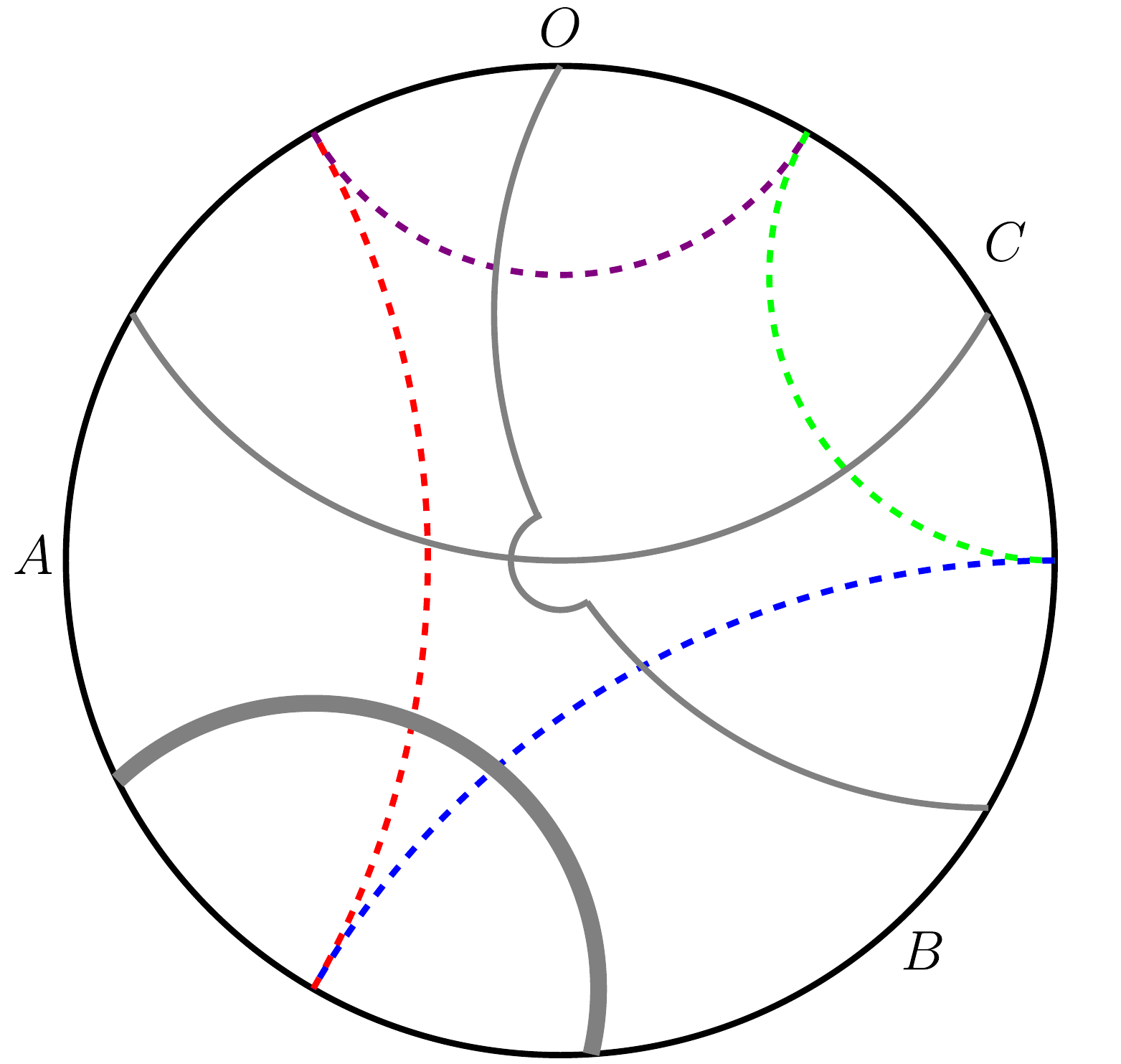}
         \label{flow3}
     \end{subfigure}
        \caption{Three multiflows for entropies~(\ref{flowcond}), displayed as gray lines. The thickened line from $A$ to $B$ is common to all of them; it is fixed by $I(A:B)$ in equation~(\ref{mifixes}). The other lines are flabby, and collectively represent perfect tensor entanglement. Their flabbiness, or the presence of perfect tensor entanglement, is detected by inequality~(\ref{monogamy}).}
        \label{fig:flow}
\end{figure}

Figure~\ref{fig:flow} shows three multiflows for the configuration
\begin{equation}
S(A) = S(B) = 2 \qquad {\rm and} \qquad S(C) = S(O) = 1,
\label{flowcond}
\end{equation}
which we considered in example~(\ref{exsvector}). We have ignored factors of $\log 2$, denoted $ABC \equiv O$, and neglected entropies of composite regions such as $AB$ because the definition of a multiflow ignores them. 

Figure~\ref{fig:flow} distinguishes two types of bit threads. The thickened line is common to all three multiflows, and is necessarily present in any other multiflow one might draw. To see this, note that $v_{AC} + v_{AO} + v_{BC} + v_{BO}$ also satisfies the conditions of (\ref{rtflow}), which implies:
\begin{equation}
S(AB) = \max \int_{AB} v \geq \int_A v_{AC} + \int_A v_{AO} + \int_B v_{BC} + \int_B v_{BO}
\end{equation}
From the definition of the multiflow, we have
\begin{equation}
S(A) = \int_A v_{AB} + \int_A v_{AC} + \int_A v_{AO} \qquad {\rm and} \qquad 
S(B) = \int_B v_{BA} + \int_B v_{BC} + \int_B v_{BO}
\end{equation}
and therefore:
\begin{equation}
I(A:B) = S(A) + S(B) - S(AB) \leq 2 \int_A v_{AB}
\label{mifixes}
\end{equation}
Consequently, there must be \emph{at least} $(1/2) I(A:B)$ bit threads connecting $A$ and $B$---that is, at least as many as the amount of EPR-like entanglement shown in equation~(\ref{eprhere}). In example~(\ref{exsvector}) we have one fixed bit thread between $A$ and $B$ because $I(A:B) = 2$ (ignoring factors of $\log 2$). 

What about the extra bit threads in a multiflow---the ones which are not fixed in place, and which do not account for EPR-like entanglement? They are flabby, subject to rearrangement---as the three multiflows in Figure~\ref{fig:flow} illustrate. Their flabbiness suggests that a multiflow artificially imposes a bipartite structure on an entanglement pattern, which is actually $S_4$-symmetric. Reference~\cite{holomonogamy} called this \emph{perfect tensor entanglement}. It is exactly the entanglement pattern in equation~(\ref{ptvector}), whose subscript $\cdot_{\rm PT}$ foreshadowed Perfect Tensor.

As we saw around equations~(\ref{monogamyhere}) and (\ref{ptvector}), the contribution of perfect tensor entanglement to an entropy vector is exactly quantified by the non-saturation of the monogamy inequality. Just like subadditivity asserts that a quantum state cannot contain a negative number of EPR pairs, monogamy~(\ref{monogamy}) asserts that states with semiclassical bulk duals cannot contain a negative amount of perfect tensor entanglement. And, just like in the case of EPR pairs~(\ref{miflow}), the contribution of perfect tensor entanglement is represented as a flexibility or ambiguity in a choice of flow. (Both inequalities say that a `negative flexibility' in choosing a flow is impossible.) We emphasize that general quantum states such as (\ref{ghz4}) can contain a negative multiple of perfect tensor entanglement; the positivity is only necessary for a semiclassical bulk dual obeying the HRT proposal. Thus far, other fundamental inequalities have not been similarly interpreted in terms of flows, although significant progress has been made \cite{cooperativeflows}.

\subsection{Entanglement shadows}
Even if a CFT state is dual to a semiclassical bulk geometry, we may ask how much of the geometry can be derived from the Ryu-Takayanagi formula. Imagine an oracle, which can tell us the entanglement entropy of any CFT subregion $A$ in some global state $\rho$. With that data, how much of the geometry dual to $\rho$ can we derive? Under certain special assumptions, it can be the whole entanglement wedge \cite{bilson1, bilson2, spillane, uhlman}. But in generality, a large region of the bulk spacetime is not penetrated by any Ryu-Takayanagi surfaces and remains invisible to entropic considerations; see e.g.~\cite{holeographyconstr}. Such a region is called an \emph{entanglement shadow} \cite{entwinement}. 

An easy example is occasioned by phase transitions in holographic entanglement entropies. In and around equation~(\ref{abcde}), we considered entanglement entropies of intervals in the state dual to a non-rotating BTZ black hole. We observed that the relevant RT surface jumps to a configuration that girdles the horizon when the interval exceeds a critical size $\alpha_{\rm critical}$. This means that no RT surface penetrates geometry~(\ref{staticbtzmetric}) deeper than:
\begin{equation}
\rho_{\rm minimal} = \cosh^{-1} \coth (\tilde{R}_0 \alpha_{\rm critical})
\end{equation}
This is the deepest point reached by geodesic~(\ref{btzgeodesic}) with $\alpha = \alpha_{\rm critical}$, shy of the horizon $\rho = 0$. It is easy to see that disconnected regions do not help; their RT surfaces probe geometry~(\ref{staticbtzmetric}) even less deeply. In fact, a region impenetrable by RT surfaces occurs generically whenever we have positive energy densities (or mass) in the bulk. 

In special setups it is possible to generalize the concept of entanglement entropy to so-called \emph{entwinement} \cite{entwinement}---an entanglement measure for more subtly defined CFT subsystems, not associated to localized subregions. When it is well-defined, entwinement does not undergo the pesky phase transitions and allows us to extract a larger portion of the bulk geometry. This work-around may shed light on the near-horizon region of AdS black holes, but it cannot explain what happens inside a black hole. The latter problem, which has motivated an entirely novel direction of research, is discussed in Section~\ref{sec:complexity}. For further work on entwinement, see \cite{entw1, entw2, entw3, entw4, entw5}. 

\section{Black holes}
\label{sec:blackholes}

This review started from the physics of black holes, and it is only fitting that it should end with them. 
We introduce firewalls as a novel perspective on the black hole information paradox, which was motivated by careful consideration of quantum entanglement. We also describe islands and replica wormholes---a new approach to resolving the paradox, which has recently dominated the holographic news cycle and whose starting point is a quantum-corrected version of the Ryu-Takayanagi proposal. For starters, however, we dispel any notion that quantum entanglement will solve all outstanding problems of gravity. Quantum information theory might do that, but its toolkit must be extended to include a further concept: that of computational complexity. 

\subsection{Holographic complexity}
\label{sec:complexity}

\paragraph{Linear growth of a wormhole}
A standard motivating example \cite{firstcomplexity} is a two-sided AdS black hole dual to the thermofield double state (\ref{tfdstate}) of two CFTs.  A key observation is that the wormhole connecting the two asymptotic regions grows linearly for an exponentially long period of time. 
No entropic quantity shows a similar growth. One might try $S\big(A_L(t) \cup A_R(t)\big)$ where $A_L(t) \subset {\rm CFT}_L$ and $A_R(t) \subset {\rm CFT}_R$ are subregions drawn from individual CFTs' slices at successive times $t$. But this could only work if the holographic entropy of $A_L(t) A_R(t)$ stays in the connected phase, and anyway this quantity is upper-bounded by
\begin{equation}
S\big(A_L(t) A_R(t)\big) \leq S\big(A_L(t) \big) + S\big(A_R(t)\big)
\end{equation}
because of subadditivity, so it cannot grow linearly for too long. We conclude after \cite{eenotenough} that the growth of the wormhole is a qualitatively different phenomenon, whose study requires new concepts other than quantum entanglement. 

\begin{figure}
     \centering
     \begin{subfigure}[t]{0.4\textwidth}
         \centering
         \includegraphics[width=\textwidth]{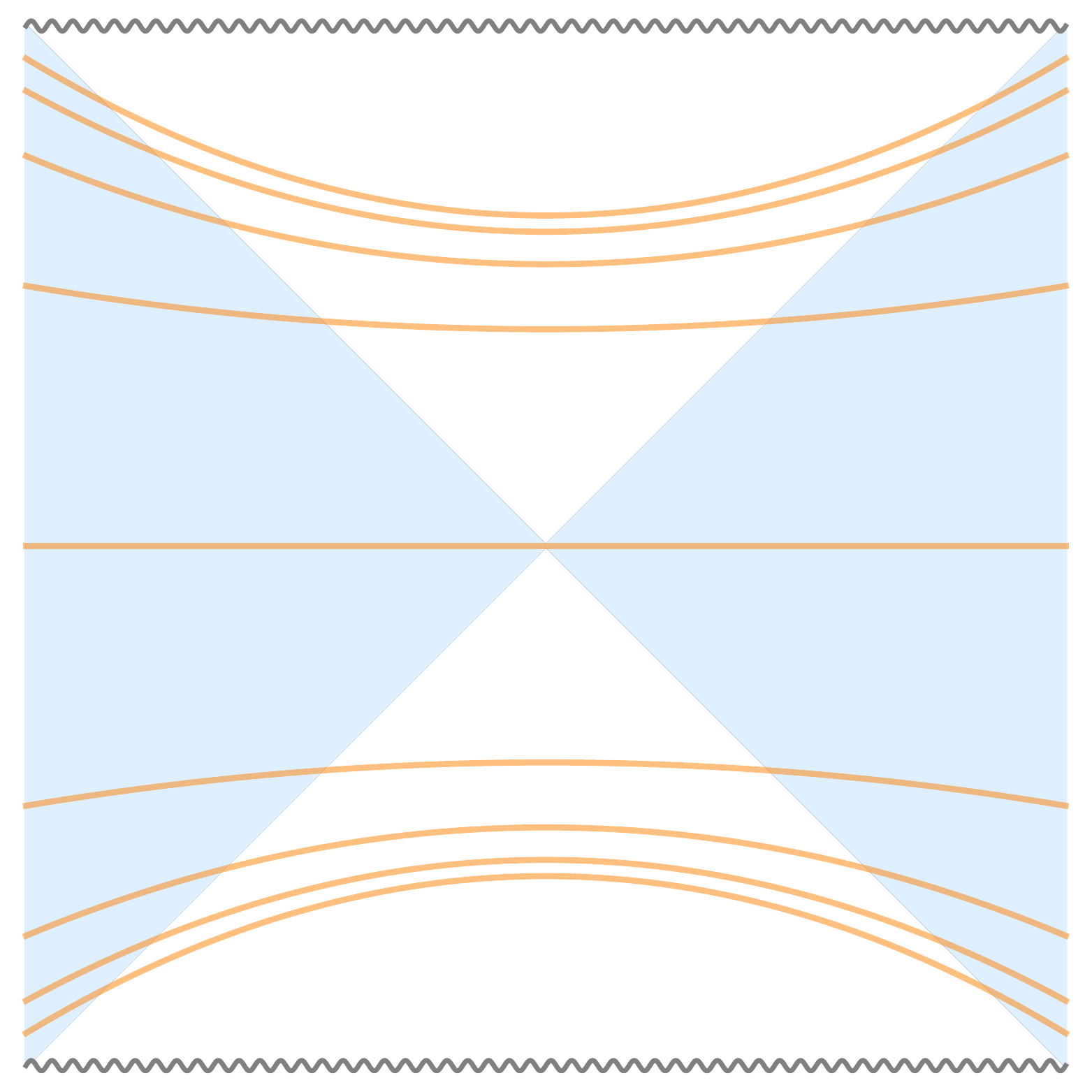}
         \label{maxslices}
     \end{subfigure}
     \hfill
     \begin{subfigure}[t]{0.4\textwidth}
        \centering
         \includegraphics[width=\textwidth]{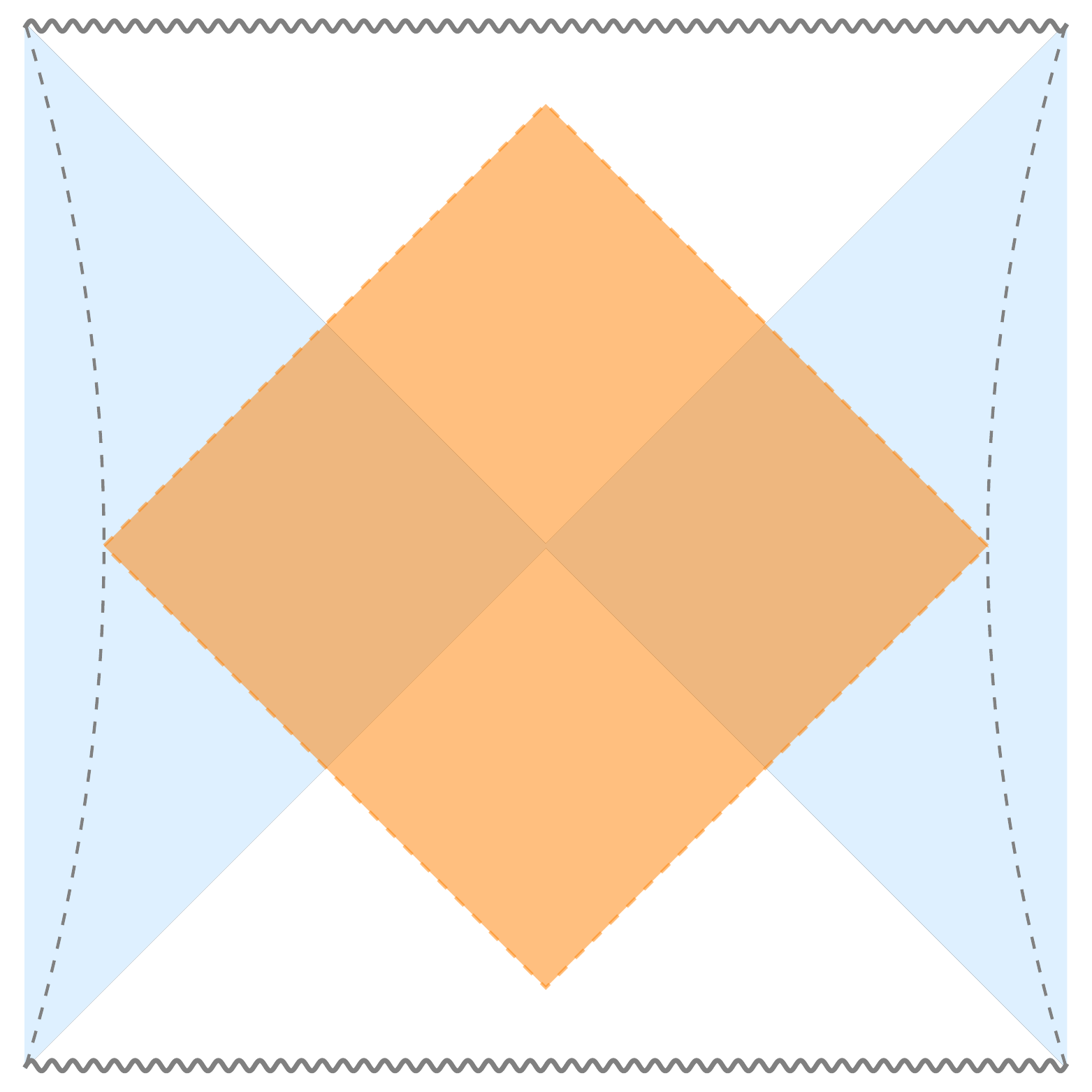}
         \label{WdW}
     \end{subfigure}
        \caption{Left: Maximal volume slices at different times. Right: A Wheeler-de Witt patch. It does not `hug' the boundary because its defining spatial slice is cut off away from the boundary by the dashed line.}
        \label{fig:penrose}
\end{figure}

We emphasize the linear growth of the wormhole, but have been vague about which exact bulk quantity captures the wormhole's size. Two methods that have gained most prominence in the literature go by the slogans `Complexity = Volume' and `Complexity = Action'; see Figure~\ref{fig:penrose}. The former uses the maximal volume of a spatial slice cutting through the spacetime \cite{complexityshocks}. For an example, take the AdS$_{d+1}$-Schwarzschild black hole
\begin{equation}
ds^2 = -f(r) dt^2 + f(r)^{-1} dr^2 + r^2 d\Omega_{d-1}^2 \qquad {\rm with}~~
f(r) = 1 + \frac{r^2}{L_{\rm AdS}^2} - \frac{\mu}{r^{d-2}},
\end{equation}
where $\mu$ is the mass of the black hole (up to a dimension-dependent factor). The volume of a bulk slice of the two-sided black hole, which does not break the $(d-1)$-dimensional rotational symmetry, is:
\begin{equation}
2\, {\rm vol}(S^{d-1})\times \int_{r_{\rm min}}^{r_{\rm max}} dr\, r^{d-1} \sqrt{\frac{1}{f(r)} - f(r) \left(\frac{dt}{dr}\right)^2}
\label{maxvol}
\end{equation}
The lower limit of integration $r_{\rm min}$ is where $dt/dr = \infty$ and the slice reaches closest to the bifurcation horizon; note that `closest' here means closest in time because $r$ is a time coordinate in the interior of the black hole. The upper limit of integration is large, though in practical applications it often suffices to set $r_{\rm max} = r_{\rm horizon}$ where $f(r_{\rm horizon}) = 0$. Looking for a maximal volume slice at time $t_c$ amounts to finding $t(r)$ that extremizes~(\ref{maxvol}) subject to the boundary condition $\lim_{r \to \infty} t(r) = t_c$. This quantity was studied in detail \cite{complexityshocks} and shown to exhibit a linear growth with $t_c$. In two dimensional models of gravity, one can even give a non-perturbative definition of `spatial volume' directly from the gravitational path integral \cite{gabormark}; it too grows linearly for an exponentially long time and then saturates. \textchg{There is also a bit thread formulation of the volume proposal \cite{volumethreads}, akin to equation~(\ref{rtflow}) for the Ryu-Takayanagi proposal, which is based on the concept of Lorentzian bit threads developed in \cite{lorentzianflows}.} 

The other common way to quantify the size of the wormhole uses so-called Wheeler-de Witt patches. A Wheeler-de Witt patch is the bulk domain of dependence (see Section~\ref{sec:ew}) of a spatial region $\mathcal{R}$. In the present context, $\mathcal{R}$ is a bulk spatial slice which asymptotes to $t=t_c$ on the asymptotic boundary, but cut off at some large radius. Removing the near-boundary region means that the Wheeler de Witt patch does not adhere to the boundary (like an entanglement wedge does) but only approaches it at $t=t_c$; see Figure~\ref{fig:penrose}.

To quantify the size of the wormhole we might take the $(d+1)$-dimensional volume of the Wheeler-de Witt patch \cite{volume20}. But a quantity more often discussed in the literature is the gravitational action contained in the patch \cite{Brown:2015bva,Brown:2015lvg}, hence `Complexity = Action.' The preference for the action is mostly because of units. Both the $d$-dimensional volume of a slice and the $(d+1)$-dimensional volume of a Wheeler-de Witt patch require extra dimensionful constants in addition to $[G_N] = {\rm (length)}^{d-1}$ to define a dimensionless quantity. On the other hand, the combination ${\rm (action)}/\hbar$ is an exponent in the path integral so it is hard-wired to be dimensionless. Nevertheless, defining the action in the Wheeler-de Witt patch is tricky because the patch can have facets and edges of various codimensions and generically reaches the black hole singularity. A detailed study of those issues is \cite{Lehner:2016vdi}. 

Per common usage, the quantities discussed above are now effectively \emph{defined} to be `holographic complexities.' All of them exhibit a linear growth, which lasts for an exponentially long time---whereafter the applicability of General Relativity is subject to subtleties. A reader unfamiliar with the subject might ask: why do we use the word `complexity' instead of `volume,' `action,' or `bigness'? To understand this, we first give one example of what complexity may mean in a physical but not necessarily holographic context.

\paragraph{A qubit model to illustrate computational complexity}
Consider a system of $S$ interacting qubits. We choose the letter $S$ for the number of qubits because their Hilbert space is $2^S$-dimensional, so generic mixed states will have entropies proportional to $S$. We let this system evolve in discrete time steps, in which arbitrary disjoint pairs of qubits are acted on by a fixed set of two-qubit gates. We will use this model below to illustrate how and why a suitably defined computational complexity also exhibits a linear growth. 

Two comments are in order. First, which gates generate the dynamics? The answer is motivated by quantum computing. A quantum computation on $S$ qubits is an $SU(2^S)$ transformation followed by a measurement. In practical settings, the $SU(2^S)$ unitary would be assembled approximately (but to arbitrary precision) from some basic building blocks, which are referred to as a universal set of gates. Many universal sets are known; one example comprises three distinguished single-qubit gates and one two-qubit gate called CNOT \cite{nielsenchuang}, which does or does not flip the second qubit depending on the state of the first qubit:
\begin{equation}
|00\rangle \to |00\rangle 
\qquad {\rm and} \qquad 
|01\rangle \to |01\rangle 
\qquad {\rm and} \qquad 
|10\rangle \leftrightarrow |11\rangle 
\end{equation}
But details of the gates are unimportant. The salient point is that we want the dynamics to be capable of generating an arbitrary transformation, so we must use a universal set of gates, which only comprises a few discrete options. 

Second, how and why is this model relevant for black holes? The motivation comes from the fast scrambling conjecture \cite{fastscramblers}, which we outline presently. It posits that black holes have the property that information dropped into one constituent of the black hole (whatever such constituents may be) will come to be shared by all of them after a \emph{scrambling time} $\log S$. This time is unusually fast, much shorter than scrambling times of more familiar systems, which are typically power-law in $S$.  The non-trivial part of the conjecture is that black holes manage to scramble information at this uncommonly fast rate. This likens the black hole to our qubit model because---as we explain shortly---it too is a fast scrambler.

Multiple reasons underlie the black hole fast scrambling conjecture. The subject could easily fill another review, so our presentation is cursory. One reason comes from studying the dissipation of perturbations in classical black holes, with the most famous example concerning the spread of electric charge on the black hole's surface. From the perspective of outside observers, the horizon of a black hole is a conductor with a characteristic surface resistivity, which in $d=3+1$ dimensions takes the surprisingly quotidian value $\rho_s = 377\Omega$ \cite{Damour:1978cg, membraneparadigm}. A localized charge dropped on the horizon spills out on its surface to cover an exponentially growing area until it covers the whole horizon, which happens at a time proportional to $\log A = \log S + {\rm const.}$ 

The second reason is supplied by a widely held resolution of the information paradox called complementarity \cite{complementarityref}, which proposes that information can simultaneously remain stored inside a black hole and get radiated away. Ordinarily, this would be a violation of the no-cloning theorem \cite{nocloning}, which states that an evolution
\begin{equation}
|\psi\rangle \to |\psi \rangle \otimes |\psi\rangle
\label{cloning}
\end{equation}
is impossible because it is non-linear. According to complementarity, however, there is no violation if the two copies of $|\psi\rangle$ cannot be brought together and compared because they are separated by a horizon. (We shall see a more modern version of complementarity in Section~\ref{sec:firewalls}.) 

A potential problem for complementarity was reported by Hayden and Preskill \cite{Hayden:2007cs}, who argued that black holes are actually information mirrors. A single qubit dropped into an old black hole (see Section~\ref{sec:page} for which black holes are old) can be retrieved outside as soon as a scrambling time $t_*$, albeit at an exceedingly high calculational expense \cite{Harlow:2013tf}. Let Alice, who carries one qubit of information, jump into a black hole and then leave a note for heroic physicists who may jump later. Such heroes of research would be motivated by a desire to witness a breakdown of quantum mechanics in the form of~(\ref{cloning}). If after time $t_*$ they can retrieve Alice's qubit outside the black hole using the Hayden-Preskill protocol and subsequently jump inside and intercept Alice's note, then they will have observed (\ref{cloning}). What saves complementarity from violating quantum mechanics is that the energy of the quanta carrying Alice's note is constrained by the mass of the black hole. By the uncertainty principle, this sets a minimal time $t_{\rm min}$ required for sending or receiving any message inside a black hole. The longer the heroic physicist remains outside the black hole to decode the first copy of Alice's qubit, the less time she has later to retrieve the second copy before hitting the singularity. Under the best circumstances, she must live for at least $t_{\rm min}$ inside the black hole, which sets an upper bound for $t_*$ if the no-cloning theorem is to be violated. Conversely, to avoid an observable violation of quantum mechanics---and so to save complementarity---we must impose a lower bound on $t_*$. A quantitative treatment of the problem reveals \cite{fastscramblers}:
\begin{equation}
t_* \geq \# \log S
\label{complconsist}
\end{equation}
This supports the fast scrambling conjecture under the assumption that principles of Nature are tightly constrained by consistency, i.e. a consistency inequality such as (\ref{complconsist}) should be saturated.

Further, independent evidence from string and M-theoretic models of black holes also supports the fast scrambling conjecture \cite{fastscramblers, Gur-Ari:2015rcq}.

\paragraph{Computational complexity of the qubit model} Our treatment follows \cite{Susskind:2018pmk} with some alterations. As a warm-up, let us confirm that the qubit model has $t_* \propto \log S$. Since time evolution is generated by discrete applications of two-qubit gates, the number of qubits which `know' about an initial single-qubit perturbation doubles with every step. Setting $2^{t_*} = S$ to spread the information among all the $S$ qubits gives the desired result. 

The evolution generated after $\log S$ steps is far from a generic $SU(2^S)$ transformation. After that, further applications of elementary gates continue to produce unseen-before $SU(2^S)$ transformations, until the time evolution visits within $\epsilon$ of every point on the $SU(2^S)$ manifold. To estimate when this happens, we first compute how many different unitaries can get generated in one step of the evolution. Ignoring single-qubit base changes, the evolution can involve
\begin{equation}
(S-1) \cdot (S-3) \cdot \ldots \cdot 3 \cdot 1
= \frac{S!}{2^{S/2}\, (S/2)!} 
\sim \frac{(S/e)^S}{2^{S/2} (S/2e)^{S/2}} = 
\left(\frac{S}{e}\right)^{S/2}
\label{baseexp}
\end{equation} 
different sets of symmetric two-qubit gates; the last expression uses Stirling's approximation. Because of the enormous volume of $SU(2^S)$, every subsequent step will increase the number of possibilities by another factor of~(\ref{baseexp}). This continues until the distinct possible evolutions compete with the number of $\epsilon$-balls in $SU(2^S)$, which is \cite{Susskind:2018pmk}:
\begin{equation}
{\rm vol}\big(SU(2^S)\big) \cdot
\frac{1}{\epsilon^{2^{2S}-1}\, {\rm vol}\big(B_{2^{2S}-1}\big)} 
= 2^{\frac{2^{S}+S-1}{2}} \frac{ \pi^{2^{S-1}}}{\sqrt{\pi}} \frac{\Gamma(2^{2S-1}+\frac{1}{2})}{1! \dots (2^{S}-1)!} \frac{1}{\epsilon^{2^{2S}-1}} \sim 
\left( \frac{\sqrt{e}}{\sqrt{2}\epsilon} \right)^{2^{2S}}
\end{equation}
The exponent of $\epsilon$ and the dimensionality of the ball in the denominator is the number of generators of $SU(2^S)$. The possible time evolutions in our $S$-qubit model cover all of $SU(2^S)$ to within $\epsilon$ error after a time $t_{\rm sat}$, which is roughly given by:
\begin{equation}
\left( \frac{S}{e} \right)^{t_{\rm sat} S/2} \sim  \left( \frac{\sqrt{e}}{\sqrt{2}\epsilon} \right)^{2^{2S}}
\qquad \Longrightarrow \qquad 
\mathcal{C}_{\rm sat} \equiv 
\frac{t_{\rm sat}S}{2} \sim 
\frac{4^S}{2} \frac{1 -  \log 2 - 2 \log \epsilon}{\log S - 1}
\label{cmax}
\end{equation}
The second equality defines the maximal \emph{computational complexity} $\mathcal{C}_{\rm sat}$ as the number of two-qubit gates, after which applying further gates becomes superfluous because all of $SU(2^S)$ has already been covered up to $\epsilon$-tolerance.

The time it takes to reach maximal complexity is exponential in $S$; here it is proportional to $4^S / \log S$. Up until that time, complexity---the number of two-qubit gates in the time-evolution operator---grows linearly with time with a slope approximately equal to $S$:
\begin{equation}
\frac{d\mathcal{C}}{dt} \propto S
\label{cgrowth}
\end{equation}

\paragraph{Growth of wormholes matches the growth of complexity in the toy model}
Equations~(\ref{cmax}-\ref{cgrowth}) mimic the growth of the wormhole in the two-sided black hole geometry. First, the linear growth proceeds for an exponentially long period of time. Second, the slope of the growth---for maximal volume slices, it is the cross-sectional area---is roughly proportional to the entropy (because $S$ is the minimal cross-sectional area by the Ryu-Takayanagi formula.) This matching motivates the proposal that the growth of wormholes reflects the growing complexity of quantum circuits, which generate time evolution in the dual conformal field theories \cite{firstcomplexity}.

Further evidence in favor of the proposals comes from shockwave geometries \cite{complexityshocks}. According to \cite{shockwaves}, they are dual to CFT states, which are related to the vacuum by injections of energy conjugated by forward and backward time evolution operators. In the bulk language, each time evolution operator extends the maximal volume slice by a certain length. On the boundary side, it adds to the circuit necessary to generate the CFT state from the vacuum, and so contributes linearly to its complexity. The agreement between the two is quantitative. 

Finally, equating spatial volumes with circuit complexities accords well with identifying bulk slices with tensor networks \cite{Miyaji:2015yva}. Just like time evolution in our qubit model, a tensor network is a sequence of gates (tensors), which assemble the state of interest. It is natural to quantify the computational complexity of a state with the count of gates (tensors) in its generating tensor network, perhaps with some weights to meaningfully combine contributions of distinct tensor types. To the extent that tensor networks model bulk spatial slices, counting tensors should be like measuring the size of a bulk slice.

\paragraph{Computational complexity in field theory} Thus far, we have only examined a toy model of computational complexity. How to define it for actual holographic conformal field theories? 

Many proposals have been put forward, but none of them bears a tractable relationship to bulk volumes or Wheeler-de Witt patches. Although reproducing volume or Wheeler-de Witt patch action has become somewhat of a Holy Grail in the holographic complexity literature, we argue that an emphasis on reproducing those specific quantities is misplaced. There is nothing sacrosanct about volume or action; they are merely the simplest bulk concepts, which quantify the size of a wormhole. A perspective often taken in the literature is that since the maximal slice volume and Wheeler-de Witt patch action are diffeomorphism-invariant concepts, there \emph{must} exist a CFT quantity dual to them; if so, we may as well \emph{define} these concepts as \emph{complexities}. But nothing guarantees that these two most often discussed bulk quantities should correspond to a \emph{simple} notion of computational complexity on the CFT side. Perhaps a different bulk measure of wormhole size is luckier in the sense of having a tractable, workable boundary dual.

We close with an incomplete list of proposals for computational complexity on the field theory side:
\begin{itemize}
\item Circuit complexity of an operator \cite{opsize1, opsize2, opsize4, opsizealex, opsizecesar, opsizejoan, johannavirasoro, opsizemichal, opsize3, michalea, opsizebrian, lamprosnew}, which builds up on work in \cite{qcgeometry1, qcgeometry2}. An example is the complexity of a generic $SU(2^S)$ transformation in (\ref{cmax}). To define the circuit complexity of an operator, such as the time-evolution operator, we must agree on a set of elementary gates and on the weights, with which to count the gates. An operator with a larger operator complexity would be interpreted holographically as probing deeper into the bulk geometry. The main difficulty with this concept are the myriad free parameters involved in choosing elementary gates and weights. 

\item Circuit complexity of a state \cite{Aaronson:2016vto, Jefferson:2017sdb, khancircuit, Chapman:2018hou, Brown:2019whu}. It is defined as the smallest complexity of an operator, which maps some reference state to the state in equation. Motivating this concept is the growth of a wormhole, which is a property of a geometry (a holographic state) and not of an operator. This concept is subject to the same arbitrariness as the circuit complexity of operators, plus the arbitrariness in choosing a reference state. The reference state is usually taken to be a CFT product state in the position basis. If geometry manifests entanglement then an unentangled state represents a disintegrated geometry with no volume, for which we can set $\mathcal{C}_{\rm initial} = 0$. 

Thus far, the circuit complexity of states has been the best studied candidate among all notions of field theory complexity. As mentioned above, counting tensors in a tensor network can be viewed as a special case. 
\item Metric distances. Multiple notions of distance between states, pure and mixed, have been defined in quantum information theory \cite{geombook}. The best known among them is the Fubini-Study metric on the Hilbert space, which is motivated by thinking about physical states as rays (defined up to normalization and overall phase) rather than points in the Hilbert space. The geodesic distance in the Fubini-Study metric between two states is
\begin{equation}
d_{\rm FS}\big(|\phi\rangle, |\psi\rangle \big) = \cos^{-1} \big| \langle \phi | \psi \rangle \big| \leq \pi/2
\end{equation}
so straight lines in the Fibini-Study metric do not tell us about holographic complexities. But distances along non-geodesic trajectories, possibly in modifications or generalizations of the Fubini-Study metric, might \cite{metriccompl, nytstory}. This approach was related to operator complexity in \cite{opsizemichal, michalea}. 
\item Path integral optimization \cite{Caputa:2017urj,Caputa:2017yrh, Takayanagi:2018pml}. We saw in equation~(\ref{rindlerdecomp}) that the vacuum state can be prepared by an appropriate path integral. There is a general argument that most, if not all states with semiclassical bulk duals can be similarly prepared, with sources added in the interior of the path integral \cite{holosource1, holosource2, holosource3}. In conformal field theory, changing the metric of the background on which the path integral is carried out only changes the normalization of the state \cite{liouvillenormchange}. If we associate a cost to performing a local piece of the path integral, every background metric gives a differently priced preparation of the same state. We may then define the complexity of the state as the cost of its cheapest preparation---one found after optimizing the path integral with respect to its underlying metric.

In special cases, optimizing the path integral can be interpreted in the language of tensor networks, with metric changes simulated by the Tensor Network Renormalization protocol \cite{tnrref}. 
For descendant states of the CFT$_2$ vacuum---which are dual to locally pure AdS$_3$ geometries \cite{banadosgeom}---optimizing the path integral can be interpreted as imposing Einstein's equations in the bulk \cite{Czech:2017ryf}. Recent work suggests that implementing path integral optimization ends up finding constant mean-curvature slices in the bulk geometry \cite{Boruch:2021hqs}. This claim, if correct, would equate the path integral-optimizing geometry with the maximal volume slice stipulated in \cite{complexityshocks}, effectively proving the Complexity = Volume proposal. Finally, Reference~\cite{Camargo:2019isp} has related path integral optimization to circuit complexity. 

\item Query complexity \cite{Chen:2020nlj}. Any quantum state is a map from operators to expectation values:
\begin{equation}
\rho\!: \mathcal{O} \to {\rm Tr}(\rho \mathcal{O})
\label{stateasmap}
\end{equation}
One may present~(\ref{stateasmap}) in the form of an algorithm, where $\mathcal{O}$ is an input and ${\rm Tr}(\rho \mathcal{O})$ the output. If so, we may ask about the query complexity of this algorithm, i.e. the number of times it calls on some subroutine, which is the bottleneck of the calculation. In the context of tensor networks the subroutine can be a tensor contraction, so the tensor network version of query complexity is just ordinary circuit complexity. But we may ponder more general subroutines, more directly related to bulk physics. We are currently developing such an approach, with basic subroutines related to the modular parallel transport discussed in Section~\ref{sec:berry}. Reference~\cite{Chen:2020nlj} contains an early example of this formalism, applied to CFT$_2$ vacuum descendants / locally pure AdS$_3$ geometries. 
\end{itemize}

Holographic complexity has grown to be an enormous subject, deserving of a separate review. Our write-up barely scratches the surface of this fast-developing, sprawling subject. In addition to searches for a convenient CFT definition of computational complexity, much research has been devoted to extending the concept to various contexts and conditions, including the complexity of mixed states \cite{yinguncomplexity, Caceres:2019pgf}, complexity of formation \cite{Chapman:2016hwi}, binding complexity \cite{bindingcompl}, complexity of purification \cite{Ghodrati:2019hnn,Camargo:2020yfv}, subregion complexity \cite{subregioncompl1, subregioncompl2} and others. 

\subsection{Firewalls}
\label{sec:firewalls}
A pure state of matter can collapse to form a black hole, which then radiates away in the form of thermal radiation. This apparent evolution from a pure to a mixed state is an ostensible violation of unitarity, and is known as the black hole information paradox \cite{hawkingparadox}. The literature on the subject is vast; some reviews include \cite{samirbhreview1, mybhreview, samirbhreview2, danielbhreview, joebhreview, suvratbhreview}. 

In 2012, a reformulation of the paradox centered on quantum entanglement took the arXiv by storm \cite{firewallsref}. It starts by considering an observer who dives into the black hole in free fall. By the equivalence principle, the diver should see a local vacuum around her, which in terms of modes inside and outside the horizon takes the entangled form~(\ref{vacuument}). For a single quantum of radiation $B$, this means there exists an interior mode $C$ such that $S(BC) = 0$, which gets excited when the black hole ejects quantum $B$. Demanding~(\ref{vacuument}) and $S(BC) = 0$, motivated by the equivalence principle, is known as the `no drama' assumption. Note that for any additional system $A$ we have $S(ABC) = S(A)$ because of subadditivity $S(A) + S(BC) \geq S(ABC)$ and the Araki-Lieb \cite{arakilieb} inequality $S(ABC) \geq |S(A) - S(BC)|$. 

The second assumption is unitarity of time evolution. For a black hole collapsed from a pure state of matter, we may divide the radiation into early radiation $R$ and late radiation $B_{\rm total}$ and posit that $S(R B_{\rm total}) = 0$ for unitarity. In terms of our single quantum $B$ of late radiation, we must have $S(R) \geq S(RB)$ because including $B$ brings the total radiation closer to being pure.

Putting together `no drama,' unitarity, and strong subadditivity gives
\begin{equation}
S(R) \geq S(RB) = S(RB) + S(BC) \geq S(B) + S(RBC) = S(B) + S(R)
\label{firewallreasoning}
\end{equation}
so $S(B) = 0$. Thus, unitarity prevents the radiation mode $B$ from being entangled with an interior partner. This is `drama'---a violation of the `no drama' assumption. The reasoning seems to imply that free fall into a black hole is impossible; instead, a diver jumping in will encounter some surprising physics, which departs from what the equivalence principle predicts. As an obstacle to free fall, this new physics is dubbed a \emph{firewall}. 

This form of the black hole information paradox was announced in \cite{firewallsref} and further elaborated in \cite{apologia}. Closely related reasoning appeared earlier in \cite{braunstein, samirsprefirewalls}.

\paragraph{Select resolutions of the firewall paradox:}
\begin{itemize}
\item $C \subset R$ (also known as `$A = R_B$' in another notation.) As the reasoning in (\ref{firewallreasoning}) presumes that $B, C, R$ are independent systems, the quickest way to circumvent it is to posit the opposite. The simplest viable option is to imagine that an interior mode is somehow a manifestation of (a subset of) the early radiation. This idea, raised previously by many authors \cite{boussonotenough, transferofe, prproposal, verlinderserrorcorr, erepr}, was critiqued in Section~2 of \cite{apologia}. We encourage the reader to compare this proposal to `islands,' which are discussed in the next subsection.

$C \subset R$ is a version of complementarity \cite{boussonotenough}. As such, it must be subject to consistency conditions like~(\ref{complconsist}), which preclude the cloning of information. In discussing complementarity in the lead-up to (\ref{complconsist}), we considered two observers who jumped into the black hole at different times. The question was whether a bit carried by the first observer could be intercepted by the second observer fast enough so they could compare notes and see cloning. But the firewall paradox imposes a more stringent demand on complementarity. We can take a single observer who collects the radiation $R$, isolates any one qubit from it, and then jumps into the black hole to find a copy of her qubit in $C$. A time scale $t_*$, which protects the recovery of one particular bit of information, does not protect operators in $C$ and $R$ from being non-commuting, which is `drama.' 

\item State dependence, or the Papadodimas-Raju proposal \cite{prproposal, prproposal2}. It states that an observer falling into a black hole does not test the `no drama' assumption at arbitrary scales, but only uses certain low energy observables of quantum field theory. Tested using this smaller subset of observables, any mode in $B$ always has a partner (often called `mirror'), with reference to which the local state appears to be~(\ref{vacuument}). The mirror must be part of the early radiation (because total radiation is getting purer, $S(RB) \leq S(R)$), so the Papadodimas-Raju proposal is a refinement of $C \subset R$. The mirror is also state-dependent, in that different black hole microstates will have differently constructed mirrors. But all of them give rise to physics, which looks like the `drama-free' state~(\ref{vacuument}).  

In what is known as the `frozen vacuum problem,' the original proposal \cite{prproposal} appears too good to be true \cite{frozenvac}; see \cite{prproposal2} for a modification response that addresses this issue. We can construct mirrors even in setups, where the local state is not the vacuum---for example, if an observer is falling alongside a suitcase---but the bulk description purported by the proposal is still~(\ref{vacuument}). Many authors (see e.g. \cite{bornrule}) have also raised conceptual objections to the state-dependent way in which bulk physics emerges from the boundary description. The same boundary operator can mean different things in bulk language because it can be a mirror of different $B$-modes; see \cite{prproposal3} for ways to resolve this. Other subtleties with the proposal were discussed in \cite{aspectsofpr}. The Papadodimas-Raju proposal can be viewed as an early harbinger of the error-correcting property of holography; see Section~\ref{sec:error}. Both emphasize that `drama-free' low energy physics in the bulk leaves a lot of room for non-local effects on the fundamental level. 

\item Nonviolent nonlocality in gravity has been advocated before firewalls \cite{nvnl1, nvnl2}. Both $C \subset R$ and state-dependence are special types of nonviolent nonlocality, but there are others. We refer the reader to original references, and to Sections~3 and 4.4 of \cite{apologia} for a critique. 
 
\item Maybe there really are firewalls and black holes break the equivalence principle. From the point of view of General Relativity, what might be the mechanism of such a breakdown? One was proposed by Mathur \cite{saddlebreak1, saddlebreak2}, who pointed out that Einstein's equations---like all classical equations of motion---reflect the saddle-point approximation of a quantum gravity path integral. But the $e^{A/4G\hbar}$ microstates of a black hole form a vast, nearly flat region near the classical saddle, which may invalidate the saddle point approximation
\begin{equation}
\int \mathcal{D}g\, e^{-S_{EH}/\hbar} \sim e^{A/4G\hbar} \cdot e^{-S_{\rm classical}/\hbar}
\end{equation}    
if $A/4G$ competes with $S_{\rm classical}$. This possibility has been supported by calculations \cite{saddlebreakcomp} in settings, where explicit microstate geometries or fuzzballs \cite{fuzzballs} can be written down. 
\end{itemize}

\subsection{Islands and replica wormholes}
\label{sec:islands}

\subsubsection{Page curve}
\label{sec:page}
\begin{figure}[tbp]
\centering 
\includegraphics[width=.47\textwidth]{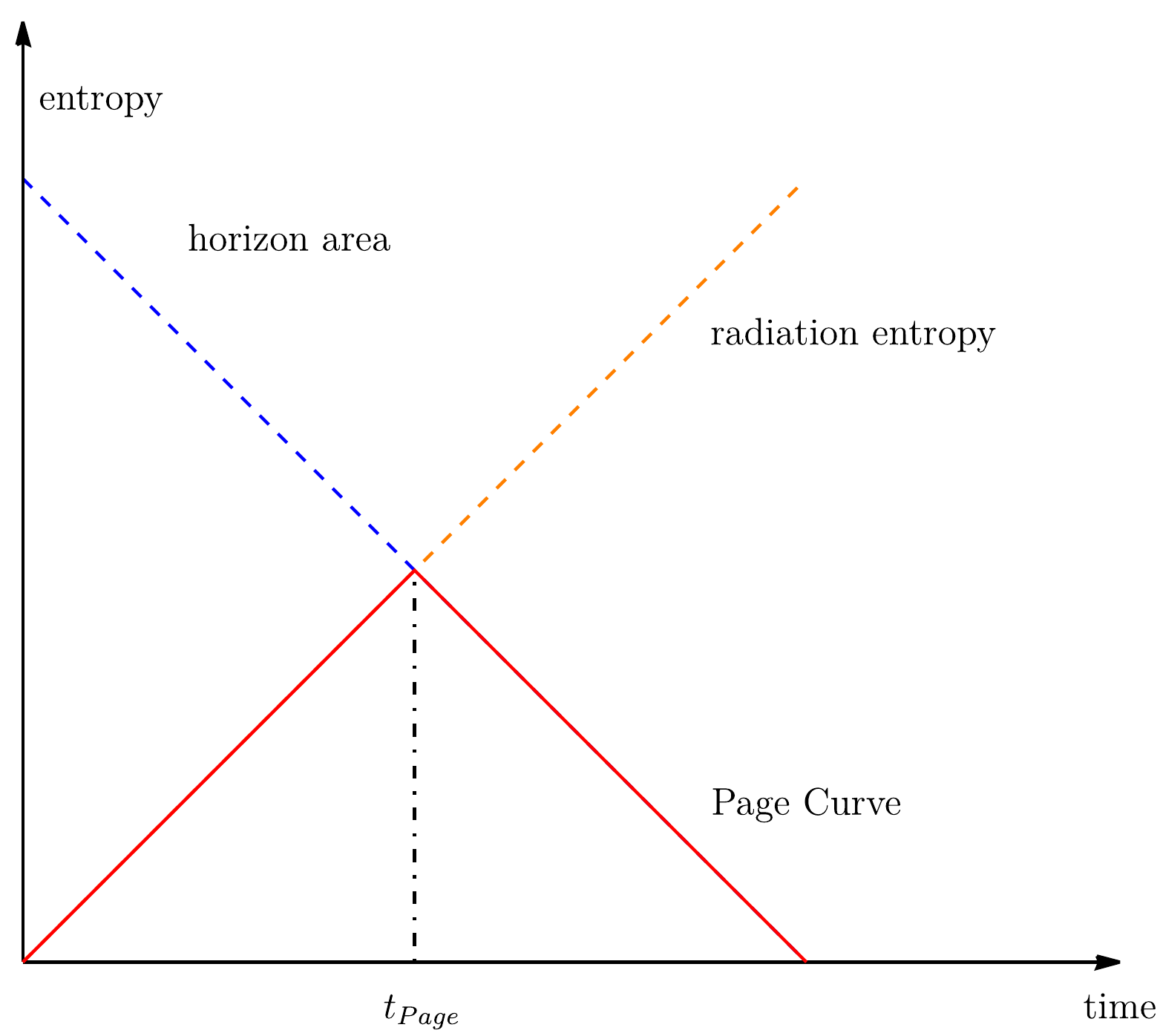}
\hfill
\caption{\label{pagecurve} Page curve for a black hole formed in an initial pure state.}
\end{figure}

A useful way to express the black hole information paradox---one which was instrumental \cite{samirbhreview1, samirsprefirewalls} in formulating its firewalls version---is in terms of the so-called \emph{Page curve} \cite{page1, page2}. Consider matter in a pure state, which collapses to form a black hole of some initial horizon area $A(t=0)$. We now plot two different entropies as a function of time. One is the coarse-grained (geometric) entropy of the black hole, which is $A(t)/4G\hbar$ and which decreases as the black hole evaporates. The other plot shows the fine-grained entropy, i.e. the entanglement entropy between the black hole and its previously emitted radiation. This quantity is initially zero because the black hole forms from a pure state of matter, but it rises as a result of Hawking emissions because every expelled radiation quantum has an entangled interior partner~(\ref{vacuument}). The plots are shown in Figure~\ref{pagecurve}.

The fact that the two plots intersect \emph{is} the black hole information paradox. In a unitary theory the fine-grained entropy must return to zero when evaporation ends, but applying Hawking's description of evaporation leads to an ever-increasing fine-grained entropy. Even before evaporation is over, how can the fine-grained entropy exceed the coarse-grained $A(t)/4G\hbar$? Evidently, something other than the $e^{A(t)/4G\hbar}$ microstates of the black hole must be entangled with the previously emitted radiation, but what can it be?

This reasoning highlights the importance of the point where the two plots in Figure~\ref{pagecurve} cross. It is here that a conflict first arises between two tenets of our understanding of black holes: the microscopic picture of Hawking radiation in~(\ref{vacuument}) and the macroscopic fact that black holes' areas shrink during evaporation. Past that point, if we insist on unitarity, some new physics must be activated. The most reasonable expectation is that, after the new physics is incorporated, the full plot of the fine-grained entropy bends and follows the coarse-grained entropy down to zero. This hypothetical time dependence of fine-grained entropy in a \emph{unitary} process of black hole evaporation is called the \emph{Page curve} after \cite{page1, page2}. 

The time of crossing in Figure~\ref{pagecurve}, often called \emph{Page time}, is special for one more reason: it is when the black hole has radiated away exactly half of its initial area: $A(t_{\rm Page}) = A(0)/2$. In terminology we already used in Section~\ref{sec:complexity}, a black hole past that point in its evaporation is called \emph{old}. For an old black hole, entanglement with previously emitted radiation fully accounts for its horizon area; the black hole information problem concerns surplus entanglement beyond $A(t)/4G\hbar$. Setting the maximal entanglement entropy to half the initial coarse-grained entropy suggests a simple idealization of unitary black hole evaporation \cite{page1}: an $(e^{A(0)/4G\hbar})$-dimensional Hilbert space $\mathcal{H}_{\rm total}$, which is divided into black hole and radiation components with a sliding division:
\begin{equation}
\mathcal{H}_{\rm total} = \mathcal{H}_{\rm BH}(t) \otimes \mathcal{H}_{\rm radiation}(t)
\qquad \textrm{such that}~~~
\mathcal{H}_{\rm BH}(0) = \mathcal{H}_{\rm total} = \mathcal{H}_{\rm radiation}(t_{\rm final})
~~~~~
\label{idealization}
\end{equation}
It is well known that such sliding divisions produce entanglement entropies, which rise and fall as dictated by the Page curve, peaking almost exactly \cite{pagethm} at $S_{\rm max} = (1/2) \log \dim \mathcal{H}_{\rm total}$. But how to realize the idealization~(\ref{idealization}) with an actual gravity calculation?

This latter question has been the Holy Grail of research in the black hole information paradox. And in 2019, the Holy Grail was found! What is extra exciting and perplexing is that the Page curve has been realized without evoking exotic ingredients, entirely within the realm of semi-classical gravity.

\subsubsection{Generalized entropy and islands}
\label{sec:genentropy}

A prerequisite for this discussion is a quantum-corrected version of the Ryu-Takayanagi proposal \cite{qes, qes2}:
\begin{tcolorbox}
{\bf Quantum-corrected Ryu-Takayanagi proposal:}
\begin{equation}
S(A) = \min_{\Xi \cup A = \partial \tilde{\Sigma}} 
\left(\frac{{\rm Area}(\Xi)}{4 G \hbar} + S_{\rm bulk}(\tilde{\Sigma})\right)
\label{qrt}
\end{equation}
We again minimize over bulk surfaces $\Xi$ homologous to $A$, which guarantees the existence of a bulk region $\tilde{\Sigma}$ between $A \subset {\rm CFT}$ and $\Xi$. The quantity being minimized, which is the sum of a Ryu-Takayanagi-like area term and the bulk entropy of $\tilde{\Sigma}$, is called \emph{generalized entropy}. The surface that realizes the minimum~(\ref{qrt}) is called a \emph{quantum extremal surface}.
\end{tcolorbox}
Proposal~(\ref{qrt}) was motivated in part by considering how quantum corrections affect the second law of thermodynamics (\ref{2ndlaw}). Another motivation comes from equations~(\ref{boostgen}) and (\ref{jlms}). On a fixed geometric background and near the RT-surface, the generator of orthogonal boosts plays the role of both the bulk and the boundary modular Hamiltonian, which highlights the role of the quantum entanglement of bulk fields $S_{\rm bulk}(\tilde{\Sigma})$.

\paragraph{Islands reproduce the Page curve}
We again consider a black hole formed from collapsing matter in a pure state. There is no second asymptotic region, so our black hole is one-sided and its interior geometry necessarily caps off. Working in an AdS/CFT setup, we can articulate this expectation by taking region $B$ to be the entire CFT and concluding by the usual Ryu-Takayanagi proposal~(\ref{rtfull}) or (\ref{hrt}) that: 
\begin{equation}
S(B) = S({\rm CFT}) = 0.
\label{naivert}
\end{equation}
This equation says that $\Xi = \emptyset$ is homologous to $B = {\rm CFT}$, i.e.~${\rm CFT} = \partial \tilde{\Sigma}$. Here $\tilde{\Sigma}$ is the entire bulk spatial slice, within and outside the black hole horizon. Writing~(\ref{naivert}) is equivalent to positing that the bulk spatial slice has no boundaries other than the asymptotic boundary where the CFT lives, so the interior geometry must cap off smoothly.

\begin{figure}
        \centering
        \includegraphics[width=0.26\textwidth]{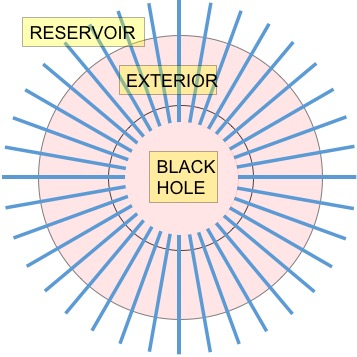}
        \hspace*{0.07\textwidth}
        \includegraphics[width=0.26\textwidth]{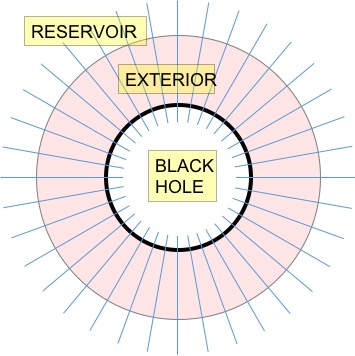}
        \hspace*{0.12\textwidth}
        \includegraphics[width=0.26\textwidth]{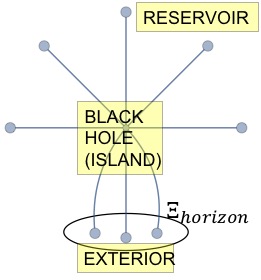}
        \caption{\textchg{A black hole whose radiation has been collected in a reservoir outside the spacetime. Blue line segments (roughly, EPR pairs) represent the entanglement between the radiation $R$ and the interior. There are two candidate quantum extremal surfaces; we highlight their corresponding bulk regions $\tilde{\Sigma}$ in pink. When $\Xi = \emptyset$, equation~(\ref{qrt}) picks up only the radiation entanglement (left panel). When $\Xi = \Xi_{\rm horizon}$ (highlighted black), (\ref{qrt}) picks up the area term but incurs no $S_{\rm bulk}$ contribution because the blue line segments do not originate in the pink $\tilde{\Sigma}$ region (middle panel). The right panel is a more schematic rendition of the setup, here showing a `black hole' with $A_{\rm horizon}=3$ and $S_{\rm bulk}({\rm radiation})=5$. It is a perfect tensor on $[A_{\rm horizon} + S_{\rm bulk}({\rm radiation})]$ legs, with $A_{\rm horizon}$ of them grouped into one region. This entanglement structure, with the ratio $A_{\rm horizon}/S_{\rm bulk}({\rm radiation})$ treated as a parameter, realizes the extremal vectors (`atomic ingredients') of the holographic cone of average entropies discussed in Section~\ref{sec:hecdetail}.}}
        \label{fig:islands}
\end{figure}

We now let the black hole radiate and observe it. Here, References~\cite{penington, princetonucsb, princeton} use one further conceptual novelty: they isolate the escaping radiation in a system separate from the CFT, which we call $R$. On a technical level, this is effected by imposing absorbing boundary conditions in AdS \cite{penington}, by coupling the CFT to an external `bath' that absorbs all outgoing radiation \cite{princetonucsb}, or by considering $(d+1)$-dimensional gravity coupled to a matter quantum field theory, which is itself holographic and has a $(d+2)$-dimensional bulk dual \cite{princeton}. The technical devices are involved but inessential; the key tactic is to treat the radiation $R$ as separate from the holographic CFT. 

After $R$ has been isolated from the CFT, equation~(\ref{naivert}) no longer holds. Indeed, $R \cup {\rm CFT}$ is pure, so $S(R) = S({\rm CFT})$, but how to compute it? We employ the quantum-corrected proposal~(\ref{qrt}) and find that there are two candidate extremal surfaces $\Xi$. \textchg{We show them in Figure~\ref{fig:islands}.}
\begin{enumerate}
\item $\Xi_0 = \emptyset$. It is still true that $\emptyset \sim {\rm CFT}$, and $\emptyset$ is certainly an extremal surface. But formula~(\ref{qrt}), after the radiation has been isolated from the CFT, gives: 
\begin{equation}
S^?_0({\rm CFT}) = 0 + S_{\rm bulk}(\tilde{\Sigma}) = S_{\rm bulk}({\rm radiation})
\label{s0option}
\end{equation}
This is the bulk entropy of the radiation, which we collected outside the CFT.
\item $\Xi_{\rm horizon}$. \textchg{There exists a quantum extremal surface, which---although it does not live exactly on the horizon of the black hole---is only slightly shifted away from it. Indeed, for any surface $\Xi$ near the horizon, $S_{\rm bulk}({\rm radiation})$ makes only a small contribution to (\ref{qrt}) because the entangled partners of the radiation quanta in $R$ sit behind the horizon, and so outside the manifold $\tilde{\Sigma}$ defined by $\partial \tilde{\Sigma} = {\rm CFT} \cup \Xi$. For such surfaces, we therefore have:
\begin{equation}
S^?_{\rm horizon}({\rm CFT}) \approx \frac{A_{\rm horizon}}{4G\hbar} + 0
\label{shorizonoption}
\end{equation}
When $A_{\rm horizon} < S_{\rm {bulk}}({\rm radiation})$, we know that at least one such surface is extremal by continuity. If we shrink $\Xi_{\rm horizon} \to \emptyset$, we get back to (\ref{s0option}), but if we expand $\Xi_{\rm horizon}$ toward the asymptotic boundary, (\ref{qrt}) also grows. This establishes that some quantum \emph{extremal} surface must approximately achieve (\ref{shorizonoption}).}
\end{enumerate}

Of course, proposal~(\ref{qrt}) takes the minimum of options~(\ref{s0option}) and (\ref{shorizonoption}): 
\begin{equation}
S(R) = S({\rm CFT}) \approx \min\{ A_{\rm horizon} / 4G\hbar, \, S_{\rm bulk}({\rm radiation}) \}
\label{scft}
\end{equation}
This is exactly the Page curve!

When $\Xi_{\rm horizon}$ has the smaller area---that is, for an old black hole---the division of the bulk spacetime into entanglement wedges is particularly interesting. The spatial manifold $\tilde{\Sigma}$ whose boundary is ${\rm CFT} \cup \Xi_{\rm horizon}$ is the exterior of the black hole---slightly shifted from its classical location by the $S_{\rm bulk}(\tilde{\Sigma})$ term in (\ref{qrt}).  The shift pushes the extremal surface `into' the black hole \cite{penington}, though Reference~\cite{islandsoutside} discussed potentially pathological cases where the shift might push the surface outside the classical horizon. 

Up until Page time, the entanglement wedge of the CFT is the domain of dependence of the full spatial slice, in and outside the horizon. But after Page time, when dominance shifts from $\Xi_0$ to $\Xi_{\rm horizon}$, the entanglement wedge of the full CFT only covers the black hole exterior. In that regime, the interior of the black hole forms the entanglement wedge of the CFT's purifying system, that is---of the radiation collected outside. This entanglement wedge looks like an \emph{island} in the middle of the spatial slice---and that is its established moniker. Combining the notations of Sections~\ref{sec:ew} and \ref{sec:firewalls}, we write:
\begin{equation}
C = \mathcal{W}(R)
\label{islandreconstr}
\end{equation}
Equation~(\ref{islandreconstr}) says that $C$ may be reconstructed from $R$ in the sense of subregion duality. Is there an ontological difference between this and $C \subset R$? 

\subsubsection{Replica wormholes}
\label{sec:replicas}
For a plurality of the community, the black hole information paradox has long been not about \emph{whether} information gets out of a black hole, but \emph{how} it does so without violating sacred tenets of gravity. The equivalence principle demands that quantum fields near the horizon take the form (\ref{vacuument}), so each emitted Hawking particle in state $|\tilde{\phi}_R\rangle$ has an entangled partner in state $|\tilde{\phi}_L\rangle$ in the interior. As the black hole radiates, the entropy of the radiation seems to grow without bound, and yet it must somehow return to 0. By what mechanism?

Equations~(\ref{qrt}) and (\ref{islandreconstr}) provide \emph{an} answer. To fully flesh out its significance, however, we need a gravitational calculation of~(\ref{scft}). For the leading order Ryu-Takayanagi formula~(\ref{rtfull}), such a calculation was done in Section~\ref{sec:renyi} using the replica trick. We presently sketch an analogous calculation of (\ref{scft}), done in the language of the gravitational path integral using the replica trick. As expected, this is a revelatory exercise, with a lesson that some readers will find unsettling. The calculation was done independently in \cite{westcoast, eastcoast} in the Jackiw-Teitelboim theory of two-dimensional gravity \cite{jt1, jt2, jt3}; see also \cite{rwreview} for a review.

We mostly follow the presentation in \cite{westcoast}. After $k$ quanta of radiation have left the black hole, we write the global state of black hole and radiation as:
\begin{equation}
|\Psi\rangle = \frac{1}{\sqrt{k}} \sum_{i=1}^k |\psi_i \rangle_C \otimes |i\rangle_R
\label{horstate}
\end{equation}
This is state~(\ref{vacuument}) which the equivalence principle demands, written in a basis that diagonalizes the matrix of component amplitudes $w_{ij} = \langle \tilde\phi_{Li} \tilde\phi_{Rj} | \Psi \rangle$. Bases that diagonalize the matrix of component amplitudes are called \emph{Schmidt bases}. We will see momentarily that finding a Schmidt basis for the near-horizon state of an old black hole is subject to a crucial subtlety, but for now we proceed. In equation~(\ref{horstate}), we label the Schmidt basis of the radiation system $|i\rangle_R$. The Schmidt basis of the interior partner system is labeled $|\psi_i\rangle_C$ for consistency with the notation of Section~\ref{sec:firewalls}. We may think of $|\psi_i\rangle_C$ as microstates of the black hole or equivalently---as done in \cite{westcoast}---as internal states of an end-of-the-world brane living inside the black hole. One final remark about (\ref{horstate}) is that it seemingly presumes the entanglement spectrum of $|\Psi\rangle$ to be flat, but the interesting part of the analysis is independent of this assumption.

In the replica trick, we compute quantities ${\rm Tr}(\rho_R^n)$ and substitute in equation~(\ref{defrenyi}). Concentrating on the $n = 1$ and $n=2$ cases, we have from (\ref{horstate}):
\begin{align}
\rho_R & = 
\frac{1}{k} \sum_{ij} \Big({}_C\langle \psi_i | \psi_j \rangle_C \Big)  |j\rangle_R {}_R\langle i|
\stackrel{?}{=} \frac{1}{k} \sum_{i=1}^k |i\rangle_R {}_R\langle i|
\label{rwrho} \\
\rho_R^2 & = \frac{1}{k^2} \sum_{ijm} 
\Big({}_C\langle \psi_i | \psi_j \rangle_C \Big)\Big({}_C\langle \psi_i | \psi_m \rangle_C \Big)^* \, 
|j\rangle_R  {}_R\langle m|
\label{rwrho2}
\end{align}
If the black hole microstates are truly orthonormal, the traces of (\ref{rwrho}) and (\ref{rwrho2}) will follow formula ${\rm Tr}(\rho_R^n) = k^{1-n}$ and $S(R) = \log k$, which is $S_{\rm bulk}({\rm radiation})$ in equation~(\ref{scft}). To have a shot at reproducing~(\ref{scft}) and the Page curve, we must test the orthogonality assumption. 

\begin{figure}[tbp]
         \raisebox{0.05\textwidth}{\includegraphics[width=0.4\textwidth]{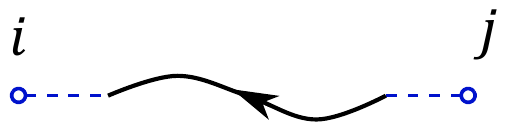}}
\hfill
         \raisebox{0cm}{\includegraphics[width=0.4\textwidth]{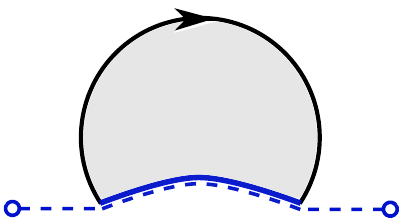}}
\caption{The boundary condition for the gravitational path integral and the saddle configuration, which compute ${}_C\langle \psi_i | \psi_j \rangle_C$. The figures are reproduced with permission from \cite{westcoast}.}
\label{fig:innprod}
\end{figure}

The traces of (\ref{rwrho}), (\ref{rwrho2}), and of higher powers of $\rho_R$, are sums of quantities such as
\begin{align}
& \phantom{a}{}_C\langle \psi_i | \psi_j \rangle_C \\
{\rm and}~~ & 
\big({}_C\langle \psi_i | \psi_j \rangle_C\big)\big({}_C\langle \psi_i | \psi_j \rangle_C\big)^*
\label{squareoverlap}
\end{align}
Each of these quantities is a gravitational amplitude, which can be computed by the gravitational path integral. Following \cite{holosource1, holosource2, holosource3}, we assume that each microstate $|\psi_j\rangle_C$ of the black hole is prepared, up to some normalization $\sqrt{\mathcal{N}}$, by a Euclidean path integral with a source inserted at Euclidean time $-\infty$. For ${}_C\langle \psi_i | \psi_j \rangle_C$, we glue the path integral preparation of ${}_C\langle \psi_i |$ with that of ${}_C\langle \psi_j |$, i.e. carry out the path integral shown in Figure~\ref{fig:innprod}. Once again, to the extent that the microstates $|\psi_i\rangle_C$ are orthogonal, we should get
\begin{equation}
{}_C\langle \psi_i | \psi_j \rangle_C = \delta_{ij}, \label{orthogonality}
\end{equation}
which leads to the conclusion that $S(R) = S_{\rm bulk}({\rm radiation})$. Although we did not use the path integral to compute (\ref{orthogonality}), we exploit it to extract the normalization $\mathcal{N}$ for later use.

\begin{figure}[tbp]
\centering 
\includegraphics[width=.8\textwidth]{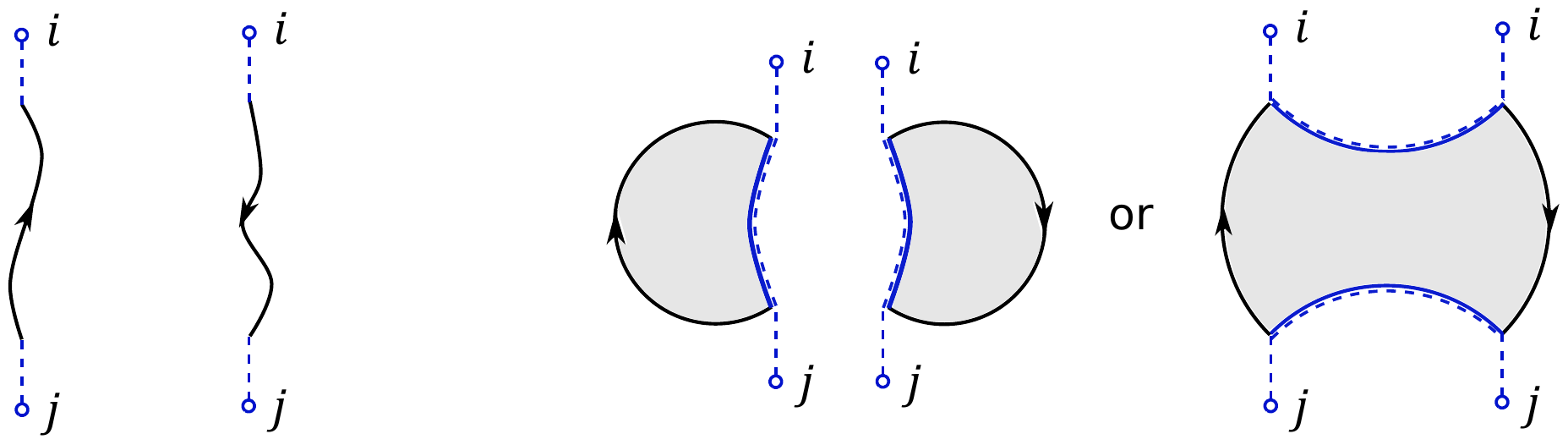}
\hfill
\caption{\label{fig:gpi} Left: The boundary conditions for the gravitational path integral, which computes $\big({}_C\langle \psi_i | \psi_j \rangle_C\big)\big({}_C\langle \psi_i | \psi_j \rangle_C\big)^*$. Right: Two saddles contribute, one disconnected and one connected. The figures are reproduced with permission from \cite{westcoast}.}
\end{figure}

In a crucial step, we do not simply substitute equation~(\ref{orthogonality}) in computing~(\ref{squareoverlap}). Rather, we set it up as a separate gravitational path integral, with boundary conditions shown in Figure~\ref{fig:gpi}. We now have two saddle point geometries:
\begin{enumerate}
\item The disconnected saddle, which is two copies of Figure~\ref{fig:innprod}. This saddle's contribution to $| {}_C\langle \psi_i | \psi_j \rangle_C |^2$ is:
\begin{equation}
\big({}_C\langle \psi_i | \psi_j \rangle_C\big)\big({}_C\langle \psi_i | \psi_j \rangle_C\big)^* 
\supset 
\frac{e^{-I_{2,{\rm disconnected}}}\, \delta_{ij} \delta_{ji}}{\mathcal{N}^2} = \delta_{ij}
\label{disconncontrib}
\end{equation}
We emphasize the correct normalization of the gravitational path integral for use below. This saddle is fully consistent with~(\ref{orthogonality}) so---used alone---it reproduces $S(R) = S_{\rm bulk}({\rm radiation})$.
\item The connected saddle, shown in the rightmost panel of Figure~\ref{fig:gpi}. It connects the two replicas of $\rho_R$ with a joint bulk, which explains the name \emph{replica wormhole}. Its contribution to $| {}_C\langle \psi_i | \psi_j \rangle_C |^2$ is:
\begin{equation}
\big({}_C\langle \psi_i | \psi_j \rangle_C\big)\big({}_C\langle \psi_i | \psi_j \rangle_C\big)^*
\supset 
\frac{e^{-I_{2,{\rm connected}}}}{\mathcal{N}^2} \approx e^{-A_{\rm horizon}/4G\hbar}
\label{rwcontrib}
\end{equation}
The crucial, qualitative feature of this expression is that it lacks a Kronecker delta, so $i$ and $j$ can be different. As we highlight below, it is also inconsistent with~(\ref{orthogonality}). This seems like bad news, but it is also good---because it reproduces the Page curve. 
\end{enumerate}

At general replica index $n$, we need similar expressions for
\begin{equation}
\big({}_C\langle \psi_i | \psi_j \rangle_C\big) 
\big({}_C\langle \psi_j | \psi_m \rangle_C\big) 
\big({}_C\langle \psi_m | \psi_i \rangle_C\big)
\label{highernbc}
\end{equation} 
and so on. For each of them, there is a disconnected saddle reproducing~(\ref{orthogonality}), and a connected replica wormhole. Collecting their contributions, we have:
\begin{equation}
{\rm Tr}(\rho_R^n) 
= k \cdot \frac{e^{-I_{n,{\rm disconnected}}}}{k^n \mathcal{N}^n} 
+ k^n \cdot \frac{e^{-I_{n,{\rm connected}}}}{k^n \mathcal{N}^n} 
= k^{1-n} + e^{(1-n) A_{\rm horizon}/4G\hbar}
\label{tracerhon}
\end{equation}
Note how the disconnected saddle contributes only $k$ terms because of the Kronecker deltas in (\ref{disconncontrib}). The disconnected saddle, on the other hand, has $k^n$ contributing choices of $i,j,m,\ldots$ in expressions like (\ref{highernbc}). Using equation~(\ref{defrenyi}), (\ref{tracerhon}) gives:
\begin{equation}
S(R) = \lim_{n \to 1^+} \frac{1}{1-n} \log {\rm Tr}(\rho_R^n) \approx 
\min\{\log k,\, A_{\rm horizon}/4G\hbar \}
\label{srholimit}
\end{equation}
This is, of course, the desired Page curve dependence. The exponentials with the smaller bases dominate because the exponents are negative since the limit in (\ref{srholimit}) approaches 1 from above.

We have not yet justified equations~(\ref{tracerhon}) and (\ref{rwcontrib}). For the disconnected contribution, $e^{-I_{n,{\rm disconnected}}} = \mathcal{N}^n$ is a tautology, but what about the contribution of the replica wormhole? Note that $\mathcal{N} = e^{-I_{1, {\rm connected}}}$, so we must have
\begin{equation}
I_{n, {\rm connected}} \approx -A_{\rm horizon}/4G\hbar
\label{ndepend}
\end{equation}
independently of $n$. This can be rationalized at various levels of rigor. A valid, if somewhat circular, argument is that \emph{if} the disconnected saddle did not exist, the connected saddle would have to reproduce $S(R) = {A}_{\rm horizon}/4G\hbar$. From this point of view, so long as the connected saddle exists at all, its action must obey~(\ref{ndepend}) at least near $n=1$ or else the proofs of the Ryu-Takayanagi proposal and generalizations would fail. Another argument \cite{donstalk} is that our calculation is being done in the microcanonical ensemble, in which the $\beta E$ term drops out from the usual relation
\begin{equation}
e^{-I} = \mathcal{Z} = e^{-\beta F} = e^{-\beta E + S}.
\end{equation}
Finally, we may follow \cite{westcoast, eastcoast} and do the explicit calculation in JT gravity, which shows that the connected saddles exist and obey~(\ref{ndepend}). 

But the neatest argument is topological \cite{westcoast}. Suppose we are only interested in the existence and qualitative mechanism of the transition between the two behaviors in (\ref{srholimit}). At that level of abstraction, it is sufficient to consider a topological theory of gravity with action $I = -S_0 \chi$, where $\chi$ is the Euler character. Because each connected saddle has $\chi = 1$, this gives the same answer as (\ref{ndepend}) except for the replacement $A_{\rm horizon}/4G\hbar \to S_0$. This exercise shows that the Page curve reflects a topological transition between the disconnected saddle with $\chi = n$ and the connected saddle with $\chi = 1$. In fact, JT gravity used in \cite{westcoast, eastcoast} contains both a topological term $-S_0 \chi$ and other, dynamical terms. The entire effect of the dynamics, insofar as it affects the Page curve, is to `renormalize' the topological coupling $S_0 \to A_{\rm horizon}/4G\hbar$. 

\paragraph{Is gravity dual to an ensemble of theories?}
The most perplexing thing about replica wormholes is what they say about the inner products of black hole microstates. Collecting contributions from (\ref{disconncontrib}) and (\ref{rwcontrib}), we have:
\begin{equation}
| {}_C\langle \psi_i | \psi_j \rangle_C |^2 = \delta_{ij} + e^{-A_{\rm horizon}/4G\hbar}
\label{overlapsq}
\end{equation}
This equation is inconsistent with (\ref{orthogonality}), although the inconsistency is exponentially small in the entropy of the black hole. This is why, throughout this subsection, we have been guarding our statements with provisos concerning the orthogonality of states $|\psi_i\rangle_C$. 

The bold idea of \cite{westcoast} is to view the gravitational path integral as computing averages in a statistical \emph{ensemble of microscopic theories} parameterized by a random variable $R_{ij}$. For definiteness, say $\overline{R_{ij}} = 0$ and $\overline{R_{ij}^2} = 1$. If, as a function of $R_{ij}$, a single microscopic theory has
\begin{equation}
{}_C\langle \psi_i | \psi_j \rangle_C = \delta_{ij} + R_{ij}\, e^{-A_{\rm horizon}/8G\hbar}
\end{equation}
then the relevant ensemble averages become:
\begin{equation}
\overline{{}_C\langle \psi_i | \psi_j \rangle_C} = \delta_{ij}
\qquad {\rm and} \qquad
\overline{| {}_C\langle \psi_i | \psi_j \rangle_C|^2} = \delta_{ij} + e^{-A_{\rm horizon}/4G\hbar}
\end{equation}
In this way, an ensemble of microscopic theories might reconcile equations~(\ref{orthogonality}) and (\ref{overlapsq}).

The idea that the gravitational path integral reflects the physics of an ensemble of theories is not entirely new. Earlier works with a similar theme, including \cite{coleman, giddingsstrominger}, were partly motivated by the observation that wormhole geometries can be sliced in various ways, with a non-constant number of asymptotic boundaries (Hilbert spaces). In recent years, much attention has been paid to the Sachdev-Ye-Kitaev (SYK) model \cite{syk1, kitaevtalk, syk3} whose low energy dynamics---the so-called Schwarzian theory---is also a limit of JT gravity \cite{kitaevtalk, jttoschwarzian, verlindeschwarzian} in which boundaries of spacetime are pushed to asymptotic infinity. As such, the SYK model is a microscopic theory of two-dimensional gravity. It is also, famously, defined as an ensemble of theories.

It is too early to conclude that gravity is (dual to) a statistical ensemble of theories. It is too early to say that an ensemble of theories is \emph{the} solution to the black hole information paradox. But the way in which considerations of quantum entanglement and black hole thermodynamics, the idea of an emergent spacetime, quantum field theory in curved space (proposal~\ref{qrt}), the black hole information paradox, and the gravitational path integral, have all converged at a single nexus, is unprecedented and very exciting!

\section*{Acknowledgments}
We thank Pawe{\l} Caputa, Yiming Chen, Patrick Hayden, Xiaoliang Qi, and G{\'a}bor S{\'a}rosi for explanations of their work during the writing of this review, and many mentors, collaborators and colleagues for multiple projects and discussions during which we learned this material. This manuscript was prepared using an adapted version of the JHEP template.


\end{document}